\documentclass
  [a4paper,UKenglish,cleveref,autoref,numberwithinsect]
  {lipics-v2021}

\title{Automata-Theoretic Characterisations of Branching-Time Temporal Logics}
\titlerunning{Automata-Theoretic Characterisations of Branching-Time Temporal
  Logics}

\author{Massimo Benerecetti}{Universit\`a di Napoli Federico II}
  {massimo.benerecetti@unina.it}{0000-0003-4664-6061}{}
\author{Laura Bozzelli}{Universit\`a di Napoli Federico II}
  {laura.bozzelli@unina.it}{0009-0004-8555-8229}{}
\author{Fabio Mogavero}{Universit\`a di Napoli Federico II}
  {fabio.mogavero@unina.it}{0000-0002-5140-5783}{}
\author{Adriano Peron}{Universit\`a di Trieste}
  {adriano.peron@units.it}{0000-0002-7111-3171}{}

\authorrunning{M. Benerecetti, L. Bozzelli, F. Mogavero, A. Peron}

\Copyright{M. Benerecetti, L. Bozzelli, F. Mogavero, A. Peron}



\acknowledgements{M. Benerecetti, F. Mogavero, and A. Peron are members of the
  Gruppo Nazionale Calcolo Scientifico-Istituto Nazionale di Alta Matematica
  (GNCS-INdAM). This work has been partially supported by the GNCS 2024 project
  ``Certificazione, monitoraggio, ed interpretabilit\`a in sistemi di
  intelligenza artificiale''.}

\usepackage[noamsthm,nothmtls,noenmtls,nofnttls]{fmocdmac}

\AtEndPreamble {

}

\DeclareRobustCommand{\ELTL}
  {{\txtname{E}}\LTL}
\DeclareRobustCommand{\STL}
  {{\txtname{S}}\TL}
\DeclareRobustCommand{\ECTLS}
  {{\txtname{E}}\CTLS}
\DeclareRobustCommand{\fin}
  {\normalfont{f}}
\DeclareRobustCommand{\cf}
  {\normalfont{cf}}

\newcommand{\langfun}{L}
\cmdmthfun{Lang}[\langfun]
\def\sttsym{q}
\def\sttset{Q}

\cmdtxtoparname{HTA}
\cmdtxtoparname{GTA}
\cmdtxtoparname{FTA}
\cmdtxtoparname{HGTA}
\cmdtxtoparname{HFTA}
\newcommand{\HGTAcf}{\HGTA[\cf]}

\cmdtxtoparname{LDL}

\cmdtxtoparname{CDL}
\cmdtxtoparname{CCDL}
\cmdtxtoparname{LCDL}
\cmdtxtoparname{FOE}
\cmdtxtoparname{MCL}
\cmdtxtoparname{OMCL}

\cmdtxtoparname{CCTLS}[CCTL*]
\cmdtxtoparname{CECTLS}[CECTL*]
\cmdtxtoparname{WCCTLS}[CCTL*]

\cmdmthset{Mod}[M]
\cmdmthset{PMod}[PM]

\cmdmthset{Hom}

\cmdmthfun{src}

\newcommand{\DefinedAs}{\defeq}
\newcommand{\tpl}[1]{\langle #1\rangle}

\newcommand{\BI}{{\mthsym{BI}}}
\newcommand{\Sing}{{\mthsym{sing}}}
\newcommand{\Empty}{{\mthsym{empty}}}
\newcommand{\Diff}{{\mthsym{diff}}}
\newcommand{\Type}{{\mthsym{t}}}
\newcommand{\TypeSet}{{\mthsym{T}}}
\newcommand{\Child}{{\mthsym{child}}}

\newcommand{\Chain}{{\mthsym{chain}}}
\newcommand{\Root}{{\mthsym{root}}}
\newcommand{\REG}{{\mthsym{REG}}}
\newcommand{\Sim}{{\mthsym{Sim}}}
\newcommand{\SEP}{{\mthsym{PATH}}}
\newcommand{\Comp}{{\mthsym{Comp}}}
\newcommand{\SUCC}{{\mthsym{Succ}}}
\newcommand{\Part}{{\mthsym{Part}}}
\newcommand{\Init}{{\mthsym{Init}}}
\newcommand{\Acc}{{\mthsym{Acc}}}
\newcommand{\Inf}{{\mthsym{Inf}}}
\newcommand{\Ext}{{\mthsym{ex}}}
\newcommand{\Strong}{{\mthsym{c}}}

\newcommand{\Lab}{\mathit{Lab}}

\newcommand{\Even}{\mathit{even}}

\newcommand{\Dom}{\mathit{Dom}}

\newcommand{\Con}{\mathsf{Con}}
\newcommand{\Dis}{\mathsf{Dis}}
\newcommand{\Ran}{\mathit{Range}}

\newcommand{\Pow}{Pow}
\newcommand{\Atoms}{\mathsf{Atoms}}
\newcommand{\Family}{\mathsf{H}}
\newcommand{\atom}{\mathsf{atom}}
\newcommand{\run}{\mathsf{r}}

\newcommand{\widt}{\mathit{width}}
\newcommand{\Buchi}{\mathsf{B}}
\newcommand{\coBuchi}{\mathsf{coB}}
\newcommand{\Test}{\tau}

\newcommand{\LT}{\mathcal{T}}
\newcommand{\Prop}{\APSet}
\newcommand{\Val}{\mathsf{V}}
\newcommand{\Var}{\VarSet}
\newcommand{\true}{\top}
\newcommand{\false}{\bot}

\newcommand{\InitState}[1]{\sttElm_{#1,I}}

\newcommand{\FOEOne}[1]{\FOE^+_{1}(#1)}

\DeclareRobustCommand{\CTLStar}
  {\CTLS}

\DeclareRobustCommand{\CCTLStar}
  {{\txtname{C}}\CTLS}

\newcommand{\Until}{{\U}}
\newcommand{\Next}{{\X}}
\newcommand{\Always}{{\G}}

\newcommand{\EQ}{{\E}}

\newcommand{\DC}{{\DSym}}
\newcommand{\Seq}[1]{\text{$\langle{#1}\rangle$}}
\newcommand{\USeq}[1]{\text{$[#1]$}}

\usepackage{tikz,tikz-qtree}
\usetikzlibrary{arrows,positioning}

\usepackage{mathtools}

\usepackage{amsmath,amsfonts,amssymb}
\usepackage{thmtools,thm-restate}

\nolinenumbers

\hypersetup {
  pdftitle  = {Automata-Theoretic Characterisations of Branching-Time Temporal Logics},
  pdfauthor = {M. Benerecetti, L. Bozzelli, F. Mogavero, A. Peron}
}

\begin{document}

  \maketitle



\begin{abstract}

\emph{Characterisations theorems} serve as important tools in model theory and
can be used to assess and compare the expressive power of temporal languages
used for the specification and verification of properties in formal methods.
While complete connections have been established for the linear-time case
between temporal logics, predicate logics, algebraic models, and automata,
the situation in the branching-time case remains considerably more fragmented.
In this work, we provide an \emph{automata-theoretic characterisation} of some
important branching-time temporal logics, namely \CTLS and \ECTLS interpreted on
arbitrary-branching trees, by identifying two variants of \emph{Hesitant Tree
Automata} that are proved equivalent to those logics.
The characterisations also apply to \emph{Monadic Path Logic} and the
bisimulation-invariant fragment of \emph{Monadic Chain Logic}, again interpreted
over trees.
These results widen the characterisation landscape of the branching-time case
and solve a forty-year-old open question.

\end{abstract}




\section{Introduction}

\emph{Temporal logics}~\cite{MP92} play a pivotal role in the \emph{formal
verification} of complex systems~\cite{MP95}.
Serving as \emph{specification languages}, they provide a framework to express
and reason about time-dependent properties, capturing the intricate behaviours
and interactions of system components over time.
Commonly, these languages are classified into two categories: \emph{linear-time
logics}, which emphasise properties spanning the entirety of a computation, and
\emph{branching-time logics}, specifically tailored to address the
non-deterministic and concurrent nature of behaviours.
Well-established representatives of the former include \emph{Linear-Time
Temporal Logic} (\LTL)~\cite{Pnu77,Pnu81}, its full $\omega$-regular extension
\ELTL~\cite{Wol83}, and the finite-horizon variant \LTL[\fin]~\cite{GV13}.
Important members of the second category, instead, belong to the families of
\emph{Dynamic Logics}~\cite{FL79,HKT00} and \emph{Computation Tree Logics},
including \CTL~\cite{EC81,CES83,EH85,CES86,EC82}, \CTLS~\cite{EH83,EH86},
\ECTLS~\cite{VW83}, \CTLS[\fin]~\cite{VS85}, and \ECTLS[\fin]~\cite{Tho87}.
Additionally, more expressive but lower-level languages, like \MC~\cite{Koz83},
have been considered, which suitably extend classic modal logic with monadic
fix-point operators, contributing to the rich tapestry of specification
languages in the field of formal verification and synthesis.
\\\indent
The semantics of these temporal logics are typically formalised, at the
meta-level, through various flavour of \emph{predicate logic}, frequently
\emph{First-Order Logic} (\FO) or \emph{Second-Order Logic} (\SO), interpreted
over either \emph{linearly-ordered structures}, such as finite and infinite
words~\cite{PP04}, or \emph{branching structures}, like Kripke
structures~\cite{Kri63}, labelled transition systems~\cite{Kel76}, and their
tree unwindings.
In tandem with this, the rich body of literature on automata-theoretic
techniques~\cite{Tho90} for words and trees, originated
from~\cite{Kle56,Moo56,Ner58,RS59}, has proven invaluable to provide effective
technical tools for the solution of related
\emph{model-checking}~\cite{CGP02,BK08,VW86,KVW00,EJS01},
\emph{satisfiability}~\cite{VW86,EJ91,Var98,Boj03,KVW00}, and
\emph{synthesis}~\cite{Chu63,PR89,Ros92} decision problems.
Predicate logics and automata theory offer, in addition, a rich and coherent
arsenal of tools to evaluate and compare the expressive power, as well as the
computational properties, of temporal languages, as witnessed by numerous
\emph{characterisation theorems}.
These results provide a dual perspective on the topic, which enhance our ability
to navigate the intricate landscape of language fragments and allow us to assess
their pros (\eg, elementary complexity of decision problems) and cons (\eg,
limitations on the expressive power).
\\\indent
The initial seminal result in this context is Kamp's
theorem~\cite{Kam68,GPSS80,Rab12,Rab14}, which establishes the equivalence of
\LTL and \FO over infinite words.
The result also extends to \LTL[\fin] and \FO on finite words~\cite{DG08a}.
A direct link has been drawn between \emph{\FO-definability} and
\emph{recognition} by \emph{counter-free finite-state automata}, in both the
finite~\cite{MP71} and infinite~\cite{Lad77,Tho79,Tho81,PP86} cases, by means of
the notions of \emph{star-free language}, \emph{aperiodic language}, and
\emph{aperiodic syntactic monoid} (see~\cite{Sch65}, for finite words,
and~\cite{Per84}, for the infinite ones).
Together these results provide a complete characterisation of the expressive
power of \LTL and \LTL[\fin] in terms of predicate logics and automata.
A parallel correspondence exists between \ELTL and \ELTL[\fin], the
\emph{Monadic Second-Order Logic} (\MSO) and its \emph{weak
(finite-quantification) fragment} (\WMSO), and regular automata on infinite and
finite words.
Notably, the equivalence between \WMSO and regular
automata~\cite{Buc60,Elg61,Tra66}, followed by the equivalence between \MSO and
$\omega$-regular automata~\cite{Buc62,Buc66,McN66,Cho74}, stands among the first
results connecting the two fields of model theory and automata theory.
\\\indent
The landscape for branching-time temporal logics is considerably more intricate,
due to the complex topology of the models and additional factors, such as
\emph{bisimulation invariance}~\cite{Ben77} and \emph{counting
quantifiers}~\cite{Fin72}, and it is not as clear and complete as the
linear-time counterpart.
A significant milestone in this setting is the full correspondence between \MC,
the \emph{bisimulation-invariant} fragment of \MSO interpreted over trees, and
\emph{(Symmetric) Alternating Parity Tree Automata}~\cite{JW96}.
This result generalises the already known connection between the latter two
formalisms~\cite{Rab69}.
Another noteworthy connection has been shown to exist between the
\emph{alternation-free} fragment of \MC (\AFMC), the bisimulation-invariant
fragments of \WMSO over bounded-branching trees, and \emph{(Symmetric)
Alternating Weak Tree Automata}~\cite{AN92,JL04}
(see~\cite{FVZ13,CFVZ14,CFVZ20,CFVZ22}, for the unbounded-branching case), which
extends previous partial results~\cite{KV98a,Rab70}.
The above equivalences lift also to the general case, by incorporating counting
quantifiers into the temporal logics~\cite{JL04,Jan05}.
The scenario in other cases appears significantly more fragmented.
In recent developments, the equivalence between \CTL and \emph{(Symmetric)
Hesitant Linear Tree Automata}~\cite{BS18} was proved.
Nonetheless, as of today, no corresponding fragment of \MSO has been identified.
By contrast, several variants of \CTLS have been linked to the \emph{path} and
\emph{chain} fragments of \MSO since the eighties, although no automata
characterisation has been provided thus far.
For instance, it was shown in~\cite{HT87} that, on binary trees, \CTLS is
equivalent to \emph{Monadic Path Logic} (\MPL)~\cite{GS85}.
Similar correspondences have been established in~\cite{Tho87} for \CTLS[\fin],
\ECTLS, and \ECTLS[\fin], which equate, respectively, to \FO, \emph{Monadic
Chain Logic} (\MCL), and its weak fragment (\WMCL).
The result concerning \CTLS was later extended to arbitrary-branching trees,
addressing both bisimulation-invariance~\cite{MR99} and counting
quantifiers~\cite{MR03}.
As far as we know, no similar results are available for the other three logics.
Finally, the recently introduced \emph{Monadic Tree Logic} (\MTL)~\cite{BBMP23}
together with its variants have yet to find a correspondence either with
temporal logics or with automata.
\\\indent
The objective of this work is to provide an \emph{automata-theoretic
characterisation} of \CTLS and \ECTLS, by identifying two specific classes of
alternating tree automata that are expressively equivalent to those logics (the
used technique extends seamlessly to the finite-horizon variants).
A first result is the proof of the equivalence of the \emph{symmetric variant}
of classic ranked \emph{Hesitant Tree Automata} (\HTA)~\cite{KVW00} with both
\ECTLS and the bisimulation-invariant fragment of \MCL.
To this end, for technical convenience, we employ two intermediate formalisms.
On the one hand, to prove the equivalence between \HTA and \ECTLS, we use a
\emph{syntactic variant} of \ECTLS, called \emph{Computation Dynamic Logic}
(\CDL), alongside its counting version (\CCDL).
In \ECTLS temporal operators are specified by means of right-linear grammars,
while \CDL uses finite automata on finite words for the same purpose
incorporated into the dynamic modalities.
Moreover, while the path subformulae in \ECTLS are part of the alphabet of the
grammar, in \CDL they are specified by means of a testing function over the set
of states of the automaton.
It is straightforward to move from one formalism to the other by means of a
linear-time translation.
This logic essentially lifts to the branching-time realm the \emph{Linear
Dynamic Logic} (\LDL) proposed in~\cite{GV13,WZ18}.
On the other hand, we consider a \emph{first-order extension}~\cite{Wal02} of
{\HTA}s (\HFTA) and show them equivalent to \MCL by proving a closure property
under \emph{chain projections}.
The final result, then, follows from the equivalence between {\HTA}s and the
bisimulation-invariant fragment of \HFTA.
As a second result, we first identify the \emph{graded extension} of {\HTA}s
(\HGTA), together with its counter-free restriction (\HGTA[\normalfont{cf}]),
and then prove their equivalence with \CCDL and \CCTLS, respectively.
While for the definition of \HGTA the standard notion of counting modalities
smoothly applies, introducing \HGTA[\normalfont{cf}] proves quite more
intricate.
We show, indeed, that a naive application of counter-freeness in the context of
tree-automata leads to a class of languages that are not \CTLS definable.
To overcome this problem, we identify the crucial \emph{mutual-exclusion} property   
of a \HGTA that constrains the automaton branching-behaviours.
This property, together with counter-freeness of the automaton linear
behaviours, provides an apt definition of \HGTA[\normalfont{cf}], something
that was previously only hypothesised in~\cite{MR99,MR03}.
The above characterisations holds also under bisimulation-invariance
assumptions.
Specifically, \HTA[\normalfont{cf}] is equivalent to both \CTLS and the
bisimulation-invariant fragment of \MPL.
All these results, coupled with the algebraic characterisation of tree languages
provided in~\cite{Tho87}, brings the expressiveness landscape for branching-time
temporal logics to the same level as their linear-time counterpart, thus closing
a forty-year-old open question posed in~\cite{Tho84,Tho87}.




\section{Preliminaries}
\label{sec:prl}

Let  $\SetN$ be the set of natural numbers.
For $i,j\in\SetN$ with $i\leq j$, $[i,j]$ denotes
the set of natural numbers $k$ such that
$i\leq k\leq j$. For a  finite or infinite word $\rho$ over some alphabet, $|\rho|$ is the length of $\rho$
($|\rho|=\omega$ if $\rho$ is infinite) and for all $0\leq i<|\rho|$, $\rho(i)$ is the
$(i+1)$-th letter of $\rho$. \vspace{0.1cm}

\noindent \textbf{Kripke Trees and Tree Languages.}
 Given a non-empty set of directions $\DSet$, a tree $\TSet$ (with set of directions in $\DSet$) is a non-empty subset of $\DSet^{*}$ which is prefix closed (i.e., for each
$w\cdot d\in \TSet$ with $d\in\DSet$,
$w\in \TSet$). Elements of $\TSet$ are called nodes and
the empty word $\varepsilon$ is the root of $\TSet$.  For $w\in \TSet$, a \emph{child} of
$w$ in $\TSet$ is a node in $\TSet$ of the form $w\cdot d$ for some $d\in \DSet$.
For $w\in\TSet$, the \emph{subtree of $\TSet$ rooted at  node $w$} is the tree
consisting of the nodes of the form $w'$ such that $w\cdot w'\in \TSet$.
A \emph{subtree of $\TSet$} is a tree $\TSet'$ such that for some $w\in \TSet$,
 $\TSet'$ is a subset of the subtree of $\TSet$ rooted at $w$.
  A  \emph{path} of $\TSet$ is a  subtree $\pi$ of $\TSet$  which is totally ordered by the child-relation (i.e., each node of $\pi$ has at most one child in $\pi$).  In the following, a path $\pi$ of $\TSet$ is also seen as a word over $\TSet$ in accordance to the total ordering in $\pi$ induced by the child relation.
 A \emph{chain} of $\TSet$ is a subset of a path of $\TSet$. A tree is \emph{non-blocking} if each node has some child.  A non-blocking tree $\TSet$ is infinite, and
maximal paths in $\TSet$ are infinite as well.

  For an alphabet $\Sigma$, a
$\Sigma$-labelled tree is a pair $(\TSet, \Lab)$ consisting of a tree and a
labelling $\Lab:\TSet \mapsto \Sigma$ assigning to each node in $\TSet$ a symbol in
$\Sigma$. A \emph{tree-language} over $\Sigma$ is a set of $\Sigma$-labeled trees.
In this paper, we consider formalisms whose specifications denote
\emph{tree-languages} over a given alphabet $\Sigma$.
For the easy of presentation, we assume that the labeled trees
of a tree-language are non-blocking. All the results of this paper can be easily adapted
to the general case, where the non-blocking assumption is relaxed.
 For a finite  set $\Prop$ of atomic propositions, a \emph{Kripke tree} over $\Prop$ is a  non-blocking $2^{\Prop}$-labelled tree.
\vspace{0.1cm}

\noindent \textbf{Automata over Infinite and Finite Words.}
We first recall the class of parity nondeterministic automata on infinite words (parity \NWA for short)
which are tuples $\AutName = \tuple {\Sigma} {\SttSet} {\trnFun} {\isttElm} {\Omega}$,   where
$\Sigma$ is a finite input alphabet, $\SttSet$ is a finite set of states,  $\trnFun: \SttSet\times \Sigma \mapsto 2^{\SttSet}$
is the transition function,
 $\isttElm\in \SttSet$ is an initial state,  and $\Omega: \SttSet \mapsto  \SetN$  is a parity acceptance condition over $\SttSet$ assigning to each state a natural number (color).
Given a word $\rho$ over $\Sigma$, a \emph{path} of $\AutName$ over $\rho$ is a word $\pi$ over $\SttSet$ of length $|\rho|+1$ ($|\rho|+1$ is $\omega$ if $\rho$ is infinite) such that
$\pi(i+1)\in \trnFun(\pi(i),\rho(i))$ for all $0\leq i <|\rho|$. A \emph{run} over $\rho$ is a path over $\rho$ starting at the initial state.
  The \NWA $\AutName$ is \emph{counter-free}  if for all $n>0$, states $\sttElm\in\SttSet$ and finite words $\rho$ over $\Sigma$, the following holds: if there is
  a path from $\sttElm$ to $\sttElm$ over $\rho^{n}$, then there is also a path from $\sttElm$ to $\sttElm$ over $\rho$.

A run $\pi$ of $\AutName$
over an infinite word $\rho$ is \emph{accepting} if the highest color of the states appearing infinitely often along $\pi$ is even. The $\omega$-language $\LangFun(\AutName)$ accepted by $\AutName$ is the set of infinite words $\rho$ over $\Sigma$ such that there is an accepting run $\pi$ of $\AutName$
over $\rho$.

A parity acceptance condition  $\Omega: \SttSet \mapsto  \SetN$ is a \emph{B\"{u}chi} (resp., \emph{coB\"{u}chi}) condition
if there is an even (resp., odd) color $n\in\SetN$ such that $\Omega(\SttSet)\subseteq \{n-1,n\}$.
A \emph{B\"{u}chi} (resp., \emph{coB\"{u}chi}) \NWA is a parity \NWA whose acceptance condition is B\"{u}chi (resp., coB\"{u}chi).

We also consider \NWA over finite words (\NWA[\fin] for short) 
which are defined as parity \NWA but the parity condition $\Omega$ is replaced with a set $\FSet\subseteq \SttSet$ of accepting states.
A run $\pi$ over a finite word is \emph{accepting} if its last state is accepting.




\section{Branching-Time Temporal Logics}
\label{sec:DynamicLogic}

In this section, we recall syntax and semantics of
Counting-$\CTLStar$ ($\CCTLStar$ for short)~\cite{MR03}, which extends the
classic branching-time temporal logic \CTLS~\cite{EH86} with counting operators.
Moreover, we introduce a novel branching-time temporal logic more
expressive than \CCTLStar, called \emph{Counting Computation Dynamic Logic} (\CCDL for short). \CCDL
can be viewed as a branching-time extension of Linear Dynamic Logic
(\LDL)~\cite{GV13}. However, unlike \LDL, we consider \NWA[\fin] over finite
words, instead of regular expressions, as the building blocks of formulae. This
approach is similar to the one adopted in~\cite{WZ18} for Visibly Linear
Dynamic Logic, a context-free extension of \LTL.\vspace{0.1cm}

\noindent \textbf{The Logic \CCTLStar.} The syntax of $\CCTLStar$ is given by
specifying inductively the set of
\emph{state formulae} $\varphi$ and the set of \emph{path formulae} $\psi$ over
a given finite set $\Prop$ of atomic propositions:\vspace{0.1cm}

$
\begin{array}{l}
	\varphi ::= \top \ | \ p \ | \ \neg \varphi \ | \ \varphi \wedge
	\varphi  \ | \ \EQ
	\psi \ | \ \DC^{n}
	\varphi\\
	\psi ::=   \varphi \ | \ \neg \psi \ | \ \psi \wedge
	\psi \ | \ \Next \psi\ | \ \psi \Until \psi
\end{array}
$\vspace{0.1cm}

\noindent where $p\in \Prop$, $\Next$ and $\Until$ are the standard ``next" and ``until"
temporal modalities, $\EQ$ is the existential path quantifier, and $\DC^{n}$ is
the counting operator with $n\in\SetN\setminus \{0\}$.   The language of $\CCTLStar$ consists of
the state formulae of $\CCTLStar$. Standard \CTLStar is the fragment of $\CCTLStar$ where the counting operators
$\DC^{n}$ with $n>1$ are disallowed, and standard \LTL~\cite{Pnu77} corresponds to
 the set of path formulae of $\CTLStar$ where the path quantifiers are disallowed.

Given a Kripke tree $\LT = (\TSet,\Lab)$ (over $\Prop$), a node $w$ of $\TSet$,
an infinite path $\pi$ of $\TSet$, and $0\leq i<|\pi|$, the satisfaction
relations $(\LT,w)\models \varphi$, for a state formula $\varphi$, (meaning that
$\varphi$ holds at node $w$ of $\LT$), and $(\LT,\pi,i)\models \psi$, for a path
formula $\psi$, (meaning that $\psi$ holds at position $i$ of the path $\pi$ in
$\LT$) are defined as 
usual:\vspace{0.2cm}

\noindent $ \begin{array}{ll}
	(\LT,w)\models p & \Leftrightarrow p\in \Lab(w);\\
	(\LT,w)\models \EQ \psi & \Leftrightarrow (\LT,\pi,0)\models \psi \text{
    for some infinite path $\pi$ of $T$ starting at node $w$};\\
	(\LT,w)\models \DC ^{n} \varphi & \text{$\Leftrightarrow$ there are at least $n$ distinct children $w'$ of $w$ in $\TSet$ 
    s.t. }(\LT,w')\models \varphi;\\
	(\LT,\pi,i)\models \varphi &  \Leftrightarrow (\LT,\pi(i))\models \varphi;\\
	(\LT,\pi,i)\models \Next\psi &  \Leftrightarrow (\LT,\pi,i+1)\models \psi;\\
	(\LT,\pi,i)\models \psi_1\Until \psi_2 & \Leftrightarrow \text{for some $j\geq
    i$: }
	(\LT,\pi,j)\models\psi_2 \text{ and } (\LT,\pi,k)\models\psi_1 \text{ for all } i\leq k<j.
\end{array}
$\vspace{0.1cm}

\noindent Note that $\DC^1\varphi$ corresponds to $\EQ\Next \varphi$. A Kripke tree $\LT$ satisfies (or is a model of) a state formula $\varphi$, written $\LT \models
\varphi$, if $\LT,\varepsilon\models \varphi$. The tree-language $\LangFun(\varphi)$ of $\varphi$ is the set of models of $\varphi$.
For an \LTL formula $\psi$ and an infinite word $\rho$ over $2^{\Prop}$, $\rho$ satisfies $\psi$, written $\rho\models \psi$, if $\LT_\rho \models \EQ\psi$, where $\LT_\rho$ is a trivial tree-encoding of $\rho$. For an
 \LTL formula $\psi$, 
 $\LangFun(\psi)$ denotes  the set of infinite words over $2^{\Prop}$ satisfying $\psi$.\vspace{0.2cm}

\noindent \textbf{The New Logic \CCDL.}
Like $\CCTLStar$, the syntax of $\CCDL$ is composed of \emph{state formulae} $\varphi$ and \emph{path formulae} $\psi$ over
a given finite set $\Prop$ of atomic propositions, defined as follows:\vspace{0.1cm} 

\noindent $
\begin{array}{l}
	\varphi ::= \top \ | \ p \ | \ \neg \varphi \ | \ \varphi \wedge
	\varphi  \ | \ \EQ
	\psi \ | \ \DC^{n}
	\varphi\\
	\psi ::=   \varphi \ | \ \neg \psi \ | \ \psi \wedge
	\psi \ | \ \Seq{\AutName}\psi  
\end{array}
$\vspace{0.1cm}

\noindent where $p\in \Prop$ and  $\Seq{\AutName}$ is the \emph{existential sequencing}  modality
applied to a \emph{testing \NWA[\fin]}  $\AutName$.  We define a \emph{testing} \NWA[\fin] $\AutName = \tuple {2^{\Prop}} {\SttSet} {\trnFun} {\isttElm} {\FSet}{\Test}$ as consisting of an \NWA[\fin] $\tuple {2^{\Prop}} {\SttSet} {\trnFun} {\isttElm} {\FSet}$ over finite words over $2^{\Prop}$ and a test function $\Test$ mapping states in $\SttSet$ to $\CCDL$ path formulae.
Intuitively, along an infinite path $\pi$ of a Kripke tree, the testing automaton accepts the labeling of a (possibly empty) infix $\pi(i)\ldots \pi(j-1)$ of $\pi$ if  the embedded \NWA[\fin] has an accepting run $\sttElm_i\ldots \sttElm_{j}$ over the labeling of such an infix so that, for each position $k\in [i,j]$, the formula 
$\Test(\sttElm_k)$ holds at position $k$ along $\pi$. A test function $\Test$ is \emph{trivial} if it maps each state to $\top$.
We also use the shorthand $\USeq{\AutName}\psi\DefinedAs \neg \Seq{\AutName}\neg\psi$ (\emph{universal sequencing} modality).
 The language of $\CCDL$ consists of
the state formulae of $\CCDL$. We also consider the \emph{bisimulation-invariant} fragment $\CDL$ of $\CCDL$ where the counting operators
$\DC^{n}$ with $n> 1$ are disallowed.
Given a Kripke tree  $\LT = (\TSet,\Lab)$,  an infinite path $\pi$ of $\TSet$, and $0\leq i<|\pi|$, the semantics of
modality $\Seq{\AutName}$
is defined as follows, where $\AutName = \tuple {2^{\Prop}} {\SttSet} {\trnFun} {\isttElm} {\FSet}{\Test}$:\vspace{0.1cm}

\noindent $ \begin{array}{ll}
	(\LT,\pi,i)\models \Seq{\AutName}\psi &  \Leftrightarrow \text{for some $j\geq i$, $(i,j)\in \RSet_{\AutName}(\LT,\pi)$ and $(\LT,\pi,j)\models \psi$}\\
%
\end{array}
$\vspace{0.1cm}

\noindent where $\RSet_{\AutName}(\LT,\pi)$ is the set of pairs $(i,j)$ with $j\geq i$ such that there is an accepting run
$\sttElm_i\ldots \sttElm_{j}$ of the \NWA[\fin] embedded in $\AutName$ over $\Lab(\pi(i))\ldots \Lab(\pi(j-1))$ and, for all
$k\in [i,j]$, it holds that
$(\LT,\pi,k)\models \Test(\sttElm_k)$.
The notions of a model and tree-language of a $\CCDL$ formula are defined as for $\CCTLStar$.\vspace{0.1cm}

\noindent \textbf{Embedding of \CCTLStar into \CCDL.} The logic \CCTLStar  can be easily embedded into \CCDL   as follows. Let $\AutName$ be the testing  \NWA[\fin] having trivial tests and accepting all and only the words of length $1$, and   for  \CCDL path formulae $\psi_1,\psi_2$,
 let $\AutName_{\psi_1,\psi_2}=\tuple {2^{\Prop}} {\{\sttElm_1,\sttElm_2\}} {\trnFun} {\sttElm_1} {\{\sttElm_2\}}{\Test}$  be   the testing   \NWA[\fin]
where, for all $a\in 2^{\Prop}$, $\trnFun(\sttElm_1,a)=  \{\sttElm_1,\sttElm_2\} $, $\trnFun(\sttElm_2,a)=  \emptyset$,
$\Test(\sttElm_1)=\psi_1$, and   $\Test(\sttElm_2)=\psi_2$. Then,  the  next and until formulae
$\Next\psi_1$ and $\psi_1\Until\psi_2$ can be expressed as follows: $\Next\psi_1\equiv \Seq{\AutName}\psi_1$ and $\psi_1\Until\psi_2\equiv   \psi_2\vee\Seq{\AutName_{\psi_1,\psi_2}}\true$.




\section{Alternating Tree Automata}
\label{sec:alternatingTreeAutomata}

In this section, we recall the class of parity \emph{alternating tree automata
with first-order constraints} (\FTA for short), introduced in~\cite{Wal02} to
provide an automata-theoretic characterization of \MSO interpreted on arbitrary
labeled trees.  Moreover, we also recall the class of \emph{graded alternating
tree automata} (\GTA for short), a subclass of \FTA, which was introduced
in~\cite{KSV02} and allows for expressing counting modal requirements on the
child relation of an input tree.  The transition relation of both \FTA and \GTA
is based on constraints on the set of states $\SttSet$ written as formulae in a
suitable language, called \emph{one-step logic}. The \emph{one-step
interpretations} of such formulae over $\SttSet$ are pairs $(\SSet,\ISet)$,
where $\SSet$ is an arbitrary (possibly infinite) non-empty set and $\ISet$ is a
mapping $\ISet: \SSet\mapsto 2^{\SttSet}$, assigning to each element of $\SSet$
a subset of $\SttSet$. Intuitively, the pair $(\SSet,\ISet)$ describes the local
behaviour of the automaton on reading a node $w$ of the input tree.  The set
$\SSet$ corresponds to the set of children of the current input node $w$ and,
for each $w'\in \SSet$, $\ISet(w')$ is the set of states associated with the
copies of the automaton which are sent to the child $w'$ of $w$.
 \vspace{0.1cm}


\noindent \textbf{One-Step Logic for \GTA}.
The  one-step relation of \GTA is specified by means of
 formulae $\theta$ of one-step positive graded modal logic over
$\SttSet$, we call \emph{graded $\SttSet$-constraints}, defined as:\vspace{0.1cm}

 $
\theta::= \true \mid \false  \mid   \theta \vee \theta \mid \theta \wedge \theta \mid
 \DMod[k]\alpha \mid \BMod[k]\alpha
$\vspace{0.1cm}

\noindent where $k \in \SetN\setminus \{0\}$ and $\alpha$ is a \emph{positive} Boolean formula over $\QSet$. The atomic formulae  $\DMod[k]\alpha$ and $\BMod[k]\alpha$ are called \emph{$\SttSet$-atoms}.
The atom $\DMod[1]\alpha$ (resp., $\BMod[1]\alpha$) is also denoted
by  $\DMod\alpha$ (resp., $\BMod\alpha$).
A formula $\theta$ is \emph{symmetric} if the atoms occurring in $\theta$ are of the form $\DMod\alpha$ or $\BMod\alpha$.

The satisfaction relation $(\SSet,\ISet)\models \theta$ for a one-step interpretation $(\SSet,\ISet)$ over $\SttSet$  is inductively defined as follows
(we omit the clauses for positive Boolean connectives which are standard):
\begin{itemize}
  \item $(\SSet,\ISet)\models \DMod[k]\alpha$ if $|\{s\in \SSet\mid  \ISet(s)\models \alpha  \}|\geq k$;
  \item $(\SSet,\ISet)\models \BMod[k]\alpha$ if $|\{s\in \SSet\mid \ISet(s)\not\models \alpha\}|< k$.
\end{itemize}
If $(\SSet,\ISet)\models \theta$, we say that $(\SSet,\ISet)$ is a model of $\theta$.
Intuitively, for an alternating automaton $\AutName$ with set of states $\SttSet$, the atom $\DMod[k]\alpha$  requires that at the current input node $w$, there are at least $k$ children  of  $w$ and, for each of such nodes $w'$, (**) there is a subset $\SttSet'\subseteq \SttSet$ satisfying $\alpha$ such that a copy of $\AutName$ is sent to node $w'$ in  state $\sttElm$, for each $\sttElm\in \SttSet'$.  For an
atom $\BMod[k]\alpha$, the previous condition~(**) is required to hold  for all but at most $k-1$  children $w'$ of $w$.
\vspace{0.1cm}

\noindent \textbf{One-Step Logic for \FTA}.
The one-step language $\FOEOne{\SttSet}$    of positive first-order formulae with equality and monadic predicates over $\SttSet$ and
first-order variables in $\Var_1$
is given by the sentences (formulae without free variables) generated by the following grammar:\vspace{0.1cm}

$
\theta::= \true \mid \false  \mid  \sttElm(x) \mid x=y \mid x\neq y \mid  \theta \vee \theta \mid \theta \wedge \theta \mid
 \exists x.\,\theta \mid \forall x.\,\theta
$\vspace{0.1cm}

\noindent where $\sttElm\in\SttSet$
and $x,y\in\Var_1$.
An $\FOEOne{\SttSet}$-sentence $\theta$ is called \emph{first-order $\SttSet$-constraint}; $\theta$ is \emph{symmetric}
if it does not contain equality and inequality atomic formulae.
In \FTA, these constraints allow formulae that refer to the children of a node of a tree 
by means of explicit first-order variables.

Given a one-step interpretation $(\SSet,\ISet)$ over $\SttSet$ and an assignment
 $\Val: \Var_1 \rightarrow \SSet$ of the first-order variables, the satisfaction relation
 $(\SSet,\ISet),\Val\models \theta$ is defined in a standard way.
For sentences $\theta$,
 this relation is independent of $\Val$, and we simply write
 $(\SSet,\ISet)\models \theta$.
  Note that  graded $\SttSet$-constraints can be trivially expressed in $\FOEOne{\SttSet}$, and
  first-order $\SttSet$-constraints $\theta$ are \emph{monotonic}, i.e., for all one-step interpretations $(\SSet,\ISet)$ and $(\SSet,\ISet')$ such that $\ISet(s)\subseteq \ISet'(s)$ for each $s\in\SSet$, it holds that
   $(\SSet,\ISet)\models \theta$ entails $(\SSet,\ISet')\models \theta$.
  A \emph{minimal model} of
   $\theta$ is a model $(\SSet,\ISet)$ of $\theta$ such that there is no model  $(\SSet,\ISet')$ of $\theta$ with
    $\ISet'\neq \ISet$ and $\ISet'(s)\subseteq  \ISet(s)$ for each $s\in\SSet$ .
   \vspace{0.1cm}

\noindent \textbf{Parity \GTA and Parity \FTA.} A \emph{parity}  \GTA
$\AutName$ is a tuple
$\AutName = \tuple {\Sigma} {\SttSet} {\trnFun} {\isttElm} {\Omega}$,   where
$\Sigma$, $\SttSet$, $\isttElm$, and $\Omega$ are defined as for parity \NWA, while the transition function $\trnFun$ is a mapping from $\SttSet\times \Sigma$ to the set
of graded $\SttSet$-constraints.
The set $\Atoms(\AutName)$
is the set of $\SttSet$-atoms occurring in the transition function of $\AutName$.
 Parity \FTA $\AutName = \tuple {\Sigma} {\SttSet} {\trnFun} {\isttElm} {\Omega}$ are defined similarly but the transition
function $\trnFun$ is of the form $\trnFun: \SttSet\times \Sigma \mapsto \FOEOne{\SttSet}$.
A \GTA (resp., \FTA) $\AutName = \tuple {\Sigma} {\SttSet} {\trnFun} {\isttElm} {\Omega}$ is \emph{symmetric} if for all $(\sttElm,a)\in\SttSet\times \Sigma$, the constraint $\trnFun(\sttElm,a)$ is symmetric.

\noindent   \GTA (resp., \FTA) $\AutName$ operate over non-blocking $\Sigma$-labeled trees $(\TSet, \Lab)$.
A run of $\AutName$ over 
$(\TSet, \Lab)$ is a   $(\SttSet\times \TSet)$-labeled  tree
$\run=(\TSet_\run,\Lab_\run)$, where each node  of $\TSet_\run$ labelled by $(\sttElm,w)$ describes a copy of $\AutName$ 
that is in state $\sttElm$ reading the node $w$ of
$\TSet$. Moreover, we require that:
\begin{itemize}
  \item $\Lab_\run(\varepsilon)=(\isttElm,\varepsilon)$ (initially, the automaton is in state $\isttElm$ reading the root 
  of the input $\TSet$);
  \item for each node $y\in \TSet_\run$ with $\Lab_\run(y)=(\sttElm,w)$ and
    denoted by $\SSet_w$ the set of children of node $w$ in the input $\TSet$,
    there is a one-step interpretation $(\SSet_w,\ISet)$ over $\SttSet$
    satisfying $\delta(\sttElm,\Lab(w))$ such that the set of labels associated
    with the children of $y$ in $\TSet_\run$ consists of the pairs
    $(\sttElm',w')$ with $w'\in \SSet_w$ and $\sttElm'\in \ISet(w')$.
\end{itemize}
The run $\run$ is accepting if, for all infinite paths $\pi$ starting from the root, the infinite sequence of states in
$\Lab_\run(\pi(0))\Lab_\run(\pi(1))\ldots$ satisfies the parity acceptance condition $\Omega$.
The language $\LangFun(\AutName)$ accepted by $\AutName$ is the tree-language over $\Sigma$ consisting of the non-blocking $\Sigma$-labeled trees $(\TSet, \Lab)$ such that there is an accepting run of $\AutName$
over $(\TSet, \Lab)$.
\vspace{0.1cm}

\noindent  \textbf{Dualization.}  For a graded $\SttSet$-constraint $\theta$, the \emph{dual $\widetilde{\theta}$} of $\theta$
is obtained from $\theta$ by exchanging $\vee$ with $\wedge$, $\true$ with $\false$, and $\SttSet$-atoms   $\DMod[k] \alpha$   with $\BMod[k] \widetilde{\alpha} $, and vice versa, where  $\widetilde{\alpha}$ is obtained from $\alpha$ by exchanging $\vee$ with $\wedge$.
For example, the dual of 
$\DMod[k_1] (\sttElm_0\vee \sttElm_1)  \wedge \BMod[k_2] \sttElm_2$ is
$\BMod[k_1] (\sttElm_0\wedge \sttElm_1) \vee \DMod[k_2] \sttElm_2$. Similarly, the dual $\widetilde{\theta}$ of a first-order $\SttSet$-constraint $\theta$
   is obtained from $\theta$ by exchanging $\vee$ with $\wedge$, $\true$ with $\false$, $x=y$ with $x\neq y$, and existential quantification $\exists x$ with
   universal quantification $\forall x$.
For a parity \GTA (resp., parity \FTA) $\AutName = \tuple {\Sigma} {\SttSet} {\trnFun} {\isttElm} {\Omega}$, the
\emph{dual automaton} of $\AutName$ is the parity \GTA (resp., parity \FTA) $\widetilde{\AutName} = \tuple {\Sigma} {\SttSet} {\widetilde{\trnFun}} {\isttElm} {\widetilde{\Omega}}$, where for all $(\sttElm,a)\in \SttSet\times\Sigma$, $\widetilde{\Omega}(\sttElm) = \Omega(\sttElm)+1$ and
$\widetilde{\trnFun}(\sttElm,a)$ is the dual of $\trnFun(\sttElm,a)$. By \cite{Wal02,CFVZ20}, the following holds.

\begin{proposition}[\cite{Wal02,CFVZ20}]
  \label{prop:DualGATA}
Let $\AutName$ be a parity  \GTA (resp., parity \FTA). Then, the dual automaton of
$\AutName$ is a parity  \GTA (resp., parity \FTA) accepting the complement of $\LangFun(\AutName)$.
\end{proposition}




 \section{Automata Characterisations of CDL and CCTL*}
\label{sec:characterizationCDL}

 In this section, we provide effective automata-theoretic characterisations of
the logics \CCDL and \CCTLStar.
We first consider the graded version of the class of \emph{hesitant alternating tree automata} (\HTA, for short),
the latter being a well-known formalism introduced in~\cite{KVW00} as an optimal automata-theoretic framework for model checking
and synthesis of \CTLStar. We show that the graded version of \HTA (\HGTA for
short) characterises the logic
\CCDL. In order to capture the logic \CCTLStar, we consider a subclass of \HGTA
obtained by enforcing a counter-freeness
requirement on the linear-time behaviour of the automaton along an existential component together with an additional condition
(we call \emph{mutual-exclusion property}) on the alphabet of the linearization word
automaton.

In the following, for a \GTA $\AutName$ and
a set $\ASet\subseteq \Atoms(\AutName)$, we denote by $\Con(\ASet)$ (resp., $\Dis(\ASet)$)
the conjunction (resp., disjunction) of the atoms occurring in $\ASet$. As usual, the empty conjunction is $\true$ and the
empty disjunction is $\false$.\vspace{0.1cm}

\noindent \textbf{Hesitant \GTA.} An \emph{hesitant \GTA} (\HGTA for short) is a tuple
$\AutName = \tuple {\Sigma} {\SttSet} {\trnFun} {\isttElm} {\Family}{\Family_{\exists}} {\Omega}$,
where  $\tuple {\Sigma} {\SttSet} {\trnFun} {\isttElm} {\Omega}$ is a parity \GTA,
  $\Family=\tpl{\SttSet_1,\ldots,\SttSet_n}$ is an \emph{ordered} tuple of non-empty pairwise disjunct subsets $\SttSet_i$
of $\SttSet$ (called \emph{components} of $\AutName$) which form a partition of $\SttSet$, and $\Family_{\exists}$ is a subset of the components
in $\Family$ (the so called  \emph{existential components}). Thus, like \HTA~\cite{KVW00}, there is an ordered partition of
$\SttSet$ into disjoint sets $\SttSet_1,\ldots,\SttSet_n$. Moreover,
each component $\SttSet_i$ is classified as \emph{transient}, \emph{existential}, or \emph{universal}, and the following holds:
\begin{itemize}
  \item \emph{transient requirement}: for each transient component $\SttSet_i$ and $(\sttElm,a)\in\SttSet_i\times \Sigma$,  $\trnFun(\sttElm,a)$ only refers to states in components $\SttSet_j$ such that $j<i$;
  \item  \emph{existential requirement}: for each existential component $\SttSet_i$ and $(\sttElm,a)\in\SttSet_i\times \Sigma$,    $\trnFun(\sttElm,a)$ can be rewritten as
  a disjunction of conjunctions of the form $\DMod\sttElm'\wedge \Con(\ASet)$, where $\sttElm'\in\SttSet_i$ and the atoms in  $\ASet$ only refer to states in components $\SttSet_j$ such that $j<i$;
  \item  \emph{universal requirement}: for each universal component $\SttSet_i$ and $(\sttElm,a)\in\SttSet_i\times \Sigma$,    $\trnFun(\sttElm,a)$ can be rewritten as
  a conjunction of disjunctions of the form $\BMod\sttElm'\vee \Dis(\ASet)$, where $\sttElm'\in\SttSet_i$ and the atoms in  $\ASet$ only refer to states in components $\SttSet_j$ such that $j<i$;
  \item \emph{hesitant acceptance requirement}: for each existential (resp., universal) component $\SttSet_i$,  the restriction $\Omega_{\SttSet_i}$ of $\Omega$ to the set $\SttSet_i$  is a B\"{u}chi condition (resp., coB\"{u}chi condition).
\end{itemize}
The first three requirements  ensure that every infinite path of a run of $\AutName$ gets trapped within some existential or universal component $\SttSet_i$.
The existential requirement establishes that from each existential state $\sttElm\in \SttSet_i$, exactly one copy is sent to a child of the current input node in  component $\SttSet_i$ (all the other copies move to states with order lower than $i$). The universal requirement corresponds to the dual of the existential requirement.
Finally, the hesitant acceptance requirement ensures that for each infinite path $\pi$ of a run that gets trapped into an existential (resp., universal) component,
$\pi$ is accepting iff $\pi$ visits infinitely many times states with even
color (resp., $\pi$ visits finitely many times states with odd color).

\begin{example}\label{example:moduloCountingTwo} Let $\Prop=\{p\}$ and $\varphi_p$ be the \CTLStar formula $\EQ\Next p$ asserting that the root of the given Kripke
tree has a child where $p$ holds. We consider the tree-language $\LangFun_2$ consisting of the Kripke trees $\LT$ such that there is an infinite path $\pi$ from the root so that $p$ never holds along $\pi$ and at the even positions $2i$, $\varphi_p$ holds at node $\pi(2i)$. $\LangFun_2$ requires counting modulo $2$ and cannot be expressed in
\CCTLStar~(a proof is in Appendix~\ref{app:moduloCountingTwo}). The language
$\LangFun_2$ is recognised  by the \HGTA $\AutName = \tuple {\Sigma} {\SttSet}
{\trnFun} {\isttElm} {\tpl{\SttSet_1,\SttSet_2}} {\{\SttSet_2\}} {\Omega}$
consisting of three states having colour $0$: the existential states $\isttElm$
and $\sttElm$ having the same and highest order
($\SttSet_2=\{\isttElm,\sttElm\}$) and the transient state $\sttElm_p$
($\SttSet_1=\{\sttElm_p\}$). Moreover, (i) $\trnFun(\sttElm_p,\{p\})=\true$ and
$\trnFun(\sttElm_p,\emptyset)=\false$, (ii)
$\trnFun(\isttElm,\emptyset)=\DMod\sttElm \wedge \DMod\sttElm_p$ and
$\trnFun(\isttElm,\{p\})=\false$, and (iii)
  $\trnFun(\sttElm,\emptyset)=\DMod\isttElm$ and $\trnFun(\sttElm,\{p\})=\false$.
\end{example}

\noindent \textbf{Linearization.}
Fix an \HGTA $\AutName = \tuple {\Sigma} {\SttSet} {\trnFun} {\isttElm} {\Family} {\Family_\exists} {\Omega}$.
 Given a component
$\SttSet_i$ of $\AutName$ and $\ASet\subseteq \Atoms(\AutName)$,  the set $\ASet$ \emph{is lower than }$\SttSet_i$ if the atoms in $\ASet$ only refer to states with order $j<i$.
For each existential (resp., universal) component $\SttSet_i$ and
$\sttElm\in \SttSet_i$, we introduce a B\"{u}chi (resp., coB\"{u}chi) \NWA $\AutName_{\SttSet_i,\sttElm}$ over the alphabet $\Sigma\times \Atoms(\AutName)$.
Intuitively, $\AutName_{\SttSet_i,\sttElm}$ encodes the \emph{modular} behaviour of  $\AutName$ starting at state $\sttElm$, which is
composed of the behaviour along 
$\SttSet_i$ (which is linear-time when $\SttSet_i$ is existential), plus additional moves that lead to states with order lower than $i$: the input alphabet $\Sigma\times \Atoms(\AutName)$ keeps track of these additional moves. When $\SttSet_i$ is universal, then
$\AutName_{\SttSet_i,\sttElm}$ can be viewed as a universal tree automaton.

\begin{definition}[Linearization word automata]\label{def:linearization}
For each non-transient component $\SttSet_i$ of $\AutName$ and $\sttElm\in\SttSet_i$,
 we denote by $\AutName_{\SttSet_i,\sttElm}$ the parity
\NWA
$
\AutName_{\SttSet_i,\sttElm}=\tuple {\Sigma\times 2^{\Atoms(\AutName)}} {\SttSet_i} {\trnFun_{\SttSet_i}} {\sttElm} {\Omega_{\SttSet_i}}
$
 where  for all $\sttElm'\in \SttSet_i$, $a\in \Sigma$, and $\ASet\subseteq \Atoms(\AutName)$, $\trnFun_{\SttSet_i}(\sttElm',(a,\ASet))$ is defined as follow:
\begin{itemize}
  \item Case  $\SttSet_i$ is existential:
  $\sttElm''\in \trnFun_{\SttSet_i}(\sttElm',(a,\ASet))$ if there is conjunction $\xi$
   in the disjunctive normal form of $\trnFun(\sttElm',a)$ such that  $\xi= \DMod\sttElm''\wedge\Con(\ASet)$ (note that $\ASet$ is lower than $\SttSet_i$).
    \item Case  $\SttSet_i$ is universal:
  $\sttElm''\in \trnFun_{\SttSet_i}(\sttElm',(a,\ASet))$ if there is disjunction $\xi$
   in the conjunctive normal form of $\trnFun(\sttElm',a)$ such that $\xi= \BMod\sttElm''\vee \Dis(\ASet)$ (note that $\ASet$ is lower than $\SttSet_i$).
\end{itemize}
Let $\Upsilon_{\SttSet_i}$ be the set
of elements $\ASet\subseteq \Atoms(\AutName)$ s.t.~$\trnFun_{\SttSet_i}(\sttElm',(a,\ASet))\neq \emptyset$ for some
$(\sttElm',a)\in \SttSet_i\times \Sigma$.
\end{definition}

\begin{remark}  Note that the transition function of $\AutName_{\SttSet_i,\sttElm}$ is independent of $\sttElm$, and $\AutName_{\SttSet_i,\sttElm}$
is a  B\"{u}chi (resp., coB\"{u}chi) \NWA if $\SttSet_i$ is existential (resp., universal).
We can equate the parity \NWA $\AutName_{\SttSet_i,\sttElm}$ to the parity \NWA over the alphabet
$\Sigma\times \Upsilon_{\SttSet_i}$ which is obtained from  $\AutName_{\SttSet_i,\sttElm}$ by restricting the transition function
to the alphabet $\Sigma\times \Upsilon_{\SttSet_i}$. In the following, we write $\AutName_{\SttSet_i,\sttElm}$ to denote this automaton.
Observe that each set of atoms
$\ASet\in \Upsilon_{\SttSet_i}$ is lower than
$\SttSet_i$.
\end{remark}

If we consider the \HGTA $\AutName$ of Example~\ref{example:moduloCountingTwo},
the B\"{u}chi \NWA $\AutName_{\SttSet_2,\isttElm}$ associated with the
existential component $\SttSet_2$ is illustrated below. Note that
$\Upsilon_{\SttSet_2}=\{\emptyset,\{\DMod\sttElm_p\}\}$.
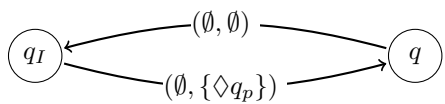
{
\begin{center}
    \begin{tikzpicture}[->,inner sep=0pt, shorten >= 1pt, shorten <= 1pt]
      \tikzstyle{state} = [draw,circle,minimum size=20pt,inner sep=2pt];
      \node[state] (s0) at (0,0) {$\isttElm$};
      \node[state] (s1) at (5,0) {$\sttElm$};

       \path[black, thick,->] (s0) edge [bend right = 15]
			node[fill=white, anchor=center, align=left, midway, sloped, font=\normalsize] {
				$(\emptyset,\{\DMod \sttElm_p\})$  } (s1);

         \path[black, thick,->] (s1) edge [bend right = 15]
			node[fill=white, anchor=center, align=center, midway, sloped, font=\normalsize] {
				$(\emptyset,\emptyset)$  } (s0);

    \end{tikzpicture}
 \end{center}
}

Let us fix an \HGTA $\AutName = \tuple {\Sigma} {\SttSet} {\trnFun} {\isttElm} {\Family} {\Family_\exists} {\Omega}$ with $\Family= \tpl{\SttSet_1,\ldots,\SttSet_n}$.
For each   graded $\SttSet$-constraint $\theta$,
 we denote by $\AutName^{\theta}$ the \HGTA
 $
 \tuple {\Sigma} {\SttSet\cup \{\theta\}} {\trnFun_{\theta}} {\theta} {\Family_{\theta}} {\Family_{\exists}} {\Omega \cup (\theta \rightarrow 0)}
  $
  where for the states in $\SttSet$, $\trnFun_{\theta}$ agrees with $\trnFun$,
  for the initial state $\theta$, $\trnFun_{\theta}(\theta,a)=\theta$ for all
  $a\in \Sigma$, and $\Family_{\theta}=
  \tpl{\SttSet_1,\ldots,\SttSet_n,\{\theta\}}$. Note that $\{\theta\}$ is a
  transient component with highest order. Thus, from the root of the input tree,
  $\AutName^{\theta}$ send copies of $\AutName$ to the children of the root
  according to the constraint $\theta$.  By construction, for each existential
  state $\sttElm$ of an \HGTA $\AutName$, we obtain the following
  characterisation of the language $\LangFun(\AutName^{\sttElm})$, where
  $\AutName^{\sttElm}$ is the \HGTA obtained from $\AutName$ by setting
  $\sttElm$ as initial state instead of $\isttElm$, in terms of the
  linearization of $\AutName$.

\begin{proposition}\label{prop:PropertiesHGATALineariztion} Let $\AutName$ be an \HGTA, $\SttSet_i$ be an existential component of $\AutName$,  and   $\sttElm\in\SttSet_i$. Then, for each input $\LT = (\TSet,\Lab)$,
$\LT\in \LangFun(\AutName^{\sttElm})$ \emph{if and only if} there is an infinite path $\pi$ of $\LT$ starting at  the root and an infinite word $\rho\in \LangFun(\AutName_{\SttSet_i,\sttElm})$ such that $\rho$ is of the form
      $\rho=(\Lab(\pi(0)),\ASet_0)(\Lab(\pi(1)),\ASet_1)\ldots$ and for each $i\geq 0$,
      $\LT_{\pi(i)}\in \LangFun(\AutName^{\Con(\ASet(i))})$, where
      $\LT_{\pi(i)}$ is the labelled subtree of $\LT$ rooted at node $\pi(i)$.
\end{proposition}

\noindent \textbf{Counter-free \HGTA.} In order to capture \CCTLStar, we
introduce a subclass of \HGTA obtained by enforcing additional conditions.
By Proposition~\ref{prop:PropertiesHGATALineariztion} and the equivalence of \LTL and B\"{u}chi \emph{counter-free}   \NWA~\cite{DG08a}, a natural condition consists in
requiring that for each non-transient component $\SttSet_i$ of the \HGTA and state $\sttElm\in\SttSet_i$,
the  \NWA $\AutName_{\SttSet_i,\sttElm}$ is counter-free (\emph{counter-freeness
requirement}).\footnote{Note that the property of an \NWA to be counter-free is
independent of the initial state.} However, this condition is not sufficient for
characterising the logic \CCTLStar. A counterexample is the \HGTA $\AutName$ of
Example~\ref{example:moduloCountingTwo} which clearly satisfies the
counter-freeness
requirement but recognises a tree-language which is not expressible in
\CCTLStar. 
We introduce an additional condition (\emph{mutual-exclusion property}) on the alphabets of the linearization automata (see Definition~\ref{def:SeparationCondition} below).
A \emph{Counter-free \HGTA} (\HGTAcf for short) is an \HGTA satisfying both the
counter-free requirement and the mutual-exclusion condition.


\begin{definition}\label{def:SeparationCondition}
An \HGTA $\AutName$ satisfies the \emph{mutual-exclusion property} if for each non-transient component $\SttSet_i$ and for all $\ASet,\ASet'\in \Upsilon_{\SttSet_i}$
such that $\ASet\neq \ASet'$, it holds that there exists an atom $\atom\in \ASet$ and an atom $\atom'\in \ASet'$ such that
$\LangFun(\AutName^{\atom})$ is the complement of  $\LangFun(\AutName^{\atom'})$. Note that if $\Upsilon_{\SttSet_i}$ is a singleton, then the previous property is fulfilled.
\end{definition}

Evidently, if $\AutName$ satisfies the mutual-exclusion condition, then
for each non-transient component $\SttSet_i$ and for all $\ASet,\ASet'\in \Upsilon_{\SttSet_i}$
such that $\ASet\neq \ASet'$, it holds that $\LangFun(\AutName^{\Con(\ASet)})\cap \LangFun(\AutName^{\Con(\ASet')})=\emptyset$. Intuitively, the mutual-exclusion condition requires that along a non-transient component $\SttSet_i$, the distinct moves $\ASet\in \Upsilon_{\SttSet_i}$ (these moves lead to components with order lower than $i$) are mutually exclusive. Let us consider again the \HGTA $\AutName$ of Example~\ref{example:moduloCountingTwo}. Since
  $\Upsilon_{\SttSet_2}=\{\emptyset,\{\DMod\sttElm_p\}\}$, by Definition~\ref{def:SeparationCondition}, $\AutName$ does not satisfy the mutual-exclusion condition. Note that  $\Con(\emptyset)=\true$ and $\Con(\{\DMod\sttElm_p\})=\DMod\sttElm_p$.  Hence,
  $\LangFun(\AutName^{\true})\cap \LangFun(\AutName^{\Con(\{\DMod\sttElm_p\})})= \LangFun(\AutName^{\Con(\{\DMod\sttElm_p\})}) =\LangFun(\EQ\Next p)\neq \emptyset$.

The dual  $\widetilde{\AutName}$ of an \HGTA $\AutName = \tuple {\Sigma} {\SttSet} {\trnFun} {\isttElm} {\Family}{\Family_\exists} {\Omega}$ is the tuple
$\tuple {\Sigma} {\SttSet} {\widetilde{\trnFun}} {\isttElm} {\Family} {\widetilde{\Family_\exists}}{\widetilde{\Omega}}$, where
$\widetilde{\trnFun}$ and $\widetilde{\Omega}$ are defined as for the dual of an arbitrary parity \GTA and
$\widetilde{\Family_\exists}$ consists of the universal components of $\AutName$. 
By construction and Proposition~\ref{prop:DualGATA}, the considered subclasses of \GTA are closed under Boolean language operations
(for details, see Appendix~\ref{app:SubclassesGATABooleanClosure}).

\begin{restatable}{proposition}{propSubclassesGATABooleanClosure}
  \label{prop:SubclassesGATABooleanClosure}
\HGTA (resp., \HGTAcf) and \HGTA satisfying the mutual-exclusion property are effectively closed
under Boolean language operations.
\end{restatable}

\noindent \textbf{Enforcing the Mutual-exclusion Property.}
By exploiting dualization, an \HGTA $\AutName$ can be converted into an equivalent \HGTA $\AutName_s$ satisfying the mutual-exclusion condition. Intuitively, $\AutName_s$ is obtained by merging in a syntactical and \emph{modular} way  $\AutName$ with a renaming of the dual \HGTA $\widetilde{\AutName}$ (see Appendix~\ref{app:SeparationConditionHGATA}).

\begin{restatable}{proposition}{propSeparationConditionHGATA}
  \label{prop:SeparationConditionHGATA}
Given an \HGTA $\AutName$, one can construct
an \HGTA $\AutName_s$ such that $\AutName_s$ satisfies the mutual-exclusion condition and $\LangFun(\AutName_s)=\LangFun(\AutName)$.
\end{restatable}

Note that the translation in Proposition~\ref{prop:SeparationConditionHGATA}  changes the second component $\Upsilon_{\SttSet_i}$ of the
alphabets of the linearization automata. Since counter-free \NWA are not closed
under inverse projection, the construction does not preserve the
counter-freeness property. For example, for the \HGTA  of
Example~\ref{example:moduloCountingTwo}, the translation replaces the edge from
$\sttElm$ to $\isttElm$ with label $(\emptyset,\emptyset)$ of the \NWA
$\AutName_{\SttSet_2,\isttElm}$ with two edges from $\sttElm$ to $\isttElm$: one
with label $(\emptyset,\{\DMod \sttElm_p\})$ and the other one
with label $(\emptyset,\{\BMod \sttElm'_p\})$ where $\LangFun(\AutName^{\Con(\{\BMod\sttElm'_p\})}) =\LangFun(\neg\EQ\Next p)$. The resulting \NWA is not counter-free.



\subsection{From Automata to Logics and Back}
\label{sec:characterizationCDL;sub:FromAutomataToLogicsAndBack}

In this section, we show that the class of  \HGTA and the logic \CCDL are effectively equivalent, and the class of \HGTAcf
effectively characterizes \CCTLStar. We start with the translations from automata to logics.

\begin{restatable}{theorem}{theoFromHGATAtoCDL}\label{theo:FromHGATAtoCDL} Let $\AutName$ be an \HGTA (resp., an \HGTAcf) over $2^{\Prop}$. Then, one can construct a  \CCDL  (resp.,  \CCTLStar) formula $\varphi_{\AutName}$ such that
  $\LangFun(\varphi_{\AutName})= \LangFun(\AutName)$. Moreover,
  $\varphi_{\AutName}$ is a  $\CDL$ (resp. a \CTLStar) formula if $\AutName$ is symmetric.
\end{restatable}
\begin{proof} We focus on the translation from \HGTAcf  $\AutName = \tuple {\Sigma} {\SttSet} {\trnFun} {\isttElm} {\Family} {\Family_\exists} {\Omega}$
  to \CCTLStar (the translation from \HGTA to \CCDL is given in Appendix~\ref{app:FromHGATAtoCDL}).
 For each $\sttElm\in \SttSet$, we construct  a  \CCTLStar   formula $\varphi_{\sttElm}$
such that $\LangFun(\varphi_{\sttElm})= \LangFun(\AutName^{\sttElm})$ and  $\varphi_{\sttElm}$ is a \CTLStar formula if $\AutName$ is symmetric.
 Thus, by setting $\varphi_{\AutName} \DefinedAs \varphi_{\isttElm}$, Theorem~\ref{theo:FromHGATAtoCDL} directly follows. The proof is by induction on the order $\ell$ of the component $\SttSet_\ell$ such that $\sttElm\in \SttSet_\ell$. We distinguish the cases where $\sttElm$ is transient, existential, or universal.
The transient case is easy
and the universal case follows from the existential case by a dualization argument (for details, see Appendix~\ref{app:FromHGATAtoCDL}). Now, assume that $\sttElm$ is existential.
Let   us consider the B\"{u}chi \NWA $\AutName_{\SttSet_\ell,\sttElm}$ over
$2^{\Prop}\times \Upsilon_{\SttSet_\ell}$ as defined in Definition~\ref{def:linearization}. Recall that
$\AutName_{\SttSet_\ell,\sttElm}$ is counter-free. Moreover, $\Upsilon_{\SttSet_\ell}\subseteq 2^{\Atoms(\AutName)}$ contains only elements $\ASet$ such that states occurring in the atoms of $\ASet$ have order $j$ lower than $\ell$. Thus, by the induction hypothesis, 
for each $\ASet\in \Upsilon_{\SttSet_\ell}$, one can build a  \CCTLStar  formula $\varphi_{\ASet}$ such that
$\LangFun(\AutName^{\Con(\ASet)})= \LangFun(\varphi_{\ASet})$. Hence, since $\AutName$ satisfies the mutual-exclusion condition,
the following holds:\vspace{0.1cm}

\noindent \emph{Claim 1.} For all
$\ASet,\ASet'\in \Upsilon_{\SttSet_\ell}$ such that $\ASet\neq \ASet'$, $\LangFun(\varphi_{\ASet})\cap \LangFun(\varphi_{\ASet'})=\emptyset$. \vspace{0.1cm}

For each $\ASet\in \Upsilon_{\SttSet_\ell}$, let $p_{\ASet}$ be a fresh atomic proposition. We denote by  $\Prop_{\Ext}$ the extension of
 $\Prop$ with these fresh propositions. Moreover, let $\AutName_{\Ext,\SttSet_\ell,\sttElm}$ be the B\"{u}chi \NWA over $2^{\Prop_{\Ext}}$ having the same set of states, initial state, acceptance condition as  $\AutName_{\SttSet_\ell,\sttElm}$ and whose transition function $\trnFun_{\Ext,\SttSet_\ell}$ is obtained from the transition function $\trnFun_{\SttSet_\ell}$ of
 $\AutName_{\SttSet_\ell,\sttElm}$ as follows: for all $\sttElm'\in\SttSet_\ell$ and  $a_{\Ext}\in 2^{\Prop_{\Ext}}$,  if $a_{\Ext}$ is of the form $a\cup \{p_{\ASet}\}$, for some $a\in 2^{\Prop}$ and $\ASet\in \Upsilon_{\SttSet_\ell}$, (i.e., $a_{\Ext}$ contains a unique proposition in $\Prop_{\Ext}\setminus \Prop$), then $\trnFun_{\Ext,\SttSet_\ell}(\sttElm',a_{\Ext})=\trnFun_{\SttSet_\ell}(\sttElm',(a,\ASet))$; otherwise, $\trnFun_{\Ext,\SttSet_\ell}(\sttElm',a_{\Ext})= \emptyset$.  Being $\AutName_{\SttSet_\ell,\sttElm}$ counter-free,
 $\AutName_{\Ext,\SttSet_\ell,\sttElm}$ is clearly counter-free as well. Thus,
by~\cite{DG08a}, one can construct an \LTL formula
 $\psi$ over $\Prop_{\Ext}$ such that $\LangFun(\psi) = \LangFun(\AutName_{\Ext,\SttSet_\ell,\sttElm})$. Since $\LangFun(\AutName^{\Con(\ASet)})= \LangFun(\varphi_{\ASet})$  for all $\ASet\in \Upsilon_{\SttSet_\ell}$, by construction and Proposition~\ref{prop:PropertiesHGATALineariztion}, we obtain   the following characterization of the
 tree-language $\LangFun(\AutName^{\sttElm})$. \vspace{0.1cm}

 \noindent \emph{Claim 2.} For each Kripke tree $\LT = (\TSet,\Lab)$,
$\LT\in \LangFun(\AutName^{\sttElm})$ iff  there is an infinite path $\pi$ of $\LT$ from the root and an infinite word $\rho$ over $2^{\Prop_{\Ext}}$ such that $\rho\models \psi$ and, for all $j\geq 0$, (i) $\rho(j)\cap \Prop = \Lab(\pi(j))$,  (ii) for all $p_{\ASet}\in \rho(j)$,
      $(\LT,\pi(j))\models \varphi_{\ASet}$, and (iii) there is a unique $\ASet\in \Upsilon_{\SttSet_\ell}$ such that $p_{\ASet}\in \rho(j)$.
\vspace{0.1cm}

Note that since $\LangFun(\psi) = \LangFun(\AutName_{\Ext,\SttSet_\ell,\sttElm})$, by construction, point~(iii) in Claim~2 follows from the fact that $\rho\models \psi$. By exploiting the always modality $\Always$ ($\Always\xi$ is a shorthand of $\neg (\true \Until \neg \xi)$) and both conjunction and disjunction, w.l.o.g.~we can assume
 that the \LTL formula $\psi$ is in negation normal form, i.e., negation is applied only to atomic propositions.
  Now, let
$f(\psi)$ be the  \CCTLStar path formula over $\Prop$ obtained from $\psi$ by replacing each literal of the form
$p_{\ASet}$ (resp., $\neg p_{\ASet}$), where $\ASet\in \Upsilon_{\SttSet_\ell}$, with the \CCTLStar state formula $\varphi_{\ASet}$
(resp., $\bigvee_{\ASet'\in \Upsilon_{\SttSet_\ell}\setminus \{\ASet\}} \varphi_{\ASet'}$). Finally, let us consider the \CCTLStar state formula $\varphi_{\sttElm}$ defined as follows:\vspace{0.1cm}

$
\varphi_{\sttElm} \DefinedAs \EQ(f(\psi) \wedge \Always \bigvee_{\ASet\in \Upsilon_{\SttSet_\ell}}\varphi_{\ASet}).
$\vspace{0.1cm}

Note that the second conjunct in the state formula $\varphi_{\sttElm}$ ensures
that, for the infinite path $\pi$ selected by the path quantifier $\EQ$ and for
each $j\geq 0$, the state formula $\varphi_{\ASet}$ holds at node $\pi(j)$ for
some $\ASet\in \Upsilon_{\SttSet_\ell}$.
We show that a Kripke tree $\LT = (\TSet,\Lab)$ satisfies $\varphi_{\sttElm}$ iff the characterization of $\LangFun(\AutName^{\sttElm})$ in Claim~2 holds. Hence, the result follows.

We shall now focus on the left-right implication of the equivalence (the right-left
implication is similar).
Thus, assume that $\LT\models \varphi_{\sttElm}$. Hence, there exists an infinite path $\pi$ of $\LT$ from the root and an infinite sequence
$\nu = \ASet_0,\ASet_1,\ldots$ over $\Upsilon_{\SttSet_\ell}$ such that $(\LT,\pi,0)\models f(\psi)$ and for each $j\geq 0$,
$(\LT,\pi(j))\models \varphi_{\ASet_j}$. Let $\Lab(\pi)\otimes \nu$ be the infinite word over $2^{\Prop_{\Ext}}$
defined as follows for all $j\geq 0$: $(\Lab(\pi)\otimes \nu)(j)= \Lab(\pi(j))\cup\{p_{\ASet_j}\}$. By Claim~2, it suffices to show that  $\Lab(\pi)\otimes \nu\models \psi$. To this purpose, we show by structural induction that for each $j\geq 0$ and subformula $\theta$ of $\psi$ if $(\LT,\pi,j)\models f(\theta)$, then
$(\Lab(\pi)\otimes \nu,j)\models \theta$. Since the formula $\psi$ is in negation normal form, by the induction hypothesis, the unique non-trivial case is when $\theta$ is either of the form $p_{\ASet}$ or of the form
$\neg p_{\ASet}$ for some $\ASet\in \Upsilon_{\SttSet_\ell}$.
\begin{itemize}
  \item $\theta= p_{\ASet}$: hence, $f(\theta)= \varphi_{\ASet}$. Since $(\LT,\pi,j)\models f(\theta)$ 
  and $(\LT,\pi(j))\models \varphi_{\ASet_j}$, by Claim~1, it follows that $\ASet=\ASet_j$, i.e.~$\theta= p_{\ASet_j}$. Hence, $(\Lab(\pi)\otimes \nu,j)\models \theta$, and the result follows.
  \item $\theta= \neg p_{\ASet}$: hence $f(\theta) = \bigvee_{\ASet'\in \Upsilon_{\SttSet_\ell}\setminus \{\ASet\}} \varphi_{\ASet'}$. Since $(\LT,\pi,j)\models f(\theta)$ 
      and $(\LT,\pi(j))\models \varphi_{\ASet_j}$, by Claim~1, 
      $\ASet\neq \ASet_j$. Hence, $(\Lab(\pi)\otimes \nu,j)\models \theta$, and we are done.\qedhere
\end{itemize}
\end{proof}

\noindent \textbf{From Logics to Automata.}  As to the translation from
\CCTLStar to \HGTAcf, in order to ensure the mutual-exclusion property of the
resulting \HGTAcf, we need a restricted syntactic form of \CCTLStar formulae,
which is still expressively complete.  A \CCTLStar formula is in \emph{simple
form} if each occurrence of the path quantifier $\EQ$ is immediately preceded by
the counter modality $\DC^{1}$ (note that $\DC^{1}$ corresponds to the standard
$\EQ\Next$ modality of \CTLStar). Formally, the set of state formulae $\varphi$
of $\CCTLStar$ in simple form is defined according to the following syntax:
$
	\varphi ::= \top \ | \ p \ | \ \neg \varphi \ | \ \varphi \wedge
	\varphi  \ | \ \DC^{1}\EQ
	\psi \ | \ \DC^{n}
	\varphi
$.
One can easily show that the simple form is indeed expressively complete (for details, see Appendix~\ref{app:SimpleFormCCTLStar}).

\begin{restatable}{theorem}{theoSimpleFromCDLtoHGATA}\label{theo:FromCDLtoHGATA}
Given a \CCDL (resp., \CCTLStar) formula $\varphi$, one can construct an equivalent \HGTA (resp., \HGTAcf)  $\AutName_\varphi$ such that $\LangFun(\AutName_\varphi)= \LangFun(\varphi)$.
Moreover, $\AutName_\varphi$ is symmetric if $\varphi$ is a  \CDL (resp., a \CTLStar) formula.
\end{restatable}
\begin{proof}
We focus on the translation from \CCTLStar to  \HGTAcf  (the translation from  \CCDL to \HGTA is given in Appendix~\ref{app:FromCDLtoHGATA}).
 Fix a  \CCTLStar  formula $\Phi$. W.l.o.g., we can assume that
 $\Phi$ is  in simple form.  As in the case of the alternating hesitant automata for \CTLStar~\cite{KVW00}, we construct the automaton  by induction on the structure of $\Phi$. With each  state subformula $\varphi$ of $\Phi$ we associate an \HGTAcf $\AutName_\varphi$ over $\Sigma=2^{\Prop}$ such that $\LangFun(\AutName_\varphi)= \LangFun(\varphi)$.
The cases where $\varphi$ is an atomic proposition, or the root operator of $\varphi$
 is the counting modality  $\DC^{n}$ are straightforward (for details, see Appendix~\ref{app:FromCDLtoHGATA}). The cases where the root operator of
 $\varphi$ is a Boolean connective directly follow from Proposition~\ref{prop:SubclassesGATABooleanClosure}.
 Now, assume that $\varphi=\EQ\psi$ for some path formula $\psi$.
Let $\max(\psi)$ be the set of state subformulae of $\psi$ of the form $\EQ\xi$ or $\DC^n\xi$ which are not preceded by the modality $\EQ$ or the counting
modality in the syntax tree of $\psi$.
Since $\psi$ is in simple form, $\max(\psi)$ is of the form $\{\DC^{n_1}\varphi_1,\ldots, \DC^{n_k}\varphi_k\}$ for some $k\geq 0$, where $\varphi_1,\ldots,\varphi_k$ are \CCTLStar formulae in simple form. Note that if $\psi$ is a \CTLStar formula, then $n_1=\ldots =n_k=1$. By the induction hypothesis, for each
$i\in [1,k]$, one can construct an \HGTAcf $\AutName_{i} = \tuple {2^{\Prop}} {\SttSet_i} {\trnFun_i} {\isttElm_i} {\Family_i} {\Family_{\exists,i}} {\Omega_i}$   such that
$\LangFun(\AutName_i)=\LangFun(\varphi_i)$. For each $i\in [1,k]$, let $\widetilde{\AutName}_{i} = \tuple {2^{\Prop}} {\widetilde{\SttSet}_i} {\widetilde{\trnFun}_i} {\widetilde{\isttElm}_i} {\widetilde{\Family}_i} {\widetilde{\Family}_{\exists,i}} {\widetilde{\Omega}_i}$ be a renaming of the dual automaton of $\AutName_i$. 

Let $\Prop_{\Ext}$ be an extension of $\Prop$ obtained by adding for each state formula $\DC^{n_i}\varphi_i$ a fresh 
 proposition $p_i$. Then, the path formula $\psi$ can be viewed as an \LTL formula $\psi_{\Ext}$ over $\Prop_{\Ext}$.
By~\cite{DG08a}, one can build a B\"{u}chi \emph{counter-free}  \NWA
$\NName_{\psi}= \tpl{2^{\Prop_{\Ext}}, {\SttSet}, {\trnFun}, {\isttElm},
{\Omega}}$ s.t.~$\LangFun(\NName_{\psi})= \LangFun(\psi_{\Ext})$.
 By construction, we easily deduce the following characterization of $\LangFun(\varphi)= \LangFun(\EQ\psi)$:\vspace{0.1cm}

\noindent \emph{Claim 1:} for each Kripke tree $\LT = (\TSet,\Lab)$, $\LT\in\LangFun(\varphi)$ \emph{iff} there exists an infinite  path $\pi$ of $\LT$ from the root and an infinite word $\rho$ over $2^{\Prop_{\Ext}}$ such that
$\rho\in \LangFun(\NName_{\psi})$ and the following holds for each $i\geq 0$: (i) $\rho(i)\cap \Prop= \Lab(\pi(i))$, (ii) for each $\ell\in [1,k]$ such that $p_\ell\in \rho(i)$, $(\LT,\pi(i))\models \DC^{n_\ell}\varphi_\ell$, and (iii)
for each $\ell\in [1,k]$ such that $p_\ell\notin \rho(i)$, $(\LT,\pi(i))\models \neg\DC^{n_\ell}\varphi_\ell$.\vspace{0.1cm}

We define $\AutName_{\varphi}$ as follows: 
$\AutName_\varphi$ simulates the B\"{u}chi \NWA $\NName_{\psi}$ along a guessed infinite path of the input tree from the root and starts additional copies of the \HGTAcf  $\AutName_1,\ldots,\AutName_k,\widetilde{\AutName}_1,\ldots,\widetilde{\AutName}_k$. According to Claim~1, these copies guarantee that whenever the \NWA $\NName_{\psi}$ assumes that  proposition $p_\ell$   labels (resp., $p_\ell$ does not label) the current node  along the guessed path, then 
$\DC^{n_\ell}\varphi_\ell$ holds (resp., $\DC^{n_\ell}\varphi_\ell$ does not hold) at this node.
%
The components of $\AutName$ consist  of the existential component $\SttSet$ (the set of states of the B\"{u}chi counter-free  \NWA $\NName_{\psi}$) and the components of the \HGTAcf  automata $\AutName_i$ and $\widetilde{\AutName}_i$ for each $i\in [1,k]$. Moreover, the existential component $\SttSet$ has highest order and the ordering of the components of $\AutName_i$ (resp., $\widetilde{\AutName}_i$) is preserved for each $i\in [1,k]$. For the transition function
 $\trnFun_\varphi$ of $\AutName_\varphi$, we have that
for states in $\SttSet_i$ (resp., $\widetilde{\SttSet}_i$), $\trnFun_\varphi$ agrees with the corresponding $\trnFun_i$ (resp., $\widetilde{\trnFun}_i$). For states  $\sttElm\in \SttSet$ and $a\in 2^{\Prop}$,
 $\trnFun_\varphi(\sttElm,a)$ is defined as follows, where for each $\ISet\subseteq [1,k]$, $\ISet(a)$ denotes the
 subset of $\Prop_{\Ext}$ given by $a\cup \bigcup_{\ell\in \ISet}\{p_\ell\}$:\vspace{0.1cm}

 $
 \trnFun_\varphi(\sttElm,a)\DefinedAs \displaystyle{\bigvee_{\ISet\subseteq [1,k]}\,\,\,\bigvee_{\sttElm'\in \trnFun(\sttElm,\ISet(a))}
 (\DMod \sttElm' \wedge \bigwedge_{\ell\in \ISet} \DMod[\ell]\isttElm_i \wedge \bigwedge_{\ell\in [1,k]\setminus\ISet} \BMod[\ell]\widetilde{\isttElm}_i)  }
 $\vspace{0.1cm}

\noindent By construction, the induction hypothesis, and Claim~1, $\AutName_\varphi$ is an \HGTA satisfying the mutual-exclusion property such that $\LangFun(\AutName_\varphi)=\LangFun(\varphi)$. 
It remains to show that for each $\sttElm\in\SttSet$, the B\"{u}chi \NWA
$\AutName_{\SttSet,\sttElm}$ over the alphabet $2^{\Prop}\times \Upsilon_{\SttSet}$ (see Definition~\ref{def:linearization}) driven by the existential component $\SttSet$ of $\AutName_\varphi$ is counter-free. Let us consider the mapping $g$ assigning to each $a_{\Ext}\in 2^{\Prop_{\Ext}}$ the pair $(a, \bigcup_{\ell\in \ISet} \{\DMod[\ell]\isttElm_i\} \cup \bigcup_{\ell\in [1,k]\setminus\ISet} \{\BMod[\ell]\widetilde{\isttElm}_i\})$, where $a=\Prop\cap a_{\Ext}$ and $\ISet$ is the set of indexes in $j\in [1,k]$ such that $p_j\in a_{\Ext}$. Clearly, $g$ is a bijection between $2^{\Prop_{\Ext}}$ and $2^{\Prop}\times \Upsilon_{\SttSet}$. Moreover, for the transition functions  $\trnFun_{\SttSet}$ and $\trnFun$ of $\AutName_{\SttSet,\sttElm}$ and $\NName_{\psi}$, respectively, it holds that, for each $(a,\ASet)\in 2^{\Prop}\times \Upsilon_{\SttSet}$ and $\sttElm'\in \SttSet$, $\trnFun_{\SttSet}(\sttElm',(a,\ASet))=\trnFun(\sttElm',g^{-1}(a,\ASet))$, where $g^{-1}$ is the inverse of $g$. Thus, since
 $\NName_{\psi}$ is counter free, $\AutName_{\SttSet,\sttElm}$ is counter free as well, and the result follows.
\end{proof}





\section{Automata Characterisation of Monadic Chain Logic (MCL)}
\label{sec:automataForMCL}

\emph{Monadic Chain Logic} (\MCL) is the fragment of \MSO over Kripke trees
where monadic second-order quantification is restricted to sets of nodes which
forms chains, \ie a subset of a path (for details
on the syntax and semantic of \MCL, see Appendix~\ref{sec:MSOFragments}).  In this section, we provide an
automata-theoretic characterisation of \MCL in terms of a subclass of parity
\FTA, called \emph{Hesitant \FTA} (\HFTA for short), which represents the \FTA
counterpart of hesitant \GTA. Moreover, we show that the bisimulation-invariant
fragment of \MCL and \CDL are expressively equivalent.\vspace{0.2cm}

\noindent \textbf{The class of \HFTA.}
 An \HFTA  is  a tuple
$\AutName = \tuple {\Sigma} {\SttSet} {\trnFun} {\isttElm} {\Family} {\Family_\exists} {\Omega}$,
where  $\tuple {\Sigma} {\SttSet} {\trnFun} {\isttElm}  {\Omega}$ is an  \FTA
and $\Family$ and $\Family_\exists$ are defined as for \HGTA. Moreover, we require that $\AutName$ satisfies the transient requirement and
the hesitant acceptance requirement of \HGTA  and the following variants of the existential and universal requirements of \HGTA:
\begin{itemize}
  \item  for each existential component $\SttSet_i$ and $(\sttElm,a)\in\SttSet_i\times \Sigma$,  $\trnFun(\sttElm,a)$ is a  disjunction of formulae of the form $\exists x.\,(\sttElm'(x)\wedge \theta(x))$ where $\sttElm'\in \SttSet_i$ and $\theta(x)$ only refers to states in lower components $\SttSet_j$ with  $j<i$  (\emph{first-order existential requirement});
  \item  for each universal component $\SttSet_i$ and $(\sttElm,a)\in\SttSet_i\times \Sigma$,  $\trnFun(\sttElm,a)$ is a  conjunction of formulae of the form $\forall x.\,(\sttElm'(x)\vee \theta(x))$  where  $\sttElm'\in \SttSet_i$ and $\theta(x)$ only refers to states in lower components $\SttSet_j$ with $j<i$ (\emph{first-order universal requirement}).
\end{itemize}

\noindent \HFTA can be easily translated into equivalent \MCL sentences (for details, see Appendix~\ref{app:FromHFATAtoMCL}). 

\begin{restatable}{theorem}{theoFromHFATAtoMCL}\label{theo:FromHFATAtoMCL}  Given an \HFTA $\AutName$ over  $2^{\Prop}$,  one can construct in polynomial time  an  \MCL  sentence $\varphi_{\AutName}$ over $\Prop$ such that  $\LangFun(\varphi_{\AutName})= \LangFun(\AutName)$.
\end{restatable}

\noindent \textbf{Chain Projection.} Like \HGTA, the tree-languages accepted by
\HFTA are closed under Boolean operations. Thus, in the translation of \MCL
formulae into equivalent parity \HFTA, the only non-trivial part concerns the
treatment of \MCL existential quantification. For this purpose, we define an
operation on tree languages that captures the semantics of \MCL existential
quantification. Let $\LangFun$ be a tree language over $2^{\Prop}$ and $p\in
\Prop$.  The \emph{chain projection of $\LangFun$ over $p$}, denoted by
$\exists^{\CSym}p.\LangFun$, is the language consisting of the Kripke trees
$(\TSet, \Lab)$ over $\Prop\setminus \{p\}$ such that there is an infinite path
$\pi$ of $\TSet$ from the root and a Kripke tree $(\TSet, \Lab')\in\LangFun$ so
that: $\Lab'(w)=\Lab(w)$, for each $w\in \TSet\setminus \pi$, and
$\Lab'(w)\setminus \{p\}=\Lab(w)$, otherwise.

We show that \HFTA are effectively closed under chain projection, i.e., for each
\HFTA $\AutName$ over $2^{\Prop}$ and $p\in\Prop$, one can construct an \HFTA
accepting $\exists^{\CSym}p.\LangFun(\AutName)$. The proof is divided in two
steps. In the first step, we define a subclass of \HFTA, called \HFTA that
\emph{are nondeterministic in one path} (see
Definition~\ref{def:HFATASeparatedOneDirection}), whose closure under chain
projection can be easily established (see
Proposition~\ref{prop:ProjectionSeparatedInOneDirection}). Then, in the second
step, we show that an \HFTA can be converted into an equivalent \HFTA that is
nondeterministic in one path.

 We now introduce this subclass of automata. By exploiting the known
notion of \emph{basic formula}~\cite{Wal02,CFVZ20},  we first define a
fragment of the one-step language $\FOEOne{\SttSet}$ for a given non-empty set
$\SttSet$. A \emph{$\SttSet$-type} is  a (possibly empty) set $\ASet\subseteq
\SttSet$. It defines the first-order constraint $\Type(\ASet)(x)
\DefinedAs\bigwedge_{\sttElm\in\ASet}\sttElm(x)$. Note that  $\Type(\ASet)(x)$
is $\true$
if $\ASet$ is empty. Let $\TypeSet_\exists$ and $\TypeSet_\forall$  be two sets of $\SttSet$-type.
The \emph{basic formula for the pair $(\TypeSet_\exists,\TypeSet_\forall)$}, denoted $\theta^{=}(\TypeSet_\exists,\TypeSet_\forall)$, is the $\FOEOne{\SttSet}$ sentence defined as follows, where $\TypeSet_\exists=\{\ASet_1,\ldots,\ASet_n\}$ for some $n\geq 0$ and for variables $z_1,\ldots,z_k$, $\Diff(z_1,\ldots,z_k)\DefinedAs \bigwedge_{i\neq j}z_i\neq z_j$:\vspace{0.05cm}

\noindent $
\displaystyle{\exists x_1\ldots \exists x_n\ldotp \Bigl( \Diff(x_1,\ldots,x_n)\wedge \bigwedge_{i=1}^{n}\Type(\ASet_i)(x_i)\,
\wedge \, \forall y\ldotp (\Diff(x_1,\ldots,x_n,y) \rightarrow \bigvee_{\ASet\in\TypeSet_{\forall}}\Type(\ASet)(y) )\Bigr). }
$\vspace{0.05cm}

\noindent Intuitively,  $\theta^=(\TypeSet_\exists,\TypeSet_\forall)$
asserts that there are $n$-distinct elements $s_1,\ldots,s_n$ of the given domain $\SSet$ such that each $s_i$ satisfies the $\SttSet$-type $\ASet_i$ of the existential part $\TypeSet_\exists$, and every other element of the domain satisfies some $\SttSet$-type in the universal part $\TypeSet_\forall$.

\begin{definition}\label{def:ConstraintsSeparatedOneDirection}Let $\SttSet'\subseteq \SttSet$ with $\SttSet'\neq\emptyset$. A basic formula
$\theta^=(\TypeSet_\exists,\TypeSet_\forall)$ is \emph{$\SttSet'$-functional in
  one direction} if there exists $\ASet\in \TypeSet_\exists$ such that $\ASet$
  is a singleton consisting of an element in $\SttSet'$, and for each $\BSet\in
  (\TypeSet_\exists\setminus \{\ASet\})\cup \TypeSet_\forall$, $\BSet$ does not
  contain elements in $\SttSet'$. A first-order $\SttSet$-constraint is
  \emph{$\SttSet'$-functional in one direction} if it is the disjunction of
  basic formulae which are $\SttSet'$-functional in one direction.
\end{definition}

Intuitively, when the local behaviour of an \HFTA $\AutName$ at the current
input node $w$ is driven by a constraint $\theta$ that is $\SttSet'$-functional
in one direction, then there is a child $w'$ of $w$ such that exactly one copy
of $\AutName$ is sent to $w'$. Moreover, the state of this copy is in $\SttSet'$
and the states of the copies sent to the children of $w$ distinct from $w'$ are
in $\SttSet\setminus\SttSet'$.

 \begin{definition}\label{def:HFATASeparatedOneDirection} An \HFTA $\AutName = \tuple {\Sigma} {\SttSet} {\trnFun} {\isttElm} {\Family} {\Family_\exists} {\Omega}$ is \emph{nondeterministic in one path} if the initial state $\isttElm$ belongs to some existential component $\SttSet_\ell$ of $\AutName$ and the following hold:
 \begin{enumerate}
   \item for each $\sttElm\in\SttSet_\ell$ and $a\in\Sigma$, $\trnFun(\sttElm,a)$ is $\SttSet_\ell$-functional in one direction;
   \item for each 
   $\LT\in \LangFun(\AutName)$ and for each infinite path $\pi$ of $\LT$ from the root, there is an accepting run $r=(\TSet_r,\Lab_r)$ of $\AutName$ over $\LT$ s.t.~for each input node $w\in \pi$, there is exactly one node $y$ of $r$ reading $w$, i.e., such that $\Lab_r(y)=(\sttElm,w)$ for some state $\sttElm$; moreover,  $\sttElm\in\SttSet_\ell$.
 \end{enumerate}
 \end{definition}

 Let $\Sigma=2^{\Prop}$, $p\in\Prop$, $\AutName = \tuple {\Sigma} {\SttSet}
 {\trnFun} {\isttElm} {\Family} {\Family_\exists} {\Omega}$ be an \HFTA that is
 nondeterministic in one path, and $\SttSet_\ell$ be the existential component
 such that $\isttElm\in\SttSet_\ell$.  We consider the \HFTA
 $\exists^{\CSym}p.\AutName = \tuple {2^{\Prop\setminus \{p\}}} {\SttSet}
 {\trnFun'} {\isttElm} {\Family} {\Family_\exists} {\Omega}$, where the
 transition function $\trnFun'$ is defined as follows for all
 $\sttElm\in\SttSet$ and $a\in 2^{\Prop\setminus \{p\}}$:
 $\trnFun'(\sttElm,a)=\trnFun(\sttElm,a)$ if $\sttElm\notin \SttSet_\ell$, and
 $\trnFun'(\sttElm,a)=\trnFun(\sttElm,a)\vee \trnFun(\sttElm,a\cup \{p\})$
 otherwise. Hence, on all the states which are not in the existential component
 $\SttSet_\ell$, $\exists^{\CSym}p.\AutName$ behaves as $\AutName$. On the
 states in $\SttSet_\ell$, the projection automaton guesses whether in the
 simulated run of $\AutName$, proposition $p$ marks the current input node or
 not, and proceeds according to the guess and the transition function of
 $\AutName$.  By Definition~\ref{def:HFATASeparatedOneDirection}, we easily
 obtain the following result (for details, see Appendix~\ref{app:ProjectionSeparatedInOneDirection}).

 \begin{restatable}{proposition}{propProjectionSeparatedInOneDirection}\label{prop:ProjectionSeparatedInOneDirection} Let $\AutName$  be an \HFTA over $2^{\Prop}$ that is nondeterministic in one path and $p\in \Prop$. Then, $\LangFun(\exists^{\CSym}p.\AutName)= \exists^{\CSym}p.\LangFun(\AutName)$.
 \end{restatable}

\noindent \textbf{From HFTA to HFTA that are nondeterministic in one path.}
We now show that \HFTA can be effectively translated into equivalent \HFTA that are nondeterministic in one path.
We first establish a preliminary result on the one-step logic $\FOEOne{\SttSet}$
for a given  non-empty set $\SttSet$.

\begin{definition}\label{def:simulatedConstraint} Let $\theta$ be a first-order $\SttSet$-constraint and
$\theta_s$ be a first-order $(\SttSet\cup 2^{\SttSet})$-constraint which is $2^{\SttSet}$-functional in one direction. We say that
$\theta_s$ simulates $\theta$ if the following hold:
\begin{itemize}
  \item for each minimal model $(\SSet,\ISet)$ of $\theta$ and for each $s\in \SSet$, 
  $(\SSet,\ISet\text{$[s \rightarrow \{\ISet(s)\}]$})$ is a model of $\theta_s$;
  \item for each minimal model $(\SSet,\ISet)$ of $\theta_s$, let $s\in \SSet$ be the unique element in $\SSet$ such that
  $\ISet(s)$ is of the form $\{\SttSet'\}$ for some $\SttSet'\in 2^{\SttSet}$.
  Then,  the pair $(\SSet,\ISet\text{$[s \rightarrow \SttSet']$})$ is a model of $\theta$;
\end{itemize}
where the mappings $\ISet\text{$[s \rightarrow \{\ISet(s)\}]$}$ and $\ISet\text{$[s \rightarrow \SttSet']$}$ are defined in the obvious way.
\end{definition}

Since each first-order $\SttSet$-constraint is effectively equivalent to a
disjunction  of basic formulae~\cite{CFVZ20}, we easily obtain the following
result (for details, see Appendix~\ref{app:ConstraintSeparationInOneDirection}).

\begin{restatable}{proposition}{propConstraintSeparationInOneDirection}\label{prop:ConstraintSeparationInOneDirection}
Let $\theta$ be a first-order $\SttSet$-constraint. Then, one can construct a first-order $(\SttSet\cup 2^{\SttSet})$-constraint $\theta_s$ which is $2^{\SttSet}$-functional in one direction and simulates $\theta$.
\end{restatable}

Fix an \HFTA   $\AutName = \tuple {\Sigma} {\SttSet} {\trnFun} {\isttElm} {\Family}{\Family_\exists} {\Omega}$ with
$\Family=\tpl{\SttSet_1,\ldots,\SttSet_n}$.
We construct in two stages  an equivalent \HFTA $\Sim(\AutName)$ that is nondeterministic in one path. First, by a kind of powerset construction, we construct
an automaton $\AutName_{\SEP}$ that is nondeterministic in one path but the acceptance condition of the existential component $\Pow_{\AutName}$
 containing the initial state is not a B\"{u}chi condition but an $\omega$-regular set over the infinite sequences on $\Pow_{\AutName}$. In the second stage of the construction, we show how the $\omega$-regular condition can be converted into a standard B\"{u}chi condition by equipping the  ``macro" states in $\Pow_{\AutName}$   with additional information.
 Intuitively, given an input tree $(\TSet,\Lab)$ accepted by $\AutName$, the automaton $\AutName_{\SEP}$ simulates the behaviour of $\AutName$ along an accepting run $r$ over $(\TSet,\Lab)$ by guessing an infinite path $\pi$ of the input tree  from the root and proceeding as follows:
\begin{itemize}
  \item in the input nodes $w\notin \pi$, $\AutName_{\SEP}$ simply simulates the behaviour of $\AutName$ along $r$;
  \item in the input nodes $w\in \pi$, $\AutName_{\SEP}$ keeps track in its ``macro" state (a state in the existential component $\Pow_{\AutName}$) of the states of $\AutName$ associated with the copies of $\AutName$ that read $w$ along $r$. Thus, in the run of $\AutName_{\SEP}$, there is a unique infinite path $\nu$ from the root associated with the guessed input path $\pi$, and $\nu$ ``collects'' the set of parallel paths $\nu_r$ of the simulated run of $\AutName$ associated with the input path $\pi$. In order to check the acceptance condition on the individual parallel paths $\nu_r$, an infinite sequence of ``macro" states $\rho$ must allow to distinguish the individual infinite paths over $\SttSet$ grouped by $\rho$.
       Thus, like in~\cite{Wal02}, a ``macro" state associated with an input node $w$
      is a set of pairs $(\sttElm_p,\sttElm)$:   the pair $(\sttElm_p,\sttElm)$ represents a copy of $\AutName$ in state $\sttElm$ at node $w$ along the simulated run $r$ which has been generated by a copy of $\AutName$ in state $\sttElm_p$ reading the parent node of $w$ in the input tree.
\end{itemize}

Formally, we denote by $\Pow_{\AutName}$ the subset of $2^{\SttSet\times \SttSet}$ consisting of the sets of pairs $(\sttElm,\sttElm')$ of $\AutName$-states such that the order of $\sttElm'$ is equal or lower than the order of $\sttElm$ (\emph{order requirement}).
A \emph{$\Pow_{\AutName}$-path} $\nu$ is an infinite word
$\nu=\PSet_0\PSet_1\ldots$ over $\Pow_{\AutName}$ such that the following
conditions are fulfilled:
(i) $\PSet_0 = \{(\isttElm,\isttElm)\}$ (\emph{initialisation}), and (ii)  for
all $i\geq 0$ and $(\sttElm_i,\sttElm_{i+1})\in \PSet_{i+1}$, there is an
element of $\PSet_i$ of the form   $(\sttElm_{i-1},\sttElm_{i})$
(\emph{consecution}).
An \emph{$\AutName$-path} of $\nu$ is a maximal  (possibly finite) non-empty sequence $\sttElm_0 \sttElm_1\ldots$ of $\AutName$-states such that   $(\sttElm_{i-1},\sttElm_{i})\in\PSet_{i}$ for all $1\leq i <|\nu|$. The $\Pow_{\AutName}$-path $\nu$ is  $\AutName$-accepting  if all infinite $\AutName$-paths of $\nu$  satisfy the parity condition $\Omega$ of $\AutName$.
The automaton $\AutName_{\SEP}$ is then given by
$
\AutName_{\SEP} = \tuple {\Sigma} {\SttSet\cup  \Pow_{\AutName}} {\trnFun_{\SEP}} {\{(\isttElm,\isttElm)\}} {\Family_{\SEP}}{\Family_\exists\cup \{\Pow_{\AutName}\}} {\Omega}
$
where $\Family_{\SEP}=\tpl{\SttSet_1,\ldots,\SttSet_n,\Pow_{\AutName}}$ (the existential component $\Pow_{\AutName}$ has highest order) and $\trnFun_{\SEP}$ is defined as follows:
\begin{itemize}
  \item for all $\sttElm\in\SttSet$ and $a\in\Sigma$, $\trnFun_{\SEP}(\sttElm,a)=\trnFun(\sttElm,a)$;
  \item for all $\PSet\in \Pow_{\AutName}$ and $a\in\Sigma$,  if $\PSet$ is empty, then
  $\trnFun_{\SEP}(\PSet,a)=\exists x.\, \PSet(x)$; otherwise, let us consider the first-order $(\SttSet\times\SttSet)$-constraint
  $\theta$ given by $\bigwedge_{(\sttElm_p,\sttElm)\in \PSet} \trnFun_{\sttElm}(\sttElm,a)$, where $\trnFun_{\sttElm}(\sttElm,a)$
  is obtained from $\trnFun(\sttElm,a)$ by replacing each predicate $\sttElm'(y)$ occurring in
  $\trnFun(\sttElm,a)$ with $(\sttElm,\sttElm')(y)$. By Proposition~\ref{prop:ConstraintSeparationInOneDirection},
  one can construct a first-order $((\SttSet\times \SttSet)\cup \Pow_{\AutName})$-constraint $\theta_s$ which is $\Pow_{\AutName}$-functional in one direction and simulates $\theta$. Then, $\trnFun_{\SEP}(\PSet,a)$ is obtained from
  $\theta_s$ by replacing each predicate $(\sttElm,\sttElm')(y)$ occurring in $\theta_s$ associated with an element of
  $\SttSet\times \SttSet$ with $\sttElm'$. Note that $\trnFun_{\SEP}(\PSet,a)$ satisfies the first-order existential requirement and is $\Pow_{\AutName}$-functional in one direction.
\end{itemize}
Note that in the definition of $\AutName_{\SEP}$, no acceptance condition is defined for the macro states in $\Pow_{\AutName}$ (the parity condition $\Omega$ inherited by $\AutName$ is defined only on the states in $\SttSet$). By construction and Proposition~\ref{prop:ConstraintSeparationInOneDirection},
  for each run $r$ of $\AutName_{\SEP}$ over an input $(\TSet,\Lab)$ and every infinite path $\pi$  of $r$ starting at the root,
 either $\pi$ is associated with a $\Pow_{\AutName}$-path $\nu$ (in this case, we say that $\pi$ is accepting if $\nu$ is accepting)
 or $\pi$ gets trapped into some non-transient component of $\AutName$ (in this case, acceptance of $\pi$ is determined by the parity condition $\Omega$).
 We denote by $\LangFun(\AutName_{\SEP})$ the set of input trees $(\TSet,\Lab)$ such that there is a run of $\AutName_{\SEP}$ over $(\TSet,\Lab)$ whose infinite paths starting at the root are all accepting. By construction and Proposition~\ref{prop:ConstraintSeparationInOneDirection}, we easily deduce the following crucial result.

\begin{lemma}\label{lemma:SimulatingLemmaFirstStage}
$\AutName_{\SEP}$ is nondeterministic in one path and $\LangFun(\AutName_{\SEP})= \LangFun(\AutName)$.
\end{lemma}

\noindent \textbf{Construction of the Automaton $\Sim(\AutName)$.}  Let
$\FSet_{\Buchi}$ (resp., $\FSet_{\coBuchi}$) be the set of states in the
existential (resp., universal) components of $\AutName$ having even (resp., odd)
color. Fix a $\Pow_{\AutName}$-path $\nu$.  By the order requirement, each
infinite $\AutName$-path of $\nu$ gets trapped into an existential or universal
component of $\AutName$. Thus, by the hesitant acceptance requirement of
$\HFTA$, the $\Pow_{\AutName}$-path $\nu$ is $\AutName$-accepting if and only if
for each infinite \emph{$\AutName$-path} $\pi$ of $\nu$, the following holds: if
$\pi$ gets trapped into an existential component, then $\pi$ visits
\emph{infinitely} many times some state in $\FSet_{\Buchi}$ (\emph{B\"{u}chi
condition}); otherwise (i.e., $\pi$ gets trapped into an universal component),
$\pi$ visits \emph{finitely} many times all the states in $\FSet_{\coBuchi}$
(\emph{coB\"{u}chi condition}).

It is known that coB\"{u}chi alternating word automata (\AWA) over infinite words can be converted in quadratic time into  equivalent
B\"{u}chi \AWA by means of the so called \emph{ranking
construction}~\cite{KV01}.
We adapt the ranking construction and the Miyano-Hayashi construction~\cite{MH84} (for converting a B\"{u}chi \AWA into an equivalent
B\"{u}chi \NWA) for providing a characterisation of acceptance of
$\Pow_{\AutName}$-paths $\nu$ by a classical B\"{u}chi condition on an extension
of $\nu$ obtained by adding to each macro-state visited by $\nu$ additional
finite-state information. Hence, we obtain the following result (for a proof,
see Appendix~\ref{app:SimulatingTheorem}).

\begin{restatable}{theorem}{theoSimulatingTheorem}\label{theorem:SimulatingTheorem}
For the given \HFTA $\AutName$, one can construct an \HFTA $\Sim(\AutName)$ that is nondeterministic in one path and such that
$\LangFun(\Sim(\AutName))= \LangFun(\AutName)$.
\end{restatable}

 By Theorem~\ref{theorem:SimulatingTheorem} and Proposition~\ref{prop:ProjectionSeparatedInOneDirection}, we obtain the following result.

 \begin{corollary}\label{cor:ClosureChainProjectionHFATA} The class of  \HFTA is effectively closed under chain projection.
 \end{corollary}

An \HFTA with transition function $\trnFun$ is in \emph{normal form} if over existential (resp., universal) components $\SttSet_\ell$, $\trnFun(\sttElm,a)$
(resp., the dual of $\trnFun(\sttElm,a)$) is $\SttSet_\ell$-functional in one direction for all $\sttElm\in \SttSet_\ell$ and $a\in\Sigma$. Since the constructions for the
Boolean language operations and the construction for the closure under chain projection (Theorem~\ref{theorem:SimulatingTheorem} and Proposition~\ref{prop:ProjectionSeparatedInOneDirection}) preserve the normal form, we deduce the following result (for a proof, see Appendix~\ref{app:FromMCLtoHFATA}).

 \begin{restatable}{theorem}{theoFromMCLtoHFATA}\label{theo:FromMCLtoHFATA}  Given an \MCL sentence $\varphi$, one can construct an \HFTA $\AutName_\varphi$ in normal form such that  $\LangFun(\AutName_{\varphi})= \LangFun(\varphi)$.
\end{restatable}

We exploit the normal form for  showing that  \CDL  (or, equivalently, the class
of symmetric \HGTA) provides a characterisation of the bisimulation-fragment of
\MCL.
It is known~\cite{Wal02,CFVZ20} that for each \FTA $\AutName$, one can construct
a symmetric \FTA $\AutName_S$ such that if $\LangFun(\AutName)$ is
bisimulation-closed, then  $\AutName$ and $ \AutName_S$ accept the same
tree-language. By adapting the construction given in~\cite{Wal02,CFVZ20},
we can show that
a similar result holds for \HFTA in normal form versus symmetric \HGTA. Hence, by Theorems~\ref{theo:FromHGATAtoCDL} and~\ref{theo:FromCDLtoHGATA} and Theorem~\ref{theo:FromMCLtoHFATA}, we deduce the following result (for details, see
Appendix~\ref{app:BisimulationInvariantMCL}).

\begin{restatable}{theorem}{theoBisimulationInvariantMCL}\label{theo:BisimulationInvariantMCL} The bisimulation-invariant fragment of \MCL,   \CDL, and the class of
    symmetric \HGTA are expressively equivalent in a constructive way.
\end{restatable}




\section{Conclusion}

This work provides automata-theoretic characterisations of branching-time
temporal logics, mainly focusing on \CTLS and \CDL, the latter being a syntactic
variant of the already known \ECTLS.
Specifically, we prove the equivalence between the symmetric variant of classic
ranked Hesitant Tree Automata (\HTA) and both \CDL and the
bisimulation-invariant fragment of Monadic Chain Logic (\MCL).
The full \MCL, instead, is proved equivalent to a first-order variant of
{\HTA}s.
In addition, we close a longstanding gap in the expressiveness landscape of
branching-time logics, by providing an automata-theoretic characterisation of
\CTLS.  This is obtained via a generalisation to tree-languages of the notion of
counter-freeness, originally introduced in the context of word languages.  The
generalisation essentially decomposes an \HTA into a number of counter-free word
automata, one for each level of the state decomposition of the \HTA.  This
decomposition, however, works correctly only when the \HTA satisfies the
additional property of mutual-exclusion.  The property requires that different
sets of automaton states, active at the same time on a given node of the input
tree, must accept different subtrees.  Both mutual-exclusion and
counter-freeness seem to be essential to capture a meaningful notion of
counter-freeness for tree automata.
Together these results bring the expressiveness landscape for branching-time
temporal logics to almost the same level as their linear-time counterparts.

There are few open questions remaining.
%
In particular, while Theorem~\ref{theo:BisimulationInvariantMCL} establishes the
equivalence between the bisimulation invariant fragment of \MCL and \CDL, the
precise relationship between \CCDL (hence, \ECTLS) and full \MCL still remains
unsettled.
In addition, techniques similar to those used in this work may also be
applicable to obtain a characterisation of Monadic Tree Logic (\MTL), a fragment
of \MSO where quantified variables range over subtrees~\cite{BBMP23}, and of
Substructure Temporal Logic (\STL), a temporal logic where one can implicitly
predicate over substructure/subtrees~\cite{BMM13,BMM15}.  The restriction that
variables range over trees, indeed, seem to be tightly connected with the notion
of counter-freeness.  The difficulty in this case is that counter-free {\HTA}s
would not suffice, since both \MTL and \STL are strictly more expressive than
\CTLS, and a meaningful definition of decomposition into word automata of a
non-hesitant tree automaton is not immediately obvious.



  \bibliographystyle{plainurl}
  \bibliography{References}

  \newpage
  \appendix



\section{Preliminaries}

\subsection{Bisimulation}\label{sec:Bisimulation}

Bisimulation is a behavioral equivalence relation  between systems.
For the class of labeled trees, it is formalized as follows.
Let $\LT=(\TSet, \Lab)$ and $\LT'=(\TSet', \Lab')$ be two $\Sigma$-labeled trees.
A \emph{bisimulation} between  $\LT$ and $\LT'$ is a binary equivalence relation $\RSet\subseteq \TSet\times \TSet'$ satisfying
the following conditions for all $(w,w')\in\RSet$: (\emph{atom}) $\Lab(w)=\Lab'(w)$, (\emph{forth}) for each child $v$ of $w$ in $\TSet$,
there is a child $v'$ of $w'$ in $\TSet'$ such that $(v,v')\in \RSet$, and (\emph{back}) for each child $v'$ of $w'$ in $\TSet'$,
there is a child $v$ of $w$ in $\TSet$ such that $(v,v')\in \RSet$. $\LT$ and $\LT'$ are bisimilar
if there exists a bisimulation $\RSet$ between $\LT$ and $\LT'$ containing $(\varepsilon,\varepsilon)$.
A tree-language $\LangFun$ over $\Sigma$ is \emph{bisimulation-closed} if for all bisimilar $\Sigma$-labeled trees
$\LT$ and $\LT'$, it holds that $\LT\in \LangFun$ iff $\LT'\in \LangFun$.
Given a formalism $\FName$ whose specifications $\xi$
denote tree-languages $\LangFun(\xi)$, $\FName$ is \emph{bisimulation-invariant}  if for each specification $\xi$ of $\FName$, $\LangFun(\xi)$ is bisimulation-closed.

 \subsection{Monadic Second-Order Logic and Monadic Chain Logic}\label{sec:MSOFragments}

In this section, we recall standard Monadic Second-order Logic (\MSO for short) interpreted over arbitrary Kripke trees.
We focus on the well-known fragment  of \MSO, namely Monadic Chain Logic (\MCL), where
second-order quantification is restricted to chains  of the given Kripke tree.
 
For a given finite set $\Prop$ of atomic propositions, \MSO  is a second-order language
defined over the signature $\{\leq\}\cup \{p \mid p\in \Prop\}$, where
second-order quantification is restricted to monadic predicates, $\leq$ is a
binary predicate, and $p$ is interpreted as a monadic predicate for each
$p\in\Prop$.

Given a tree $\TSet$ with set of directions in $\DSet$ and a node $w\in\TSet$,  a
\emph{descendant}  of $w$ in $\TSet$ is a node in $\TSet$ of the form $w\cdot
w'$ for some $w'\in \DSet^{*}$. \vspace{0.1cm}

 \noindent \textbf{Syntax of \MSO.}
  Given a finite set $\Prop$ of atomic propositions, a finite set $\Var_1$
  of first-order variables (or \emph{node} variables), and a finite set $\Var_2$
of second-order variables (or \emph{set} variables), the syntax of \MSO   is the set of formulas
built according to the following grammar:
\[
\varphi \seteq  p(x) \mid x \leq y \mid x \in
X \mid \neg \varphi \mid \varphi  \wedge \varphi
\mid \exists x \ldotp \varphi
\mid \exists  X \ldotp \varphi
\]
where $p \in \Prop$, $x,y \in \Var_1$, and $X \in \Var_2$.
We also exploit the
standard logical connectives $\vee$ and $\rightarrow$ as abbreviations, the
universal first-order quantifier $\forall x$, defined as $\forall
x.\varphi\defeq \neg\exists x. \neg\varphi$, and the universal second-order
quantifier $\forall  X$, defined as $\forall
X\ldotp\varphi\defeq \neg\exists X\ldotp \neg\varphi$.
We may also
make use of the shorthands (i) $x=y$ for $x\leq y \wedge y\leq x$, (ii) $x<y$
for $x\leq y \wedge \neg (y\leq x)$; (iii) $\exists x\in \XvarElm.\,\varphi$ for
$\exists x.\,(x\in \XvarElm\wedge \varphi)$, and (iv) $\forall x\in
\XvarElm.\,\varphi$ for $\forall x.\,(x\in \XvarElm \rightarrow \varphi)$. 
Moreover, the child relation is definable in \MSO  by the binary predicate
     $\Child(x,y) \defeq x<y \wedge \neg \exists z.\,(x<z\wedge z<y) $ which exploits
     only first-order quantification.

 As usual, a \emph{free variable} of a formula $\varphi$ is a variable
 occurring in $\varphi$ that is not bound by a quantifier.  A \emph{sentence} is
 a formula with no free variables.  The language of \MSO  consists of its
 sentences. \vspace{0.1cm}

\noindent \textbf{Semantics of \MSO.}
Formulas of $\MSO$ are interpreted over Kripke trees over $\Prop$.
A Kripke tree $\TName=(\TSet,\Lab)$ induces the relational structure with domain $\TSet$,
where the binary predicate $\leq$ corresponds to the descendant
relation in $\TSet$, and $p(x)$ denotes the set of
$p$-labeled nodes.

Given a Kripke tree   $\TName$,  a \emph{first-order
valuation for   $\TName$} is a mapping $\Val_1: \Var_1 \mapsto \TSet$ assigning to
each first-order variable a node of $\TSet$. A \emph{second-order valuation for $\TName$} is a
mapping $\Val_2: \Var_2 \mapsto 2^{\TSet}$ assigning to each
second-order variable a subset of $\TSet$.
Given an $\MSO$ formula $\varphi$, a Kripke tree
 $\TName=(\TSet,\Lab)$ over $\Prop$, a first-order valuation $\Val_1$ for $\TName$, and a
 second-order valuation $\Val_2$ for $\TName$, the satisfaction relation
 $(\TName,\Val_1,\Val_2)\models \varphi$, meaning that $\TName$ satisfies the formula
 $\varphi$ under the valuations $\Val_1$ and $\Val_2$, is
 defined as follows (the treatment of Boolean connectives is
 standard):
\[
 \begin{array}{l@{\hspace{2pt}}l}
 \text{$(\TName, \Val_1,\Val_2)\models  p(x)$}  &   \Leftrightarrow \text{$p\in \Lab(\Val_1(x))$};\\
 \text{$(\TName,\Val_1,\Val_2)\models x\leq y$} & \Leftrightarrow \text{$\Val_1(y)$ is a descendant of  $\Val_1(x)$ in $\TSet$};\\
 \text{$(\TName,\Val_1,\Val_2)\models x\in X$} & \Leftrightarrow \text{$\Val_1(x)\in \Val_2(X)$};\\
 \text{$(\TName,\Val_1,\Val_2)\models \exists x\ldotp \varphi$} & \Leftrightarrow \text{$(\TName,\Val_1[x \mapsto w],\Val_2) \models \varphi$  for some    $w\in \TSet$};\\
 \text{$(\TName,\Val_1,\Val_2)\models \exists X\ldotp \varphi$} & \Leftrightarrow
  \text{$(\TName,\Val_1,\Val_2[X \mapsto \SSet])\models \varphi$  for some set  of nodes   $\SSet \subseteq \TSet$}.
\end{array}
\]
where $\Val_1[x \mapsto w]$ denotes the
first-order valuation for $\TName$ defined as: $\Val_1[x \mapsto w](x) = w$ and
$\Val_1[x \mapsto w](y)=\Val_1(y)$ if $y\not=x$.
The meaning of notation $\Val_2[X \mapsto \SSet]$ is similar.  

Note that the satisfaction relation $(\TName,\Val_1,\Val_2)\models \varphi$, for
fixed $\TName$ and $\varphi$, depends only on the values assigned by $\Val_1$ and
$\Val_2$ to the variables occurring free in $\varphi$.  In particular, if
$\varphi$ is a sentence, we say that $\TName$ \emph{satisfies} $\varphi$, written
$\TName\models \varphi$, if $(\TName,\Val_1,\Val_2)\models \varphi$ for some valuations
$\Val_1$ and $\Val_2$. In this case, we also say that $\TName$ is a model of
$\varphi$.\vspace{0.1cm}

\noindent \textbf{The fragment \MCL.} We first consider the predicate  $\Chain(X)$ which captures the subsets of the given tree
which are chains.  It can be easily expressed in $\MSO$ by using only first-order
quantification  as $\forall x\in X.\,\forall y\in X.\,(x \leq y\vee y\leq x)$.
Moreover, the \emph{existential chain quantifier} $\exists^{\CSym} X$ ranges over chains of the given tree and is
defined  as  $\exists^{\CSym} X \ldotp \varphi  \defeq \exists  X \ldotp (\Chain(X)\wedge\varphi)$.

The logic \MCL   corresponds to the syntactical fragment of
\MSO  where the second-order existential quantification takes only the form
$\exists^{\CSym} X$. We also consider the universal chain quantifier
 $\forall^{\CSym} X$  defined as:
 $\forall^{\CSym}  X\ldotp\varphi\defeq \neg\exists^{\CSym} X\ldotp \neg\varphi$.




\section{Missing Proofs of Section~\ref{sec:characterizationCDL}}
\label{app:characterizationCDL}

\subsection{Inexpressiveness result of Example~\ref{example:moduloCountingTwo}}\label{app:moduloCountingTwo}

Let $\Prop=\{p\}$ and  $\LangFun_2$ be the tree-language of Example~\ref{example:moduloCountingTwo}. Recall that $\LangFun_2$  consists of the Kripke trees $\LT$ such that there is an infinite path $\pi$ from the root so that $p$ never holds along $\pi$ and at the even positions $2i$, the \CTLStar formula $\EQ\Next p$ holds at node $\pi(2i)$.
In this section, we show that $\LangFun_2$ cannot be expressed in
 \CCTLStar.

 For each $n\geq 1$, let $\LT_n$ be the Kripke tree over $\Prop$ such that there exists an infinite path $\pi$ from the root (called \emph{main path})
 so that the following conditions hold:
 \begin{itemize}
   \item $p$ never holds along $\pi$;
   \item $\pi(n+1)$ has a unique child in $\LT_n$ (note that this child is $\pi(n+2)$);
   \item for each $i\in \SetN\setminus \{n+1\}$, there is a unique child $w_i$ of $\pi(i)$ which is not in $\pi$. Moreover,
   the subtree rooted at $w_i$ consists of a unique infinite path encoding the infinite word $\{p\}\emptyset^{\omega}$.
 \end{itemize}

\noindent Evidently, by construction, the following holds.

 \begin{remark}\label{remark:treeFamily} For each $n\geq 1$, if $n$ is even then $\LT_n\in \LangFun_2$; otherwise, $\LT_n\notin \LangFun_2$.
 \end{remark}

 We show that for each $n\geq 1$, no $\CCTLStar$ formula $\varphi$ of length $|\varphi|$ smaller or equal to  $n$ distinguishes the Kripke trees
$\LT_n$ and $\LT_{n+1}$. Hence, by Remark~\ref{remark:treeFamily}, no \CCTLStar formula can express the language $\LangFun_2$.

\begin{lemma}\label{lemma:inexpressivenessCCTLStar} For each $n\geq 1$ and \CCTLStar formula $\varphi$ such that $|\varphi|\leq n$, it holds that
$\LT_n\models \varphi$ if and only if $\LT_{n+1}\models \varphi$.
\end{lemma}
\begin{proof} Fix $n\geq 1$.  Let
$\pi_{n+1}$ be the main path of $\LT_{n+1}$. Note that by construction the labeled subtree of $\LT_{n+1}$ rooted at node $\pi_{n+1}(1)$ is isomorphic to $\LT_n$. Thus,
the result directly follows from the following claim. \vspace{0.2cm}

\noindent \emph{Claim 1.} Let $0\leq i\leq n$ and $\psi$ be a \CCTLStar path formula such that $|\psi|\leq n-i$. Then,
$(\LT_{n+1},\pi_{n+1},i)\models \psi$ if and only if  $(\LT_{n+1},\pi_{n+1},i+1)\models \psi$. \vspace{0.2cm}

We prove Claim~1 by structural induction on $\psi$. The case where $\psi$ is the  atomic proposition $p$ is trivial, and the cases where the root operator of $\psi$ is a Boolean connective directly follows from the induction hypothesis. For the other cases, we proceed as follows.
\begin{itemize}
  \item $\psi = \Next\psi_1$. Since $0\leq |\psi_1|< |\psi|$ and $|\psi|\leq n-i$, we have that $i+1\leq n$ and $|\psi_1|\leq n-(i+1)$. By the induction hypothesis,
   $(\LT_{n+1},\pi_{n+1},i+1)\models \psi_1$ if and only if  $(\LT_{n+1},\pi_{n+1},i+2)\models \psi_1$. Hence, the result follows.
  \item $\psi = \psi_1\Until \psi_2$: let
  $(\LT_{n+1},\pi_{n+1},i)\models \psi$. By the semantics of the until modality, either $(\LT_{n+1},\pi_{n+1},i+1)\models \psi$ or $(\LT_{n+1},\pi_{n+1},i)\models \psi_2$. In the  second case, by the induction hypothesis, $(\LT_{n+1},\pi_{n+1},i+1)\models \psi_2$. Hence, $(\LT_{n+1},\pi_{n+1},i+1)\models \psi$ and the result holds.
  Now, assume that $(\LT_{n+1},\pi_{n+1},i+1)\models \psi$. Then either $(\LT_{n+1},\pi_{n+1},i+1)\models \psi_2$, or $(\LT_{n+1},\pi_{n+1},i+1)\models \psi_1$ and
    $(\LT_{n+1},\pi_{n+1},i+2)\models \psi$. In the first case, by the induction hypothesis, we obtain that $(\LT_{n+1},\pi_{n+1},i)\models \psi_2$, and the result follows.
    In the second case, by the induction hypothesis, $(\LT_{n+1},\pi_{n+1},i)\models \psi_1$. Hence, by the semantics of the until modality,
      $(\LT_{n+1},\pi_{n+1},i)\models \psi$, and we are done.
  \item $\psi = \DC^{k}\psi_1$ for some state formula $\psi_1$: by construction, there is a unique child $w_i$ of $\pi(i)$ which is not in $\pi$ and a unique child  $w_{i+1}$ of $\pi(i+1)$ which is not in $\pi$. Moreover, the labeled subtree of $\LT_{n+1}$ rooted at node $w_i$ is isomorphic to the labeled
  subtree of $\LT_{n+1}$ rooted at node $w_{i+1}$. By the induction hypothesis,
  $(\LT_{n+1},\pi_{n+1},i)\models \psi_1$ iff $(\LT_{n+1},\pi_{n+1},i+1)\models \psi_1$. Hence, the result directly follows.
  \item $\psi = \EQ\psi_1$: this case is similar to the previous one. \qedhere
\end{itemize}
\end{proof}

By Lemma~\ref{lemma:inexpressivenessCCTLStar}, we obtain the desired result.

 \begin{corollary} There is no \CCTLStar formula $\varphi$ such that $\LangFun(\varphi)=\LangFun_2$.
 \end{corollary}

\subsection{Proof of Proposition~\ref{prop:SubclassesGATABooleanClosure}}\label{app:SubclassesGATABooleanClosure}

\propSubclassesGATABooleanClosure*
\begin{proof}
Let $\AutName$ be a  \HGTA (resp., \HGTAcf). By construction and Proposition~\ref{prop:DualGATA},
the dual automaton $\widetilde{\AutName}$ of
$\AutName$ is an \HGTA (resp., an \HGTA satisfying the counter-free requirement) accepting the complement of $\LangFun(\AutName)$. Moreover, if $\AutName$
satisfies the mutual-exclusion condition, then $\widetilde{\AutName}$ satisfies the mutual-exclusion condition as well.
Hence, for the complementation language operation, the result follows.
For union and intersection, we focus on the intersection language operation (the construction for union being similar).
Let $\AutName = \tuple {\Sigma} {\SttSet} {\trnFun} {\isttElm} {\Family}{\Family_\exists} {\Omega}$ and
$\AutName' = \tuple {\Sigma} {\SttSet'} {\trnFun'} {\isttElm'} {\Family'}{\Family_\exists'} {\Omega'}$ be two \HGTA (resp., \HGTAcf) where
$\Family=\tpl{\SttSet_1,\ldots,\SttSet_n}$ and $\Family'=\tpl{\SttSet'_1,\ldots,\SttSet'_{n'}}$. We can assume that
$\SttSet$ and $\SttSet'$ are disjoint. We define
$\AutName = \tuple {\Sigma} {\SttSet\cup \SttSet'\cup\{\isttElm_{\cap}\}} {\trnFun_{\cap}} {\isttElm_{\cap}} {\Family_{\cap}}{\Family_{\exists,\cap}} {\Omega_{\cap}}$,
where $\isttElm_{\cap}$ is a fresh  state, $\Family_{\cap}= \tpl{\SttSet_1,\ldots,\SttSet_n,\SttSet'_1,\ldots,\SttSet'_{n'},\{\isttElm_{\cap}\}}$, $\Family_{\exists,\cap}= \Family'_{\exists}\cup \Family''_{\exists}$, $\Omega_\cap = \Omega'\cup \Omega'' \cup (\isttElm_{\cap} \rightarrow 0)$  and $\delta_\cap$ is defined as follows. For states in $\SttSet$ and $\SttSet'$, $\trnFun_{\cap}$ agrees with $\trnFun$ and $\trnFun'$, respectively (recall that $\SttSet$ and $\SttSet'$  are disjoint). For the state $\isttElm_{\cap}$ and for each $a\in\Sigma$, $\trnFun_{\cap}(\isttElm_{\cap},a)=\trnFun(\isttElm,a)\,\wedge\,\trnFun'(\isttElm',a)$. Thus, from the root of the input tree and in the initial state $\isttElm_{\cap}$, $\AutName_\cap$ sends all the copies sent (initially) by both $\AutName$ and $\AutName'$. The singleton $\{\isttElm_{\cap}\}$ constitutes a transient component with the highest order. Hence, the mutual-exclusion property is preserved and the result easily follows.
\end{proof}

\subsection{Proof of Proposition~\ref{prop:SeparationConditionHGATA}}\label{app:SeparationConditionHGATA}

\propSeparationConditionHGATA*
\begin{proof}
Let $\AutName = \tuple {\Sigma} {\SttSet} {\trnFun} {\isttElm} {\Family}{\Family_\exists} {\Omega}$ with
$\Family=\tpl{\SttSet_1,\ldots,\SttSet_n}$ and   $\widetilde{\AutName} = \tuple {\Sigma} {\widetilde{\SttSet}} {\widetilde{\trnFun}} {\widetilde{\isttElm}} {\widetilde{\Family}}{\widetilde{\Family}_\exists} {\widetilde{\Omega}}$ with $\widetilde{\Family}=\tpl{\widetilde{\SttSet}_1,\ldots,\widetilde{\SttSet}_n}$ be a renaming of the dual automaton
of $\AutName$ which is still an \HGTA. For each state $\sttElm$ of $\AutName$, $\widetilde{\sttElm}$ denotes the renaming of
$\sttElm$. For each atom $\atom\in \Atoms(\AutName)$, the \emph{strong dual} $\Strong(\atom)$ of $\atom$  is the atom defined as follows:
  if $\atom$ is of the form $\DMod[k]\alpha$ (resp., $\BMod[k]\alpha$), then $\Strong(\atom)$ is obtained from the
  dual $\BMod[k]\widetilde{\alpha}$ (resp., $\DMod[k]\widetilde{\alpha}$) by replacing each occurrence of a state $\sttElm$ in $\widetilde{\alpha}$
  with its copy $\widetilde{\sttElm}$. Evidently,  $\Strong(\atom)\in\Atoms(\widetilde{\AutName})$.
  The \emph{strong dual of an atom} in $\Atoms(\widetilde{\AutName})$ is defined similarly. Note that $\Atoms(\widetilde{\AutName})$ is the
  set of strong duals of the atoms in $\Atoms(\AutName)$, and vice versa.
Evidently, by construction, for each $\atom\in \Atoms(\AutName)$, it holds that $\LangFun(\AutName^{\atom})$ is the complement of  $\LangFun((\widetilde{\AutName})^{\,\Strong(\atom)})$. Dually, for each $\atom\in \Atoms(\widetilde{\AutName})$,
$\LangFun((\widetilde{\AutName})^{\,\atom})$ is the complement of $\LangFun(\AutName^{\Strong(\atom)})$.

For each $i\in [1,n]$, an atom $\atom$ in $\Atoms(\AutName)\cup \Atoms(\widetilde{\AutName})$ has order at most $i$ if $\atom$ refers only to states
having  order at most $i$.
For a set of atoms $\ASet\subseteq \Atoms(\AutName)\cup \Atoms(\widetilde{\AutName})$, $\ASet$ has order at most $i$, if each atom in
$\ASet$ has order at most $i$. We say that $\ASet$ is \emph{$i$-complete} if for each atom $\atom\in \Atoms(\AutName)$ of order at most $i$, \emph{exclusively}, either $\atom$ or its strong dual is in $\ASet$.
For each $\ASet \subseteq \Atoms(\AutName)$ (resp., $\ASet \subseteq \Atoms(\widetilde{\AutName})$), we denote by $\Comp(\ASet,i)$ the family of $i$-complete subsets
$\CSet \subseteq \Atoms(\AutName)\cup \Atoms(\widetilde{\AutName})$ such that $\ASet\subseteq \CSet$.

 Then, the  \HGTA $\AutName_s$ equivalent to $\AutName$ and satisfying the mutual-exclusion property is defined as follows:
\[
\AutName_s = \tuple {\Sigma} {\SttSet\cup \widetilde{\SttSet}} {\trnFun_s} {\isttElm} {\Family_s}{\Family_\exists\cup \widetilde{\Family}_\exists } {\Omega\cup \widetilde{\Omega}}
\]
where $\Family_s=\tpl{\widetilde{\SttSet}_1,\SttSet_1,\ldots,\widetilde{\SttSet}_n,\SttSet_n}$ and the transition function $\trnFun_s$ is defined as follows. We focus on the definition of $\trnFun_s(\sttElm,a)$  when $\sttElm$ is in $\SttSet$. The definition of $\trnFun_s$ for the dual states is similar. We distinguish the following cases:
\begin{itemize}
  \item $\sttElm$ is transient: $\trnFun_s(\sttElm,a)=\trnFun(\sttElm,a)$;
  \item $\sttElm$ is in some existential component $\SttSet_i$: recall that $\trnFun(\sttElm,a)$ can be written as a disjunction
  of constraints of the form $\DMod\sttElm'\wedge \Con(\ASet)$, where $\sttElm'\in\SttSet_i$ and $\ASet$ has order at most $i-1$ (note that $\ASet$ may be $\emptyset$). Then, $\trnFun_s(\sttElm,a)=\trnFun(\sttElm,a)$ if $i=1$ (in this case, $\ASet=\emptyset$); otherwise, $\trnFun_S(\sttElm,a)$ is obtained from $\trnFun(\sttElm,a)$ by replacing each conjunct $\DMod\sttElm'\wedge \Con(\ASet)$ with
  $\DMod\sttElm'\wedge \bigvee_{\CSet\in\Comp(\ASet,i-1)}\Con(\CSet)$.
    \item $\sttElm$ is in some universal component $\SttSet_i$: recall that $\trnFun(\sttElm,a)$ can be written as a conjunction
  of constraints of the form $\BMod\sttElm'\vee \Dis(\ASet)$, where $\sttElm'\in\SttSet_i$ and $\ASet$ has order at most $i-1$ (note that $\ASet$ may be empty). Then, $\trnFun_s(\sttElm,a)=\trnFun(\sttElm,a)$ if $i=1$ (in this case, $\ASet=\emptyset$); otherwise, $\trnFun_s(\sttElm,a)$ is obtained from $\trnFun(\sttElm,a)$ by replacing each disjunct $\BMod\sttElm'\vee \Dis(\ASet)$ with
  $\BMod\sttElm'\vee \bigwedge_{\CSet\in\Comp(\ASet,i-1)}\Dis(\CSet)$.
\end{itemize}

\noindent By a straightforward induction on $i\in [1,n]$, one can show that the following conditions hold:
 \begin{enumerate}
   \item for each $\atom\in \Atoms(\AutName)$, $\LangFun(\AutName_s^{\atom})$ is the complement of  $\LangFun(\widetilde{\AutName}^{\,\Strong(\atom)})$;
   \item for each set $\ASet\subseteq \Atoms(\AutName)$ of order at most $i$,
   \[
    \LangFun(\AutName^{\Con(\ASet)}) = \LangFun(\AutName_s^{\bigvee_{\CSet\in\Comp(\ASet,i)}\Con(\CSet)}) \quad\quad\LangFun(\AutName^{\Dis(\ASet)}) =  \LangFun(\AutName_s^{\bigwedge_{\CSet\in\Comp(\ASet,i)}\Dis(\CSet)})
    \]
   \item the variants of the previous two conditions with $\AutName$ (resp., $\widetilde{\AutName}$) replaced with $\widetilde{\AutName}$ (resp., $\AutName$).
 \end{enumerate}

\noindent By construction and Conditions~(1)--(3), it easily follows that $\AutName_s$ satisfies the mutual-exclusion property and $\LangFun(\AutName_s)= \LangFun(\AutName)$.
\end{proof}

\subsection{Remaining cases in the proof of Theorem~\ref{theo:FromHGATAtoCDL}}\label{app:FromHGATAtoCDL}

In this section, we complete the proof of Theorem~\ref{theo:FromHGATAtoCDL}. Recall that
 an \NWA with transition function $\trnFun$  is \emph{deterministic} if for all states $\sttElm$ and input symbols $a$, $\trnFun(\sttElm,a)$ is a singleton $\{\sttElm'\}$ (in this case, we write $\trnFun(\sttElm,a)= \sttElm'$).   We use the acronym \DWA for the subclass of deterministic \NWA.

Let $\AutName = \tuple {\Sigma} {\SttSet} {\trnFun} {\isttElm} {\Family} {\Family_\exists} {\Omega}$ be an \HGTA (resp., an \HGTAcf) over $\Sigma=2^{\Prop}$ satisfying the mutual-exclusion property with $\Family=\tpl{\SttSet_1,\ldots,\SttSet_n}$.
By Proposition~\ref{prop:SeparationConditionHGATA}, if $\AutName$ is an arbitrary \HGTA, we can assume that $\AutName$ satisfies the mutual-exclusion property.
 For each $\sttElm\in \SttSet$, we show that one can construct  a \CCDL  (resp.,  \CCTLStar)  formula $\varphi_{\sttElm}$
such that $\LangFun(\varphi_{\sttElm})= \LangFun(\AutName^{\sttElm})$. Moreover,  
$\varphi_{\sttElm}$ is a $\CDL$ (resp. a \CTLStar) formula if $\AutName$ is symmetric.
 Thus, by setting $\varphi_{\AutName} \DefinedAs \varphi_{\isttElm}$, Theorem~\ref{theo:FromHGATAtoCDL} directly follows. The proof is by induction on the order $\ell$ of the component $\SttSet_\ell$ such that $\sttElm\in \SttSet_\ell$. We distinguish the cases where $\sttElm$ is transient, existential, or universal. Note that by hypothesis, $\AutName^{\sttElm}$ satisfies the mutual-exclusion property.
 For each $a\in 2^{\Prop}$, we denote by $\theta(a)$ the propositional formula given by $\bigwedge_{p\in a} p\wedge \bigwedge_{p\in \Prop \setminus a}\neg p$.  \vspace{0.2cm}

\noindent \textbf{Case where $\sttElm$ is transient:} by construction for each $a\in \Sigma$, $\trnFun(\sttElm,a)$ contains only states having an order smaller than $\sttElm$ (note that for the base case, where the order of $\sttElm$ is $1$, $\trnFun(\sttElm,a)\in\{\true,\false\}$). Thus, by the induction hypothesis, for each state $\sttElm'$ occurring in $\trnFun(\sttElm,a)$, one can construct a  \CCDL (resp., \CCTLStar) formula $\varphi_{\sttElm'}$
such that $\LangFun(\varphi_{\sttElm'})= \LangFun(\AutName^{\sttElm'})$. Let $\psi_{\sttElm,a}$ be the \CCDL (resp., \CCTLStar) formula obtained from
$\trnFun(\sttElm,a)$ by replacing each atom $\DMod[k]\alpha$ (resp., $\BMod[k]\alpha$) occurring in
$\trnFun(\sttElm,a)$  with $\DC^{k}\varphi_{\alpha}$ (resp., $\neg \DC^{k}\neg\varphi_{\alpha}$), where $\varphi_{\alpha}$ is obtained from $\alpha$
by replacing each state $\sttElm'$ occurring in $\alpha$ with $\varphi_{\sttElm'}$. Note that for $k=1$,
$\DC^{1}$ corresponds to the modality $\EQ\Next$.
 Then,
$\varphi_{\sttElm}$ is given by $\bigvee_{a\in\Sigma}(\theta(a)\wedge \psi_{\sttElm,a})$. Correctness of the construction easily follows.\vspace{0.2cm}

\noindent \textbf{Case where $\sttElm$ is existential:} this case has been already illustrated in Section~\ref{sec:characterizationCDL;sub:FromAutomataToLogicsAndBack}  when $\AutName$ is an \HGTAcf.
Now, assume that $\AutName$ is an arbitrary \HGTA satisfying the mutual-exclusion property.
 Let $\SttSet_\ell$ be the existential component such that
$\sttElm \in \SttSet_\ell$ and let us consider the B\"{u}chi \NWA $\AutName_{\SttSet_\ell,\sttElm}$ over
$2^{\Prop}\times \Upsilon_{\SttSet_\ell}$ defined in Definition~\ref{def:linearization}. Recall that
 $\Upsilon_{\SttSet_\ell}\subseteq 2^{\Atoms(\AutName)}$ contains only elements $\ASet$ such that states occurring in the atoms of $\ASet$ have order $j$ lower than $\ell$. Thus, by the induction hypothesis and by proceeding as for the case where $\sttElm$ is a transient state, for each $\ASet\in \Upsilon_{\SttSet_\ell}$, one can construct a \CCDL   formula $\varphi_{\ASet}$ such that
$\LangFun(\AutName^{\Con(\ASet)})= \LangFun(\varphi_{\ASet})$. Therefore, since $\AutName$ satisfies the mutual-exclusion condition,
the following holds:\vspace{0.2cm}

\noindent \emph{Claim 1.} For all
$\ASet,\ASet'\in \Upsilon_{\SttSet_\ell}$ such that $\ASet\neq \ASet'$, $\LangFun(\varphi_{\ASet})\cap \LangFun(\varphi_{\ASet'})=\emptyset$. \vspace{0.2cm}

 By~\cite{Saf88}, one can construct
a parity $\DWA$
$
\DName_{\sttElm}=  \tuple {2^{\Prop}\times \Upsilon_{\SttSet_\ell}} {\SttSet_{D}} {\trnFun_{D}} {\InitState{D}} {\Omega_{D}}
$
such that $\LangFun(\DName_{\sttElm})=\LangFun(\AutName_{\SttSet_\ell,\sttElm})$.
For each $\ASet\in \Upsilon_{\SttSet_\ell}$, let us consider the parity \NWA  $\NName_{\ASet} = \tuple {2^{\Prop}} {\SttSet_{N}} {\trnFun_{N}} { (\InitState{D},\ASet)} {\Omega_{N}}$ over $2^{\Prop}$ with initial state
$(\InitState{D},\ASet)$
  which simulates $\DName_{\sttElm}$ by keeping track in the current state of the
guessed second component of the next input symbol. Formally $\SttSet_{N}= \SttSet_{D}\times \Upsilon_{\SttSet_\ell}$,
  $\trnFun_{N}((\sttElm',\ASet'),a)=\displaystyle{\bigvee_{\ASet''\in \Upsilon_{\SttSet_\ell}}}(\trnFun_{D}(\sttElm',(a,\ASet')),\ASet'')$ and
$\Omega_{N}(\sttElm',\ASet')= \Omega_{D}(\sttElm')$ for all $\sttElm'\in \SttSet_{D}$, $a\in 2^{\Prop}$, and
$\ASet'\in \Upsilon_{\SttSet_\ell}$. Note that for all distinct $\ASet,\ASet'\in \Upsilon_{\SttSet_\ell}$, the parity \NWA
  $\NName_{\ASet}$ and $\NName_{\ASet'}$ differ only for the initial state.
  Moreover, let $\tau$ be the testing function assigning
  to each state $(\sttElm',\ASet')\in \SttSet_{N}$ the \CCDL formula $\varphi_{\ASet'}$.
  Since $\LangFun(\AutName^{\Con(\ASet)})= \LangFun(\varphi_{\ASet})$ for all $\ASet\in \Upsilon_{\SttSet_\ell}$, by Proposition~\ref{prop:PropertiesHGATALineariztion}, we obtain the following characterization of the   language $\LangFun(\AutName^{\sttElm})$.\vspace{0.2cm}

\noindent \emph{Claim 2.} For each Kripke tree $\LT = (\TSet,\Lab)$,
$\LT\in \LangFun(\AutName^{\sttElm})$ iff for some $\ASet\in \Upsilon_{\SttSet_\ell}$, there exists an accepting run $\nu$ of
$\NName_{\ASet}$ over $\Lab(\pi(0))\Lab(\pi(1))\ldots$ such that  $(\LT,\pi(i))\models \Test(\nu(i))$ for all $i\geq 0$.\vspace{0.2cm}

  We now show that the characterization of the language $\LangFun(\AutName^{\sttElm})$ in Claim~2 can be captured by a \CCDL formula. For all states
  $(\sttElm',\ASet')\in \SttSet_{N}$ and set $\PSet \subseteq \SttSet_{N}$, we denote by $_{(\sttElm',\ASet')}\NName_{\PSet}$ the testing \NWA[\fin] with test function $\Test$ and whose embedded \NWA[\fin] is obtained from the automata $\NName_{\ASet}$ by setting a fresh copy of   $(\sttElm',\ASet')$ as initial state,  and $\PSet$ as set of accepting states. This fresh copy behaves as $(\sttElm',\ASet')$ and has the same test as $(\sttElm',\ASet')$, and ensures that the automaton cannot accept the empty word.
  Finally, let $\SttSet_{N,\Even}$ be the set of states in $\SttSet_{N}$ having even color, and for each
  $(\sttElm',\ASet')\in \SttSet_{N}$, let $\SttSet_{N}>(\sttElm',\ASet')$ be the set of states in $\SttSet_{N}$ having color greatest than the color of $(\sttElm',\ASet')$. We consider the  \CCDL formula $\varphi_{\sttElm}\DefinedAs \EQ\psi_{\sttElm}$ where the path \CCDL formula $\psi_{\sttElm}$  is defined as follows:
\[
\begin{array}{l}
\text{$\psi_{\sttElm} \DefinedAs  \displaystyle{\bigvee_{\ASet\in \Upsilon_{\SttSet_\ell}}\bigvee_{(\sttElm',\ASet')\in \SttSet_{N,\Even}}}(\psi_1(\ASet,\sttElm',\ASet')\wedge \psi_2(\ASet,\sttElm',\ASet'))$}\\
\text{$\psi_1(\ASet,\sttElm',\ASet')\DefinedAs \Seq{_{(\InitState{D},\ASet)}\NName_{\{(\sttElm',\ASet')\}}}\USeq{_{(\sttElm',\ASet')}\NName_{\SttSet_N>(\sttElm',\ASet')}}\neg\top$}\\
\text{$\psi_2(\ASet,\sttElm',\ASet')\DefinedAs \USeq{_{(\InitState{D},\ASet)}\NName_{\{(\sttElm',\ASet')\}}}
\Seq{_{(\sttElm',\ASet')}\NName_{\{(\sttElm',\ASet')\}}}\top$}
\end{array}
\]
By Claim~2, correctness of the construction directly follows from the following claim whose proof relies on the mutual-exclusion condition expressed in Claim~1. \vspace{0.2cm}

\noindent \emph{Claim 3.} For each Kripke tree $\LT = (\TSet,\Lab)$ and infinite  path $\pi$ from the root,
$(\LT,\pi,0)\models \psi_{\sttElm}$ iff for some $\ASet\in \Upsilon_{\SttSet_\ell}$, there exists an accepting run $\nu$ of
$\NName_{\ASet}$ over $\Lab(\pi(0))\Lab(\pi(1))\ldots$ such that  $(\LT,\pi(i))\models \Test(\nu(i))$ for all $i\geq 0$.\vspace{0.1cm}

\noindent \emph{Proof Claim 3.} The left-right implication easily follows from construction. For the right-left implication,
assume that for some $\ASet \in \Upsilon_{\SttSet_\ell}$, there exists an accepting run $\nu= (\sttElm_0,\ASet_0)(\sttElm_1,\ASet_1)\ldots $ of $\NName_{\ASet}$  over $\rho=\Lab(\pi(0))\Lab(\pi(1))\ldots$ with
$(\sttElm_0,\ASet_0)= (\InitState{D},\ASet)$ such that $(\LT,\pi(i))\models \varphi_{\ASet_i}$ for all $i\geq 0$.
Since $\nu$ is accepting there exists a state $(\sttElm',\ASet')\in \SttSet_{N,\Even}$ having an even color $n$ such that
$n$ is the maximum color associated to the states which occur infinitely many times along $\nu$.
We show that  $(\LT,\pi,0)\models \psi_1(\ASet,\sttElm',\ASet')\wedge \psi_2(\ASet,\sttElm',\ASet')$. Hence, the result follows.
We focus on the conjunct $\psi_2(\ASet,\sttElm',\ASet')$ (the proof for the conjunct $\psi_1(\ASet,\sttElm',\ASet')$ is similar).
By construction of $\psi_2(\ASet,\sttElm',\ASet')$, it suffices to show that for all $j\geq 0$ and accepting runs $\nu_f$ of $_{(\InitState{D},\ASet)}\NName_{\{(\sttElm',\ASet')\}}$ over $\rho[0,j]$ whose states satisfy the associated tests, then
$\nu_f$ is a prefix of $\nu$.
Let $\nu_f= (\sttElm'_0,\ASet'_0)\ldots (\sttElm'_{j+1},\ASet'_{j+1})$ be such a finite run over $\rho[0,j]$
 such that $(\sttElm'_{0},\ASet'_{0})=(\InitState{D},\ASet)$  and for all $i\in [0,j+1]$, $(\LT,\pi(i))\models \varphi_{\ASet'_i}$. Since $(\LT,\pi(i))\models \varphi_{\ASet_i}$ for all $i\geq 0$, by Claim~1,
 it follows that $\ASet'_i=\ASet_i$ for all $i\in [0,j+1]$. Thus, since $\DName_{\sttElm}$ is deterministic, we deduce that
 $\sttElm'_i=\sttElm_i$ for all $i\in [0,j+1]$, and the result follows. \vspace{0.2cm}

\noindent \textbf{Case where $\sttElm$ is universal:} Let $\widetilde{\AutName^{\sttElm}}$ be the dual automaton of $\AutName^{\sttElm}$.
By Proposition~\ref{prop:SeparationConditionHGATA}, $\widetilde{\AutName^{\sttElm}}$ is an \HGTA satisfying the mutual-exclusion condition (resp., an \HGTAcf)  which accepts the complement
of $\LangFun(\AutName)$. Moreover, $\sttElm$ is an existential state in $\widetilde{\AutName^{\sttElm}}$, and  the order of $\sttElm$ in $\widetilde{\AutName^{\sttElm}}$ coincides with the order of
 $\sttElm$ in $\AutName$. Thus, by the case for the existential states, one can construct a \CCDL (resp., \CCTLStar) formula $\widetilde{\varphi_{\sttElm}}$ such that $\LangFun(\widetilde{\varphi_{\sttElm}})=\LangFun(\widetilde{\AutName^{\sttElm}})$. Thus, we set $\varphi_{\sttElm}\DefinedAs \neg \widetilde{\varphi_{\sttElm}}$, and the result directly follows. This concludes the proof of Theorem~\ref{theo:FromHGATAtoCDL}.

\subsection{Expressive completeness of CCTL* in simple form}\label{app:SimpleFormCCTLStar}

In this section, we prove the following result.

\begin{restatable}{proposition}{propSimpleFormCCTLStar}\label{prop:SimpleFormCCTLStar}
Given a \CCTLStar (resp., \CTLStar) formula, one can construct an equivalent \CCTLStar (\CTLStar) formula in simple form.
\end{restatable}
\begin{proof}
Evidently, it suffices to show that a \CCTLStar (resp., \CTLStar) formula of the form $\EQ\psi$ can be converted into an equivalent
\CCTLStar (resp., \CTLStar) formula in simple form. The proof is by induction on the nesting depth of the path quantifier $\EQ$ in
$\EQ\psi$. By the induction hypothesis, we can assume that $\psi$ has an equivalent path formula $\widehat{\psi}$
in simple form such that for each Kripke tree $\LT$, infinite path $\pi$, and position $i\geq 0$, $(\LT,\pi,i)\models \psi$
iff $(\LT,\pi,i)\models \widehat{\psi}$. By exploiting De Morgan's laws and the equivalence $\psi_1\Until \psi_2\equiv \psi_2 \vee \Next(\psi_1\wedge (\psi_1\Until \psi_2))$, the path formula $\widehat{\psi}$ can be rewritten into an equivalent disjunction
of conjuncts $\CSet$ of the form
\[
\CSet = \displaystyle{\bigwedge_{j=1}^{m}\DC^{m_j}\varphi_j \,\wedge\, \bigwedge_{j=1}^{n}\neg\DC^{n_j}\theta_j \,\wedge \,
\bigwedge_{j=1}^{p} \Next\xi_j}\,\wedge \,\chi_p
\]
where $\chi_p$ is a propositional formula.
Note that $m_j=1$ and $n_i=1$ if  $\widehat{\psi}$ is in \CTLStar. Now, we observe that
\[
\EQ\CSet \equiv \displaystyle{\bigwedge_{j=1}^{m}\DC^{m_j}\varphi_j \,\wedge\, \bigwedge_{j=1}^{n}\neg\DC^{n_j}\theta_j} \,\wedge \,
\DC^{1}\EQ(\bigwedge_{j=1}^{p} \xi_j)\,\wedge \chi_p
\]
Thus, since the path quantifier  $\EQ$ is distributive with respect to disjunction, the result follows.
\end{proof}

\subsection{Remaining cases in the proof of Theorem~\ref{theo:FromCDLtoHGATA}}\label{app:FromCDLtoHGATA}

For completing the proof of Theorem~\ref{theo:FromCDLtoHGATA}, it remains to consider the cases where
the given  \CCDL (resp., \CCTLStar) state formula $\varphi$ is either an atomic proposition or a formula of the form $\DC^{n}\varphi'$, and the case
where $\varphi$ is a \CCDL formula of the form $\varphi=\EQ\psi$. The cases where the root modality of
 $\varphi$ is a Boolean connective directly follow from Proposition~\ref{prop:SubclassesGATABooleanClosure}. \vspace{0.2cm}

  \noindent \textbf{Case where $\varphi=p\in\Prop$:} in this case $\AutName_{\varphi}$ has a unique state $\isttElm$ with color $0$
  and transition function  $\trnFun_p$ defined as follows:   $\trnFun_p(\isttElm,a)=\true$ if $p\in a$, and $\trnFun_p(\isttElm,a)=\false$ otherwise. Note that $\isttElm$ is a transient state, and the result easily follows.\vspace{0.2cm}

  \noindent \textbf{Case where $\varphi=\DC^{n}\varphi'$ for some $n\geq 1$ and state formula $\varphi'$:} by the induction hypothesis one can construct an \HGTA (resp., \HGTAcf) $\AutName_{\varphi'} = \tuple {2^{\Prop}} {\SttSet} {\trnFun} {\isttElm} {\Family} {\Family_\exists} {\Omega}$ with   $\Family=\tpl{\SttSet_1,\ldots,\SttSet_n}$
  such that $\LangFun(\AutName_{\varphi'})=\LangFun(\varphi')$. Then, the \HGTA (resp., \HGTAcf) $\AutName_{\varphi}$ is given by $\AutName_{\varphi}=\tuple {2^{\Prop}} {\SttSet\cup \{\isttElm_\varphi\}} {\trnFun_\varphi} {\isttElm_\varphi} {\Family_\varphi} {\Family_\exists} {\Omega_\varphi}$, where $\isttElm_{\varphi}$ is a fresh  state, $\Family_{\varphi}= \tpl{\SttSet_1,\ldots,\SttSet_n,\{\isttElm_{\varphi}\}}$,  $\Omega_\varphi = \Omega \cup (\isttElm_{\varphi} \rightarrow 0)$  and $\delta_\varphi$ is defined as follows. For states in $\SttSet$, $\trnFun_{\varphi}$ agrees with $\trnFun$. For the state $\isttElm_{\varphi}$ and for each $a\in 2^{\Prop}$, $\trnFun_{\varphi}(\isttElm_{\varphi},a)= \DMod[n]\isttElm$. Thus, from the root  of the input tree and in the initial state $\isttElm_{\varphi}$, $\AutName_\varphi$ sends at least $n$  copies to $n$ distinct children of the root and from each of such children $w$, $\AutName_{\varphi}$ simulates the behaviour of $\AutName_{\varphi'}$ over the subtree rooted at node $w$. Note that the  singleton $\{\isttElm_{\varphi}\}$ constitutes a transient component with the highest order. Hence, the mutual-exclusion property for \HGTAcf is preserved and the result easily follows.\vspace{0.2cm}

\noindent \textbf{Case where $\varphi=\EQ\psi$ for some \CCDL path formula $\psi$:}
Let $\max(\psi)=\{\varphi_1,\ldots, \varphi_k\}$ for some $k\geq 0$
and $\Prop_{\Ext}$ be an extension of $\Prop$ obtained by adding for each state formula $\varphi_i$ a fresh
atomic proposition $p_i$. Then, the \CCDL path formula $\psi$ can be seen as a formula $\psi_{\Ext}$ over $\Prop_{\Ext}$ in the linear-time fragment of \CCDL, denoted \LCDL, obtained from \CCDL by disallowing the path quantifiers and the counting modality $\DC^{n}$. Since \LCDL corresponds to a fragment of Visibly Linear Dynamic Logic,
by~\cite{WZ18}, one can construct a  B\"{u}chi \NWA $\NName_{\psi}$ such that
$\LangFun(\NName_{\psi})= \LangFun(\psi_{\Ext})$.
For the rest, the proof proceeds as for the case of a \CCTLStar formula of the form $\EQ\psi$ (see Section~\ref{sec:characterizationCDL;sub:FromAutomataToLogicsAndBack}).




\section{Missing Proofs of Section~\ref{sec:automataForMCL}}
\label{app:automataForMCL}

\subsection{Proof of Theorem~\ref{theo:FromHFATAtoMCL}}
\label{app:FromHFATAtoMCL}

\theoFromHFATAtoMCL*
\begin{proof} Let $\AutName = \tuple {\Sigma} {\SttSet} {\trnFun} {\isttElm} {\Family} {\Family_\exists} {\Omega}$ be an \HFTA  over $\Sigma=2^{\Prop}$ with $\Family=\tpl{\SttSet_1,\ldots,\SttSet_n}$. For the syntax and semantics of \MCL, see Appendix~\ref{sec:MSOFragments}.  For each $\sttElm\in \SttSet$, we show that one can construct  an \MCL  formula $\varphi_{\sttElm}(x)$ with one  free first-order variable $x$ and no free second-order variable  such that for all Kripke trees $\LT$ and nodes $w$ of $\LT$, $(\LT,\, x\rightarrow w)\models \varphi_{\sttElm}(x)$ \emph{iff} $\LT_w\in \LangFun(\AutName^{\sttElm})$, where $\LT_w$ denotes the labelled subtree  of $\LT$ rooted at node $w$.
 Thus, by setting $\varphi_{\AutName} \DefinedAs \exists x\ldotp (\Root(x)\wedge \varphi_{\isttElm}(x))$ with
 $\Root(x)\DefinedAs \neg\exists y\ldotp y<x$, Theorem~\ref{theo:FromHFATAtoMCL} directly follows. The proof is by induction on the order $\ell$ of the component $\SttSet_\ell$ such that $\sttElm\in \SttSet_\ell$. We distinguish the cases where $\sttElm$ is transient, existential, or universal.
 For each $a\in 2^{\Prop}$ and first-order variable $x$, we denote by $\theta(a,x)$ the first-order formula given by $\bigwedge_{p\in a} p(x)\wedge \bigwedge_{p\in \Prop \setminus a}\neg p(x)$.  \vspace{0.2cm}

\noindent \textbf{Case where $\sttElm$ is transient:} by construction for each $a\in 2^{\Prop}$, $\trnFun(\sttElm,a)$ contains only states having an order smaller than $\sttElm$ (note that for the base case, where the order of $\sttElm$ is $1$, $\trnFun(\sttElm,a)\in\{\true,\false\}$). Thus, by the induction hypothesis, for each state $\sttElm'$ occurring in $\trnFun(\sttElm,a)$, one can construct an  \MCL formula $\varphi_{\sttElm'}(x)$
such that for all Kripke trees $\LT$ and nodes $w$ of $\LT$, $(\LT,\, x \rightarrow w)\models \varphi_{\sttElm'}(x)$ \emph{iff} $\LT_w\in \LangFun(\AutName^{\sttElm'})$. Let $\psi_{\sttElm,a}(x)$ be the \MCL formula obtained from
$\trnFun(\sttElm,a)$ (we can assume that $\trnFun(\sttElm,a)$ does not contain occurrences of variable $x$)  by replacing each predicate
$\sttElm'(y)$  occurring in
$\trnFun(\sttElm,a)$  with $\Child(x,y)\wedge \varphi_{\sttElm'}(y)$. Then,
$\varphi_{\sttElm}(x)$ is given by $\bigvee_{a\in 2^{\Prop}}(\theta(a,x)\wedge \psi_{\sttElm,a}(x))$. Correctness of the construction easily follows.\vspace{0.2cm}

\noindent \textbf{Case where $\sttElm$ is existential:} let $\SttSet_\ell$ be the existential component such that
$\sttElm \in \SttSet_\ell$. By the first-order existential  requirement, for each $\sttElm'\in\SttSet_\ell$ and $a\in 2^{\Prop}$,
$\trnFun(\sttElm',a)$ is a disjunction of first-order constraints of the form $\exists x\ldotp (\sttElm'(x) \wedge \psi(x))$
where $\sttElm'\in \SttSet_\ell$ and the states occurring in $\psi(x)$ have order lower than $\ell$. We denote by
$\Upsilon_{\SttSet_\ell}$ the finite set of these  $\FOEOne{\SttSet}$ formulas $\psi(x)$, where $x$ may occurs free.
For each $\psi(x)\in \Upsilon_{\SttSet_\ell}$ and first-order variable $z$ which does not occur in $\psi(x)$, we denote
by $\widehat{\psi}(z,x)$ the $\MCL$ formula obtained by replacing each predicate
$\sttElm'(y)$  occurring in
$\psi(x)$  with $\Child(z,y)\wedge \varphi_{\sttElm'}(y)$. Note that the existence of $\varphi_{\sttElm'}(y)$ follows from the induction hypothesis  since the order of state $\sttElm'$ is lower than $\ell$.
Given a Kripke tree $\LT = (\TSet,\Lab)$, an infinite path $\pi$ of $\LT$, and an infinite word $\rho=(\sttElm_0,\psi_0(x))(\sttElm_1,\psi_1(x))\ldots$ over $\SttSet_\ell\times \Upsilon_{\SttSet_\ell}$, we say that $\rho$ is \emph{$\sttElm$-consistent with $\pi$} iff the following conditions hold:
\begin{itemize}
  \item $\sttElm_0=\sttElm$ (\emph{initialization});
  \item for each $i\geq 0$, $\exists x\ldotp (\sttElm_{i+1}(x)\wedge \psi_{i}(x))$ is a conjunct of $\trnFun(\sttElm_{i},\Lab(\pi(i)))$
  and $(\LT,z \rightarrow \pi(i),x \rightarrow \pi(i+1))\models \widehat{\psi_i}(z ,x)$ (\emph{consecution});
  \item for infinitely many $i\geq 0$, the color of state $\sttElm_i$ is even (\emph{acceptance}).
\end{itemize}

 By construction and the induction hypothesis, we easily deduce the following characterization of the tree-language $\LangFun(\AutName^{\sttElm})$.\vspace{0.2cm}

\noindent \emph{Claim 1.} For all Kripke trees $\LT$ and nodes $w$ of $\LT$, $\LT_w\in \LangFun(\AutName^{\sttElm})$ \emph{iff}
there is an infinite path $\pi$ of $\LT$ starting at node $w$ and an infinite word $\rho$ over $\SttSet_\ell\times \Upsilon_{\SttSet_\ell}$ which is $\sttElm$-consistent with $\pi$.
\vspace{0.2cm}

We now define an \MCL formula $\varphi_{\sttElm}(z)$ capturing the characterization of Claim~1. For each
$n\geq 1$, we consider the predicate $\Part(z,X_1,\ldots,X_n)$ expressing that the chains $X_1,\ldots,X_n$
form a partition of an infinite path from node $z$, and the predicate $\Inf(X)$ expressing that the chain $X$
is infinite. Both the predicates can be easily specified by using only first-order quantification.
Fix an ordering $(\sttElm_1,\psi_1(x)),\ldots,(\sttElm_n,\psi_n(x))$ of the set $\SttSet_\ell\times \Upsilon_{\SttSet_\ell}$
and for each $i\geq 1$, let $\tau_i= (\sttElm_i,\psi_i(x))$. Moreover, let $\Init$ (resp., $\Acc$) be the set of elements
$(\sttElm_i,\psi_i(x))$ such that
$\sttElm_i=\sttElm$ (resp., $\sttElm_i$ has even color), and for each element $(\sttElm_i,\psi_i(x))$ and $a\in 2^{\Prop}$,
let $\SUCC((\sttElm_i,\psi_i(x)),a)$ be the set of elements $(\sttElm_k,\psi_k(x))$ such hat $\exists x\ldotp (\sttElm_k(x)\wedge \psi_i(x))$ is a conjunct of $\trnFun(\sttElm_i,a)$. Then, formula $\varphi_{\sttElm}(z)$ is defined as follows:
\[
\begin{array}{l}
\text{$\varphi_{\sttElm}(z) \DefinedAs
  \exists^{\CSym} X_{\tau_1}\ldots \exists^{\CSym} X_{\tau_n}\ldotp \Bigl(
\Part(z,X_{\tau_1},\ldots, X_{\tau_n}) \,\,\wedge\,\,
\displaystyle{\underbrace{\bigvee_{\tau_i\in \Init}z\in X_{\tau_i}}_{\text{Initialization}}
\wedge \,\, \underbrace{\bigvee_{\tau_i\in \Acc} \Inf( X_{\tau_i})}_{\text{Acceptance}} \,\,\wedge
} $} \vspace{0.2cm}\\

\text{$\displaystyle{
\underbrace{\bigwedge_{i=1}^{n} \forall y\in X_{\tau_i} \ldotp \bigvee_{a\in 2^{\Prop}}\,\,\bigvee_{\tau_k\in \SUCC(\tau_i,a)} \exists x\in X_{\tau_k}\ldotp\, (\theta(y,a)\wedge \Child(y,x)\wedge \widehat{\psi_i}(y,x))}_{\text{Consecution}}\Bigr) }$}
\end{array}
\]
Intuitively, for each $i\in [1,n]$, $X_{\tau_i}$ represents the (possibly empty) set of nodes of the guessed infinite path $\pi$ in Claim~1 which are associated with the positions $k$ of the guessed  infinite word $\rho$ over $\SttSet_\ell\times \Upsilon_{\SttSet_\ell}$  such that $\rho(k)=\tau_i$. Thus, by Claim~1, correctness of the construction easily follows.\vspace{0.2cm}

\noindent \textbf{Case where $\sttElm$ is universal:} as for the proof of Theorem~\ref{theo:FromHGATAtoCDL},
we consider the dual automaton of $\AutName^{\sttElm}$ which is an \HFTA accepting the complement of $\LangFun(\AutName^{\sttElm})$ and apply the case for the existential states. This concludes the proof of Theorem~\ref{theo:FromHFATAtoMCL}.
\end{proof}

\subsection{Proof of Proposition~\ref{prop:ProjectionSeparatedInOneDirection}}
\label{app:ProjectionSeparatedInOneDirection}

 \propProjectionSeparatedInOneDirection*
 \begin{proof}
 Let $\SttSet_\ell$ be the existential component of $\AutName$ containing the initial state, and $\trnFun$ be the transition function of
 $\AutName$.

 First we show that $\LangFun(\exists^{\CSym}p.\AutName)\subseteq  \exists^{\CSym}p.\LangFun(\AutName)$.  Let $(\TSet, \Lab)$ be a Kripke tree over $\Prop\setminus \{p\}$ such that $(\TSet, \Lab)\in \LangFun(\exists^{\CSym}p.\AutName)$.
 By definition of  $\exists^{\CSym}p.\AutName$ and Property~1 in Definition~\ref{def:HFATASeparatedOneDirection}, there is an
 accepting run $r=(\TSet_r, \Lab_r)$ of $\exists^{\CSym}p.\AutName$ and an infinite path $\pi$ of $\TSet$ from the root such that the following holds:
 \begin{itemize}
   \item for each node $w$ of $\pi$, there is exactly one node $y$ of $r$ such that $\Lab_r(y)$ is of the form $(\sttElm,w)$. Moreover,
   $\sttElm\in\SttSet_\ell$ and the one-step interpretation  $(\SSet_w,\ISet)$ used from node $y$ along $r$ for labelling the children of $y$ in $r$ is a model of either $\trnFun(\sttElm,\Lab(w))$ or  of $\trnFun(\sttElm,\Lab(w)\cup\{p\})$
   \item for each node $w\in \TSet\setminus \pi$ and node $y$ of $r$ reading $w$, the state labelling $y$ is not in
    $\SttSet_\ell$ and
    the one-step interpretation  $(\SSet_w,\ISet)$ used from node $y$ along $r$ for labelling the children of $y$ is a model of $\trnFun(\sttElm,\Lab(w))$.
 \end{itemize}
 Hence, there is a Kripke tree
 $(\TSet, \Lab')$ over $2^{\Prop}$ such  that (i) $\Lab'(w)=\Lab(w)$ for each $w\in \TSet\setminus \pi$,
and $\Lab'(w)\setminus \{p\}=\Lab(w)$ otherwise, and (ii) $r$ is an accepting run of $\AutName$ over
 $(\TSet, \Lab')$. This means that $(\TSet, \Lab)\in \exists^{\CSym}p.\LangFun(\AutName)$, and the result follows.\vspace{0.1cm}

 For the converse implication, assume that there is  a Kripke tree  $(\TSet, \Lab)$  over $\Prop\setminus \{p\}$ such that $(\TSet, \Lab)\in \exists^{\CSym}p.\LangFun(\AutName)$. Hence, there is a Kripke tree
 $(\TSet, \Lab')$ over $2^{\Prop}$ and  an infinite path $\pi$ of $\TSet$ from the root such  that (i) $(\TSet, \Lab')\in \LangFun(\AutName)$ and  (ii) $\Lab'(w)=\Lab(w)$ for each $w\in \TSet\setminus \pi$, and $\Lab'(w)\setminus \{p\}=\Lab(w)$ otherwise. By Property~2 in Definition~\ref{def:HFATASeparatedOneDirection}, there is an accepting run $r=(\TSet_r,\Lab_r)$ of $\AutName$ over $(\TSet,\Lab')$ such that for each input node $w\in \pi$, there is exactly one node $y$ of $r$ reading $w$, i.e., such that $\Lab_r(y)=(\sttElm,w)$ for some state $\sttElm$; moreover, state $\sttElm$ is in $\SttSet_\ell$. Note that by Property~1 in Definition~\ref{def:HFATASeparatedOneDirection}, the fact that $\AutName$ is an \HFTA and the initial state is in $\SttSet_\ell$, for each node $y$ of the run $r$ which is not associated to an input node in $\pi$, the state in the label of $y$ is not in $\SttSet_\ell$. Thus, by definition of the automaton $\exists^{\CSym}p.\AutName$, it follows that
 $r$ is an accepting run of  $\exists^{\CSym}p.\AutName$ over $(\TSet,\Lab)$. Hence, $(\TSet, \Lab)\in \LangFun(\exists^{\CSym}p.\AutName)$, and we are done.
 \end{proof}

\subsection{Proof of Proposition~\ref{prop:ConstraintSeparationInOneDirection}}
\label{app:ConstraintSeparationInOneDirection}

\propConstraintSeparationInOneDirection*
\begin{proof}
It is known~\cite{CFVZ20} that each  first-order $\SttSet$-constraint is
equivalent to a disjunction  of basic formulas.
Thus, it suffices to show the result for a basic formula
$\theta^=(\TypeSet_\exists,\TypeSet_\forall)$.
Recall that $\theta^=(\TypeSet_\exists,\TypeSet_\forall)$ is defined as follows,
where $\TypeSet_\exists =\{\ASet_1,\ldots,\ASet_n\}$ for some $n\geq 0$.
\[
\displaystyle{\exists x_1\ldots \exists x_n\ldotp \Bigl( \Diff(x_1,\ldots,x_n)\wedge \bigwedge_{i=1}^{n}\Type(\ASet_i)(x_i)\,
\wedge \, \forall y\ldotp (\Diff(x_1,\ldots,x_n,y) \rightarrow \bigvee_{\ASet\in\TypeSet_{\forall}}\Type(\ASet)(y) )\Bigr) }
\]
Let $\theta_s$ be the first-order $(\SttSet\cup 2^{\SttSet})$-constraint defined as follows:
\[
\theta_s \DefinedAs \displaystyle{\bigvee_{i=1}^{n} \exists x_i\ldotp (\ASet_i(x_i)\, \wedge\,  \theta_i(\TypeSet_\exists,\TypeSet_\forall)) \vee
\bigvee_{\ASet\in \TypeSet_\forall} \exists z\ldotp (\ASet(z) \wedge \theta_z(\TypeSet_\exists,\TypeSet_\forall))}
\]
where $z$ is a fresh variable, $\theta_i(\TypeSet_\exists,\TypeSet_\forall)$ is obtained from $\theta(\TypeSet_\exists,\TypeSet_\forall)$ by removing the quantifier $\exists x_i$ and the conjunct $\Type(\ASet_i)(x)$, and $\theta_z(\TypeSet_\exists,\TypeSet_\forall)$ is obtained from $\theta(\TypeSet_\exists,\TypeSet_\forall)$ by replacing $\Diff(x_1,\ldots,x_n)$ and $\Diff(x_1,\ldots,x_n,y)$ with
$\Diff(x_1,\ldots,x_n,z)$ and $\Diff(x_1,\ldots,x_n,y,z)$ respectively. By construction $\theta_s$  is $2^{\SttSet}$-functional in one direction and simulates $\theta^=(\TypeSet_\exists,\TypeSet_\forall)$.
\end{proof}

\subsection{Proof of Theorem~\ref{theorem:SimulatingTheorem}}
\label{app:SimulatingTheorem}

Let $\AutName = \tuple {\Sigma} {\SttSet} {\trnFun} {\isttElm} {\Family}{\Family_\exists} {\Omega}$ be an \HFTA with
$\Family=\tpl{\SttSet_1,\ldots,\SttSet_n}$. Let $\FSet_{\Buchi}$ (resp., $\FSet_{\coBuchi}$) be the set of states in the existential (resp., universal) components of $\AutName$ having even (resp., odd) color. In order to prove Theorem~\ref{theorem:SimulatingTheorem} for the \HFTA $\AutName$,
we need some preliminary results.

For the definition of $\Pow_{\AutName}$ and $\Pow_{\AutName}$-path, see Section~\ref{sec:automataForMCL}.
For each $\PSet\in  \Pow_{\AutName}$, $\Ran(\PSet)$ denotes the range of $\PSet$, i.e., the set of $\AutName$-states $\sttElm$ such that $(\sttElm',\sttElm)\in \PSet$ for some $\sttElm'\in \SttSet$, and $\Dom(\PSet)$ denotes the domain of $\PSet$, i.e., the set of  $\AutName$-states $\sttElm$ such that $(\sttElm,\sttElm')\in \PSet$ for some $\sttElm'\in \SttSet$.

Fix a $\Pow_{\AutName}$-path $\nu=\PSet_0 \PSet_1\ldots$. First, we associate to the $\Pow_{\AutName}$-path $\nu$ an infinite \emph{acyclic} graph $\GSet_\nu = (\VSet,\ESet)$, defined as follows:
\begin{itemize}
   \item The set $\VSet$ of vertices consists of the pairs $(i,\sttElm)$ such that $i\geq 0$ and $\sttElm\in \Ran(\PSet_i)$ (we say that
   $(i,\sttElm)$ is a \emph{vertex of level $i$}).
   \item $\ESet$ consists of the edges $((i,\sttElm),(i+1,\sttElm'))$ such that $(\sttElm,\sttElm')\in\PSet_{i+1}$.
 \end{itemize}
 Note that there is a unique vertex of level $0$ (\emph{initial vertex}) and it is given by $(0, \isttElm)$. Moreover, every vertex is reachable from the initial vertex  and the paths from the initial vertex correspond to the $\AutName$-paths of $\nu$.
 For a set $\SttSet'\subseteq \SttSet$,  a $\SttSet'$-vertex is  a vertex  whose $\SttSet$-component is in $\SttSet'$.
 A vertex is universal if it is associated with an universal state of $\AutName$.
 The graph $\GSet_\nu$ is \emph{accepting} if $\nu$ is accepting. We now recall
the notion of even ranking function~\cite{KV01}.

\begin{definition}[Ranking functions]
  \label{definition:Ranking}
  Let $n_{\USet}$ be the number of universal states in $\AutName$.
  For a $\Pow_{\AutName}$-path $\nu$,   a \emph{ranking function for the graph $\GSet_\nu$} is a function $f_{\nu}: \VSet
  \rightarrow \{1,\ldots,2n_{\USet}+1\}$ satisfying the following:
  \begin{enumerate}
  \item for all $(j,\sttElm)\in \VSet$ such that $\sttElm\in \FSet_{\coBuchi}$, $f_{\nu}$ is
    odd;
  \item for all \emph{universal} vertices $(j,\sttElm),(j',\sttElm')\in \VSet$ such that $(j',\sttElm')$ is a
    successor of $(j,\sttElm)$ in $\GSet_\nu$, it
    holds that $f_\nu(j',\sttElm')\leq f_\nu(j,\sttElm)$.
  \end{enumerate}
Thus, since the image of $f_\nu$ is bounded, for every infinite
path $\pi=v_0,v_1,\ldots$ of $\GSet_\nu$ (note that this path corresponds to a suffix of an $\AutName$-path of $\nu$) that gets trapped into an universal component of $\AutName$, $f_\nu$ converges
to a value: there is a number $\ell$ such that
$f_\nu(v_{\ell'})=f_\nu(v_\ell)$ for all $\ell'\geq \ell$.
We say that $f_\nu$ is \emph{even} if for all such infinite
paths $\pi$ of $\GSet$, $f_\nu$ converges to an even value (or,
equivalently, any of such paths  visits infinitely many times
vertices $v$ such that $f_\nu(v)$ is even).
\end{definition}

The following Lemma~\ref{lemma:Ranking} is a trivial adaptation
of the ranking construction in~\cite{KV01}.
 
\begin{lemma}
  \label{lemma:Ranking}
  Let $\nu$ be a $\Pow_{\AutName}$-path. Then $\nu$  is accepting \emph{iff}
  \begin{enumerate}
  \item there exists an even ranking function  for the graph $\GSet_\nu$;
  \item every infinite path of $\GSet_\nu$ which gets trapped in the component
    of an existential component of $\AutName$ visits infinitely many times $\FSet_{\Buchi}$-vertices.
  \end{enumerate}
\end{lemma}
 \begin{proof}
By Definition~\ref{definition:Ranking}, it follows that
  Conditions~1--2 in the lemma imply that $\GSet_\nu$ is
  accepting. For the converse implication, assume that $\GSet_\nu=(\VSet,\ESet)$ is accepting. Clearly, Conditions~2 in the lemma holds. For Conditions~1,  we construct a ranking function for the graph
  $\GSet_\nu$ and show that it is even.

  Let $\GSet'=\tpl{\VSet',\ESet'}$ be a sub-graph of $\GSet_\nu$ and $v\in \VSet'$.
  The vertex $v$ is \emph{finite in $\GSet'$} if the set of vertices which
  are reachable from $v$ in $\GSet'$ is finite.
  The vertex $v$ is \emph{$\FSet_{\coBuchi}$-free in $\GSet'$} if no $\FSet_{\coBuchi}$-vertex is
  reachable from $v$ in $\GSet'$.
  Moreover, for each $l\geq 1$, define
  \[\widt(\GSet',l) \DefinedAs |\{(l,\sttElm)\in \VSet'\}| \]
  which is the number of vertices in $\GSet'$ associated with level
  $l$. First, we inductively define an infinite sequence
  $(\GSet_i=\tpl{\VSet_i,\ESet_i})_{i\geq 1}$ of sub-graphs of $\GSet_\nu$ as follows:

  \begin{itemize}
  \item $\VSet_1$ is the set of  universal vertices of $\GSet_\nu$ and $\ESet_1=E\cap
    \VSet_1\times \VSet_1$.
  \item $\VSet_{2i}\DefinedAs \VSet_{2i-1}\setminus \{v\mid v$ is finite in $\GSet_{2i-1}\}$
    and $\ESet_{2i}=\ESet_{2i-1}\cap \VSet_{2i}\times \VSet_{2i}$.
  \item $\VSet_{2i+1}\DefinedAs \VSet_{2i}\setminus \{v\mid v$ is $\FSet_{\coBuchi}$-free in
    $\GSet_{2i}\}$ and $\ESet_{2i+1}=\ESet_{2i}\cap \VSet_{2i+1}\times \VSet_{2i+1}$.
  \end{itemize}

  Since $\GSet_{2i}$ is obtained from $\GSet_{2i-1}$ by removing all the
  vertices that can only access finitely many vertices and the number
  of successors of any vertex is finite, it follows that every maximal
  path in the graph $\GSet_{2i}$ is infinite.
  We claim that if $\GSet_{2i}$ is not empty, then $\GSet_{2i}$ contains
  some $\FSet_{\coBuchi}$-free vertex.
  By contradiction, we assume the contrary, which implies that there
  is an infinite path $\pi$ of $\GSet_{2i}$ which visits $\FSet_{\coBuchi}$-vertices
  infinitely many times.
  Since $\GSet_{2i}$ is a sub-graph of  $\GSet_\nu$, it follows that $\pi$
  is an infinite path of the graph $\GSet_\nu$ which gets trapped into an universal component and does not satisfy
  the  coB\"{u}chi acceptance condition $\FSet_{\coBuchi}$.
  This is a contradiction since $\GSet_\nu$ is  accepting.
  Hence, the claim follows.
  Since every maximal path in the graph $\GSet_{2i}$ is infinite, the
  claim implies that if $\GSet_{2i}$ is not empty, then there is an
  infinite path $\pi$ of $\GSet_{2i}$ which visits only $\FSet_{\coBuchi}$-free
  vertices. Hence, there is some position $\ell$ such
  that for all  positions  $h\geq \ell$,
  $\pi$ visits some vertex associated with the position $h$.
  Since all the vertices of $\pi$, which are $\FSet_{\coBuchi}$-free vertices, are
  removed in $\GSet_{2i+1}$, we obtain that for some $\ell\geq 1$ and all the
   positions $h\geq \ell$,
  \[\widt(\GSet_{2i+1},h) \leq \widt(\GSet_{2i},h) -1 \]
  Since each step only removes vertices, we obtain that for some $\ell\geq 1$
  and all  positions $h\geq l$,
  \[\widt(\GSet_{2i+1},h) \leq \widt(\GSet_{1},h) - i \]
 Let $n_{\USet}$ be the number of universal $\AutName$-states. Since $\GSet_1$ contains only  universal vertices, we deduce that for some $\ell\geq 1$ and for all  positions $h\geq \ell$, $\GSet_{2n_{\USet}+1}$ does \emph{not} contain
  vertices associated with the  position $h$.
  Since  every infinite path of $\GSet_\nu$ visits some  position $h\geq \ell$, it follows that
  every vertex of $\GSet_{2n_{\USet}+1}$ is finite in $\GSet_{2n_{\USet}+1}$.
  Thus, we obtain the following result.\vspace{0.2cm}

\noindent \emph{Claim 1.}  $\GSet_{2n_{\USet}+2}$ is empty.
 \vspace{0.2cm}

We define a  function $f_\nu: \VSet \rightarrow
 \{1,\ldots,2n_{\USet}+1\}$  for the
 graph $\GSet_\nu$ as follows:
  \[
  \begin{array}{l}
    f_\nu(v)  \text{$\DefinedAs$}  \left\{
      \begin{array}{ll}
        2i
        &    \text{ if   $i\leq n_{\USet}$,  $v\in \VSet_{2i}$,   and  $v\notin \VSet_{2i+1}$}
        \\
        2i-1
        &    \text{ if   $i\leq  n_{\USet}+1$,  $v\in \VSet_{2i-1}$,  and $v\notin \VSet_{2i}$}
        \\
        1
        &    \text{ otherwise }
      \end{array}
    \right.
  \end{array}
 \]
  Note that $f_\nu$ is well-defined since $\VSet_i\supseteq \VSet_{i+1}$ for all
  $i\geq 1$ and $\GSet_{2n_{\USet}+2}$ is empty.   We show that
  $f_\nu$ is an even ranking function of the  graph $\GSet_\nu$. Hence, the result follows.

 First, we show that $f_\nu$ is a ranking function of $\GSet_\nu$, i.e. $f_\nu$ satisfies Properties~1 and~2 of
 Definition~\ref{definition:Ranking}.
 For Property~1, let $(j,\sttElm)\in \VSet$ such that $\sttElm\in \FSet_{\coBuchi}$. We need to prove
 that $f_\nu(j,\sttElm)$ is odd. Since $\sttElm$ is universal, $(j,\sttElm)$ is a vertex of
 $\GSet_1$.
 Moreover, since $(j,\sttElm)$ is not $\FSet_{\coBuchi}$-free, there is no $i\geq 1$ such
 that $(j,\sttElm)\in \VSet_{2i}$ and $(j,\sttElm)\notin \VSet_{2i+1}$.
 Thus, by Claim~1, there is $i\leq n_{\USet}+1$ such that $(j,\sttElm)\in \VSet_{2i-1}$ and
 $(j,\sttElm)\notin \VSet_{2i}$.
 By definition of $f_\nu$, we obtain that $f_\nu(j,\sttElm)=2i-1$
 and the result follows.
 For Property~2 of Definition~\ref{definition:Ranking}, let us consider two universal
vertices $(j,\sttElm),(j',\sttElm')\in \VSet$ such that $(j',\sttElm')$ is a successor of $(j,\sttElm)$ in
 $\GSet_\nu$.
 We need to show that $f_\nu(j',\sttElm')\leq f_\nu(j,\sttElm)$.
 Note that $(j,\sttElm)$ and $(j',\sttElm')$ are vertices of $\GSet_1$. By Claim~1,
 there are $1\leq i,i'\leq 2n_{\USet}+1$ such that $(j,\sttElm)\in \VSet_i$, $(j,\sttElm)\notin
 \VSet_{i+1}$, $(j',\sttElm')\in \VSet_{i'}$, and $(j',\sttElm')\notin
 \VSet_{i'+1}$.
 Moreover, either $i$ is odd and $(j,\sttElm)$ is finite in $\GSet_i$ or $i$ is
 even and $(j,\sttElm)$ is $\FSet_{\coBuchi}$-free in $\GSet_i$.
 Since $(j',\sttElm')$ is a successor of $(j,\sttElm)$ in $\GSet_1$, it follows that
 $i'\leq i$. %
 Thus, by definition of $f_\nu$, we obtain that
 $f_\nu(j',\sttElm')\leq f_\nu(j,\sttElm)$, and the result holds.

 Finally, we show that $f_\nu$ is even.
  Let $\pi=v_0,v_1,\ldots$ be an infinite path of $\GSet_\nu$ that gets
  trapped in some universal component of $\AutName$.
  We need to show that $f_\nu$ converges to an even value along
  $\pi$.
  We assume the contrary and derive a contradiction.
  Since $f_\nu$ is a ranking function of $\GSet_\nu$, there is $k\geq 0$ such that for all $h\geq k$,
  $f_\nu(v_h)=f_\nu(v_k)$, $v_h$ is universal, and $f_\nu(v_k)$ is odd.
  By definition of $f_\nu$, this means that there is $i\leq n_{\USet}+1$
  such that $v_h\in \VSet_{2i-1}$ and $v_h\notin \VSet_{2i}$ for all $h\geq
  k$. This entails that $v_h$ is finite in $\GSet_{2i-1}$ for all $h\geq
  k$.
  Hence, $v_k$ is finite in $\GSet_{2i-1}$ and there is an infinite path
  from $v_k$ in $\GSet_{2i-1}$, which is a contradiction.
  This concludes the proof of Lemma~\ref{lemma:Ranking}.
\end{proof}

Now, based on Lemma~\ref{lemma:Ranking} and the classical breakpoint
construction~\cite{MH84}, we give a characterization of the accepting $\Pow_{\AutName}$-paths $\nu$  in terms
of extensions of $\nu$ which satisfy classical acceptance conditions a la B\"{u}chi.

For a set $\OSet\subseteq \SttSet$ of $\AutName$-states and $\PSet\in\Pow_{\AutName}$, we denote by
$\Ran_{\OSet}(\PSet)$ the set of states $\sttElm'\in \SttSet$ such that $(\sttElm,\sttElm')\in \PSet$ for some
$\sttElm\in \OSet$. Note that $\Ran_{\OSet}(\PSet)\subseteq \Ran(\PSet)$.
Let $n_{\USet}$ be the number of universal states of $\AutName$. A \emph{region} of $\AutName$ is a triple $(\PSet,\OSet,f)$ where
$\PSet\in\Pow_{\AutName}$, $\OSet \subseteq \Ran(\PSet)$, and $f:\Ran(\PSet) \rightarrow \{1,\ldots,2n_{\USet}+1\}$ is a function assigning to each $\AutName$-state $\sttElm$ in $\Ran(\PSet)$ an integer in $\{1,\ldots,2n_{\USet}+1\}$ such that $f(\sttElm)$ is odd if $\sttElm\in \FSet_{\coBuchi}$. A state $\sttElm$ of $\AutName$ is \emph{accepting with respect to $f$} if $\sttElm\in \Ran(\PSet)$ and (i)
either $\sttElm$ is existential and $\sttElm\in \FSet_{\Buchi}$, or (ii) $\sttElm$
is universal and  $f(\sttElm)$ is even. Intuitively, in the second component $\OSet$ of the region $(\PSet,\OSet,f)$, we keep track of the states
reached by some finite path in the graph $\GSet_\nu$ of the given $\Pow_{\AutName}$-path that does not visit accepting states. When the second component of
the current region becomes empty, a new phase is started by initializing the second component of the next region to the set of non-accepting states
associated with the next level of $\GSet_\nu$.

We denote by $\REG_{\AutName}$ the set of regions of $\AutName$. For each region $(\PSet,\OSet,f)\in\REG_{\AutName}$ and $\PSet'\in \Pow_{\AutName}$ such that $\Dom(\PSet')\subseteq \Ran(\PSet)$, let $\SUCC((\PSet,\OSet,f),\PSet')$ be the set of regions of the form $(\PSet',\OSet',f')$ such that the following conditions hold, where $\Acc$ (resp., $\Acc'$) denotes the set of accepting states of
$\AutName$ with respect to $f$ (resp., $f'$):
\begin{itemize}
\item \emph{Ranking requirement:} for each $(\sttElm,\sttElm')\in \PSet'$ such that $\sttElm$ and $\sttElm'$ are universal,
     $f(\sttElm)\geq f'(\sttElm')$;
\item \emph{Miyano-Hayashi requirement:} either (i) $\OSet= \emptyset$ and $\OSet'= \Ran(\PSet')\setminus \Acc'$, or (ii) $\OSet\neq\emptyset$ and $\OSet'= \Ran_{\OSet}(\PSet')\setminus  \Acc'$.
\end{itemize}
Let $\nu= \PSet_0\PSet_1\ldots$ be a $\Pow_{\AutName}$-path. An \emph{extension of $\nu$} is an infinite sequence $\xi_{\nu}$
of $\AutName$-regions of the form $\xi_{\nu}= (\PSet_0,\OSet_0,f_0)(\PSet_1,\OSet_1,f_1)\ldots$ satisfying the following \emph{local} conditions:
%
\begin{itemize}
\item \emph{Initialization:} $\OSet_0=\emptyset$.
\item  for each $i\geq 0$, $(\PSet_{i+1},\OSet_{i+1},f_{i+1})\in \SUCC((\PSet_{i},\OSet_{i},f_{i}),\PSet_{i+1})$.  
\end{itemize}
We say that $\xi_\nu$ is \emph{good} iff for infinitely many $i$, $\OSet_i= \emptyset$.

Intuitively,  the ranking requirement along the extension $\xi_{\nu}$ of the $\Pow_{\AutName}$-path $\xi_{\nu}$ ensures that  there is a ranking function $f_{\nu}$ for the acyclic graph $\GSet_\nu$ associated with $\nu$.
By Lemma~\ref{lemma:Ranking}, $\nu$ is accepting if $f_{\nu}$
is even and Condition~2 in Lemma~\ref{lemma:Ranking} holds.
This, in turn, is equivalent to require that every infinite path of
$\GSet_{\nu}$ visits infinitely many vertices in $\Acc$, where $\Acc$ is the set
of $\GSet_\nu$-vertices $(i,\sttElm)$ such that $\sttElm$ is an accepting state of $\AutName$ with respect to $f_i$.
This last condition is satisfied iff there is an infinite sequence of  positions
$0=h_1<h_2<\ldots$  such that for all $i\geq 0$, any finite path
of $\GSet_\nu$ that starts at position $h_{i}+1$ and ends at position
$h_{i+1}$ visits a vertex in $\Acc$.
Thus, the Miyano-Hayashi and the goodness requirement  on the sets
$\OSet_i$ ensure the existence of such an infinite sequence of
positions $h_j$. Formally, the following holds.

\begin{lemma}
  \label{lemma:CharacterizationPowerPaths}
  Let $\nu$ be a $\Pow_{\AutName}$-path. Then $\nu$  is accepting \emph{iff}
  there exists an extension of $\nu$ which is good.
\end{lemma}
\begin{proof} Let $\nu=\PSet_0\PSet_1\ldots$ and $\GSet_\nu=(\VSet,\ESet)$ be the acyclic graph associated with $\nu$.
First, assume that there is an extension $\xi_{\nu}= (\PSet_0,\OSet_0,f_0)(\PSet_1,\OSet_1,f_1)\ldots$ which is good.
For each $i\geq 0$, let $\Acc_i$ be the set of $\AutName$-states which are accepting with respect to $f_i$. Moreover, let $\Acc$ be the set of vertices $(i,\sttElm)\in\VSet$ such that $\sttElm \in\Acc_i$, and $f:\VSet \rightarrow \{1,\ldots,2n_{\USet}+1\}$ be the mapping associating to each vertex $(i,\sttElm)\in\VSet$ the rank $f_i(\sttElm)$. The ranking requirement of the extension $\xi_{\nu}$ of $\nu$ ensures that $f$ is a ranking function for the graph $\GSet_\nu$. Thus, by Lemma~\ref{lemma:Ranking}, in order to prove that $\GSet_\nu$ is accepting, it suffices to show that each infinite path in $\GSet_\nu$ visits infinitely many times vertices in $\Acc$. We assume on the contrary that there is an infinite path in $\GSet_\nu$ which does not visit infinitely many times vertices in $\Acc$, and derive a
contradiction. Then, since  $\xi_{\nu}$ is good, there must be an infinite path $\pi$ of $\GSet_\nu$ of the form $\pi=(k,\sttElm_k)(k+1,\sttElm_{k+1})\ldots$ such that
$\OSet_k= \emptyset$ and for all $i\geq k$, $\sttElm_i\in \Ran(\PSet_i)\setminus \Acc_i$.   Since $\OSet_k=\emptyset$, by the Miyano-Hayashi requirement, we deduce that $\sttElm_i\in \OSet_i$ for each $i>k$. This means that $\OSet_i\neq \emptyset$
for each $i>k$, which contradicts the goodness of $\xi_{\nu}$. Hence, the result follows.
\vspace{0.2cm}

For the converse implication, assume that $\GSet_\nu$ is accepting. By Lemma~\ref{lemma:Ranking}, there is an even ranking function
$f: \VSet \rightarrow \{1,\ldots,2n_{\USet}+1\}$ such that each infinite path of $\GSet_\nu$ visits infinitely many times vertices in $\Acc$, where $\Acc$ is the set of vertices $(i,\sttElm)$ such that either (i) $\sttElm$ is existential and $\sttElm\in \FSet_{\Buchi}$, or (ii) $\sttElm$ is universal and $f(\sttElm)$ is even.
Let $K\geq 0$ be an arbitrary  natural number. We first show that there exists $i_K>K$ such that
each $\GSet_\nu$-path from a $(K+1)$-level vertex to a $i_K$-level vertex visits some vertex in $\Acc$.
We assume the contrary and derive a contradiction. Hence, for each $i>K$, the set $\TSet_i$ of vertices $(i,\sttElm)$ of level $i$
 in $\GSet_\nu$ such that there is some finite path from a vertex of level $K+1$ to $(i,\sttElm)$ which does not visit   $\Acc$-vertices is not empty. Let $\GSet'$ be the subgraph of $\GSet_\nu$ whose set of vertices is the infinite set $\bigcup_{i>K}\TSet_i$. Note that $\GSet'$ does not contain $\Acc$-vertices.
 By construction, every vertex in $\GSet'$ is reachable in
  $\GSet'$ from a vertex of the form $(K+1,\sttElm)$.
  Moreover, each vertex of $\GSet'$ has only finitely many
  successors.  Since $\GSet'$ is infinite and the set of vertices of the form $(K+1,\sttElm)$
  is finite, by K\"{o}nig's Lemma, $\GSet'$ contains an infinite path
  $\pi$.
Thus, since $\pi$ does not visit vertices in
  $\Acc$ and $\pi$ is also an infinite path of $\GSet_\nu$, we obtain a contradiction, and the result follows. Therefore, since $K$ is arbitrary, the following holds:\vspace{0.2cm}

  \noindent \emph{Claim:} there is an infinite sequence of positions $\eta=\ell_0<\ell_1<\ldots$ such that $\ell_0=0$ and for each
each $i\geq 0$ and finite paths $\pi$ of $\GSet_\nu$ of the form $\pi= (\ell_{i}+1,\sttElm),
  \ldots,(\ell_{i+1},\sttElm')$, $\pi$ visits some state in $\Acc$.\vspace{0.2cm}

Let $\eta= \ell_0<\ell_1<\ldots$ be the infinite sequence of positions in the previous claim, and for each $i>0$, let $\LSet(i)$ be the greatest index $\ell_j$ of $\eta$ such that $\ell_j<i$. We define the infinite sequence of regions  $\xi_\nu= (\PSet_0,\OSet_0,f_0) (\PSet_1,\OSet_1,f_1)\ldots$ as follows:
\begin{itemize}
  \item $\OSet_0=\emptyset$ and for all $i>0$, $\OSet_i$ is the set of states associated with the vertices $v$ of level $i$ such that some path from a vertex of level $\LSet(i)+1$ to vertex $v$ does not visit vertices in $\Acc$.
  \item for each $i\geq 0$ and $\sttElm\in\Ran(\PSet_i)$, $f_i(\sttElm)=f(i,\sttElm)$.
\end{itemize}
By construction, it easily follows that $\nu_\xi$ is an extension of $\nu$. Moreover, by the previous claim, $\OSet_{\ell_i}= \emptyset$ for all $i\geq 0$. Hence, $\nu_\xi$ is good and the result follows. This concludes the proof of Lemma~\ref{lemma:CharacterizationPowerPaths}.
\end{proof}

By Lemmata~\ref{lemma:SimulatingLemmaFirstStage} and~\ref{lemma:CharacterizationPowerPaths}, we deduce the main result of this section.

\theoSimulatingTheorem*
\begin{proof}
Without loss of generality, we can assume that the initial state of $\AutName$ is transient.
The \HFTA $\Sim(\AutName)$ is defined as:
\[
\Sim(\AutName) = \tuple {\Sigma} {\SttSet\cup  \REG_{\AutName}} {\trnFun_{\REG}} { \isttElm_{\REG}} {\Family_{\REG}}{\Family_\exists\cup \{\REG_{\AutName}\}} {\Omega_{\REG}}
\]
where the initial state is  $(\{(\isttElm,\isttElm)\},\emptyset,f_0)$
 with $f_0(\isttElm)= 1$, and  $\Family_{\REG}=\tpl{\SttSet_1,\ldots,\SttSet_n,\REG_{\AutName}}$ (the existential component $\REG_{\AutName}$ has highest order).
For the parity condition $\Omega_{\REG}$, it coincides with $\Omega$ over the $\AutName$-states, and for a region
$(\PSet,\OSet,f)$, $\Omega_{\REG}$ assigns to it the color $2$ if $\OSet=\emptyset$, and the color $1$ otherwise. Finally,
the transition function $\trnFun_{\REG}$ is obtained form the transition function $\trnFun_{\SEP}$ of $\AutName_{\SEP}$ (for the definition of $\AutName_{\SEP}$, see Section~\ref{sec:automataForMCL}.) as follows:
\begin{itemize}
  \item for all $\sttElm\in\SttSet$ and $a\in\Sigma$, $\trnFun_{\REG}(\sttElm,a)=\trnFun(\sttElm,a)$ (recall that on the $\AutName$-states $\sttElm$, $\trnFun_{\SEP}(\sttElm,a)=\trnFun(\sttElm,a)$);
  \item for all regions $(\PSet,\OSet,f)\in \REG_{\AutName}$ and $a\in\Sigma$, recall that  $\trnFun_{\SEP}(\PSet,a)$ is a disjunction of constraints of the form
  $\exists x. (\PSet'(x)\wedge \theta(x))$ where $\PSet'\in\Pow_{\AutName}$. Then $\trnFun_{\REG}((\PSet,\OSet,f),a)$ is obtained
  from $\trnFun_{\SEP}(\PSet,a)$ by replacing each disjunct $\exists x. (\PSet'(x)\wedge \theta(x))$ with
  \[
  \displaystyle{\bigvee_{(\PSet',\OSet',f')\in \SUCC((\PSet,\OSet,f),\PSet')}}\exists x. ((\PSet',\OSet',f')(x)\wedge \theta(x))
  \]
\end{itemize}
By Lemmata~\ref{lemma:SimulatingLemmaFirstStage} and~\ref{lemma:CharacterizationPowerPaths}, correctness of the construction easily follows.
\end{proof}

\subsection{Proof of Theorem~\ref{theo:FromMCLtoHFATA}}
\label{app:FromMCLtoHFATA}

In this section, we provide a proof of Theorem~\ref{theo:FromMCLtoHFATA} by showing that each \MCL sentence can be translated into an equivalent \HFTA.
For the syntax and semantics of \MCL, see Appendix~\ref{sec:MSOFragments}.
In the translation, it is convenient to consider a one-sorted variant of  \MCL where first-order variables are encoded as
second-order variables which are singletons.
In particular, for $p\in \Prop$ and second-order variables $X,Y$, we consider the predicates (i) $X\subseteq Y$ of expected meaning,
(ii) the predicate $X\subseteq p$ asserting that for each node $x$ of $X$, $p(x)$ holds, (iii) the predicate $\Child(X,Y)$
asserting that each node of $X$ has some child in $Y$, and (iv) the predicate $\Sing(X)$ asserting that  $X$ is a singleton.
The predicates $X\subseteq Y$, $X\subseteq p$, and $\Child(X,Y)$ can be expressed in $\MCL$ by using only first-order quantification, while the predicate $\Sing(X)$ can be expressed in \MCL as follows:
\[
  \Sing(X)\equiv \neg\Empty(X)\wedge\forall^{\CSym} Y\ldotp (Y\subseteq X \rightarrow (\Empty(Y)\vee X\subseteq Y)) \quad \Empty(Y)\DefinedAs \forall^{\CSym} Z\ldotp Y\subseteq Z
\]
The one-sorted fragment of \MCL (\OMCL for short) uses only chain second-order quantification and is defined by the following syntax for the given finite set $\Prop$ of atomic propositions.
\[
\varphi \seteq  X\subseteq p \mid X\subseteq Y \mid \Child(X,Y) \mid \neg \varphi \mid \varphi  \wedge \varphi
\mid \exists^{\CSym}  X \ldotp \varphi
\]
where $p \in \Prop$ and $X,Y \in \Var_2$. For each first-order variable $x$, let $\widehat{x}$ be a fresh second-order variable.
We observe that (i)  $p(x)$ can be expressed as $\widehat{x}\subseteq p \wedge \Sing(\widehat{x})$,
(ii) $x\in X$ corresponds to $\widehat{x}\subseteq X \wedge \Sing(\widehat{x})$, (iii) $\Child(x,y)$ corresponds to
$\Child(\widehat{x},\widehat{y})\wedge \Sing(\widehat{x})\wedge \Sing(\widehat{y}) $, and  (iv)   $\exists x\ldotp \varphi$ can be reformulated as $\exists \widehat{x}\ldotp (\Sing(\widehat{x})\wedge \widehat{\varphi})$, where $\widehat{\varphi}$ is obtained from $\varphi$ by replacing each
free occurrence of $x$ with $\widehat{x}$. Moreover, the atomic formula $x<y$ can be expressed in $\MCL$ by using only the predicate $\Child(x,y)$ and \emph{path quantification}:
\[
x<y \equiv \exists^{\PSym}  X \ldotp (y\in X \wedge x\notin X \wedge \exists z\in X\ldotp \Child(x,z)  )
\]
where the operator $\exists^{\PSym}  $ is the existential path quantifier which restricts the second-order existential quantification to paths of the given tree.  Note that the path quantifier $\exists^{\PSym}$ can be expressed in terms of the chain quantifier $\exists^{\CSym}$ by using only the predicate
$\Child(x,y)$ and first-order quantification.   Hence, we obtain the following result.

\begin{proposition}\label{prop:OMCLequivalentMCL} \MCL  and  \OMCL are expressively equivalent.
\end{proposition}

We can now prove Theorem~\ref{theo:FromMCLtoHFATA}.

\theoFromMCLtoHFATA*
\begin{proof}
By Proposition~\ref{prop:OMCLequivalentMCL}, we can assume that the given \MCL sentence is in \OMCL.
Fix a  \OMCL formula $\psi(X_1,\ldots,X_n)$  with free second-order variables in $\{X_1,\ldots,X_n\}$.
Define $\Prop_n \DefinedAs \Prop \cup \{X_1,\ldots,X_n\}$, i.e., $\Prop_n$ is obtained from $\Prop$ by adding the
second-order variables $X_1,\ldots,X_n$ (interpreted as atomic propositions). For each Kripke tree $\LT=(\TSet, \Lab)$ over
$\Prop$, an  \emph{$\Prop_n$-extension} of $\LT$ is a Kripke tree $\LT'$ over $\Prop_n$ of the form $\LT'=(\TSet,\Lab')$ such that
$\Lab'(w)\cap \Prop = \Lab(w)$ for each $w\in\TSet$ (i.e., $\LT'$ is obtained from $\LT$ by enriching the label of each node with some variables in $\{X_1,\ldots,X_n\}$). For each second-order valuation $\Val_2$ and Kripke tree $\LT=(\TSet, \Lab)$, we denote by
$\LT_{\Val_2}$ the $\Prop_n$-extension of $\LT$ given by $(\TSet,\Lab')$, where for each node $w\in \TSet$ and $i\in [1,n]$, $X_i\in \Lab'(w)$ iff $w\in \Val_2(X_i)$. Then, the  result directly follows from the following claim.\vspace{0.2cm}

\noindent \emph{Claim.} One can construct an \HFTA $\AutName_\psi$ in normal form over $2^{\Prop_n}$ such that for each Kripke tree $\LT$ over $\Prop$ and second-order valuation $\Val_2$, it holds that $(\LT,\Val_2)\models \psi(X_1,\ldots,X_n)$ iff
$\LT_{\Val_2}\in\LangFun(\AutName_\psi)$.\vspace{0.2cm}

We prove the previous claim by structural induction on $\psi(X_1,\ldots,X_n)$:
\begin{itemize}
  \item $\psi(X_1,\ldots,X_n)= X_i \subseteq X_j$ for some $i,j\in [1,n]$: the \HFTA $\AutName_{X_i\subseteq X_j}$ in normal form consists of two universal states which have the same order: the initial state $\isttElm$ which  has color $0$, and the  rejecting state $\sttElm_{r}$ with color $1$. The transition function $\trnFun$ is defined as follows for each $a\in 2^{\Prop_n}$: $\trnFun(\sttElm_{r},a )= \forall x\ldotp \sttElm_{r}(x)$, $\trnFun(\isttElm,a )=\forall x\ldotp \sttElm_{r}(x)$ if $X_i\in a$ and $X_j\notin a$, and $\trnFun(\isttElm,a )=\forall x\ldotp \isttElm(x)$ otherwise. Correctness of the construction
  easily follows.
  \item  $\psi(X_1,\ldots,X_n)= X_i \subseteq p$ for some $i\in [1,n]$ and $p\in \Prop$: this case is similar to the previous one.
   \item $\psi(X_1,\ldots,X_n)= \Child(X_i,X_j)$ for some $i,j\in [1,n]$: we build an \HFTA $\AutName_{\neg \Child(X_i,X_j)}$ in normal form
   for the formula $\neg\Child(X_i,X_j)$. Thus, by taking the dual of $\AutName_{\neg \Child(X_i,X_j)}$, the result follows (note that dualization of \HFTA preserves the normal form).
   Evidently,  $\neg\Child(X_i,X_j)$ holds iff there is a $X_i$-node $w$ in the given Kripke tree such that each child of $w$ is not a $X_j$-node. Then, $\AutName_{\neg \Child(X_i,X_j)}$ has a transient state $\sttElm_{c}$ with lower order and color $0$, and three existential states with the same order: $\isttElm$ (the initial one) with color $1$, and   $\sttElm_{\exists}$ and $\sttElm'_{\exists}$ with color $2$. The transition function $\trnFun(\sttElm,a)$ is defined as follows:
  \[
  \begin{array}{l}
      \text{$\trnFun(\sttElm,a)  \DefinedAs$}  \left\{
      \begin{array}{ll}
        \text{$\exists x\ldotp \isttElm(x)$}
        &    \text{ if   $\sttElm= \isttElm$  and  $X_i\notin a$}
        \\
        \text{$\exists x\ldotp \isttElm(x) \vee [\exists x\ldotp (\sttElm_\exists(x)\wedge \forall y\ldotp y\neq x \rightarrow  \sttElm_c(y))]$}
        &    \text{ if   $\sttElm= \isttElm$  and  $X_i\in a$}
        \\
        \text{$\exists x\ldotp \sttElm_\exists'(x)$}
         &   \text{ if   $\sttElm= \sttElm_\exists$ and $X_j\notin a$}
        \\
        \false
         &   \text{ if   $\sttElm= \sttElm_\exists$ and $X_j\in a$}
        \\
        \text{$\exists x\ldotp \sttElm'_\exists(x)$}
         &   \text{ if   $\sttElm= \sttElm'_\exists$}
        \\
        \true
       & \text{ if   $\sttElm= \sttElm_c$  and  $X_j\notin a$}
        \\
        \false &    \text{ otherwise }
      \end{array}
    \right.
  \end{array}
 \]
By construction, $\AutName_{\neg \Child(X_i,X_j)}$ is in normal form and accepts a Kripke tree over $2^{\Prop_n}$ iff there is a $X_i$-node with no
 $X_j$-child. Hence, the result follows.
  \item $\psi(X_1,\ldots,X_n)= \neg \psi_1(X_1,\ldots,X_n)$ or $\psi(X_1,\ldots,X_n)=  \psi_1(X_1,\ldots,X_n)\wedge \psi_2(X_1,\ldots,X_n)$: these cases directly follow from the induction hypothesis and the effective closure of \HFTA in normal form under Boolean language operations. Indeed the dualization (for handling language complementation) preserves the normal form and the construction for 
      conjunction simply adds a transient state with highest order and preserves the normal form.
  \item $\psi(X_1,\ldots,X_n) =\exists^{\CSym} X_{n+1}\ldotp \theta(X_1,\ldots,X_n,X_{n+1})$: by the induction hypothesis,
  one can construct an \HFTA  $\AutName_\theta$ over $2^{\Prop_{n+1}}$ satisfying the claim with $n$ replaced with $n+1$ and $\psi$ replaced with $\theta$. By Corollary~\ref{cor:ClosureChainProjectionHFATA}, one can construct an
  \HFTA $\AutName_\psi$ over $2^{\Prop_{n}}$ such that $\LangFun(\AutName_\psi)= \exists^{\CSym} X_{n+1}.\LangFun(\AutName_\theta)$. Note that the construction of
  $\AutName_\psi$ (see Section~\ref{sec:automataForMCL}) preserves the normal form. Hence, the result follows.
\end{itemize}
\end{proof}

\subsection{Proof of Theorem~\ref{theo:BisimulationInvariantMCL}}
\label{app:BisimulationInvariantMCL}

In order to prove Theorem~\ref{theo:BisimulationInvariantMCL}, we need additional definitions and preliminary results.
For the notions of bisimulation, see Appendix~\ref{sec:Bisimulation}.

\noindent Fix a non-empty finite set $\SttSet$.
Given a one-step interpretation  $(\SSet,\ISet)$ over $\SttSet$ the \emph{complement
of $(\SSet,\ISet)$} is the one-step interpretation  $(\SSet,\ISet^c)$ over $\SttSet$ where $\ISet^c(s)=\SttSet\setminus \ISet(s)$ for each
$s\in\SSet$. Dual formulas of the considered one-step logics satisfy the following property.

\begin{proposition}[\cite{Wal02,CFVZ20}]
  \label{prop:CompOneStep}
  Given a graded $\SttSet$-constraint (resp., a
first-order $\SttSet$-constraint) $\theta$ and one-step interpretation
$(\SSet,\ISet)$ over $\SttSet$, it holds that $(\SSet,\ISet)\models \theta$ iff
$(\SSet,\ISet^{c})\not\models \widetilde{\theta}$.
 \end{proposition}

Recall that a $\SttSet$-type is  a (possibly empty) set $\ASet\subseteq \SttSet$ and
$\Type(\ASet)(x) \DefinedAs\bigwedge_{\sttElm\in\ASet}\sttElm(x)$.
Moreover, for two sets  $\TypeSet_\exists$ and $\TypeSet_\forall$  of $\SttSet$-types,
the  basic formula $\theta^{=}(\TypeSet_\exists,\TypeSet_\forall)$
 for the pair $(\TypeSet_\exists,\TypeSet_\forall)$ is defined as follows, where
$\TypeSet_\exists =\{\ASet_1,\ldots,\ASet_n\}$ for some $n\geq 0$:
\[
\displaystyle{\exists x_1\ldots \exists x_n\ldotp \Bigl( \Diff(x_1,\ldots,x_n)\wedge \bigwedge_{i=1}^{n}\Type(\ASet_i)(x_i)\,
\wedge \, \forall y\ldotp (\Diff(x_1,\ldots,x_n,y) \rightarrow \bigvee_{\ASet\in\TypeSet_{\forall}}\Type(\ASet)(y) )\Bigr) }
\]
We also consider  the \emph{symmetric basic formula}
$\theta(\TypeSet_\exists,\TypeSet_\forall)$ which is the symmetric first-order $\SttSet$-constraint defined as follows.
\[
\theta(\TypeSet_\exists,\TypeSet_\forall) \DefinedAs\displaystyle{\exists x_1\ldots \exists x_n\ldotp \Bigl(   \bigwedge_{i=1}^{n}\Type(\ASet_i)(x_i)\,
\wedge \, \forall y\ldotp  \bigvee_{\ASet\in\TypeSet_{\forall}}\Type(\ASet)(y)  \Bigr) }
\]
Intuitively, $\theta(\TypeSet_\exists,\TypeSet_\forall)$ is obtained from $\theta^{=}(\TypeSet_\exists,\TypeSet_\forall)$ by removing the equality and inequality
atomic formulas.
 Recall that for each non-empty set $\SttSet'\subseteq \SttSet$, $\theta^{=}(\TypeSet_\exists,\TypeSet_\forall)$ is $\SttSet'$-functional in one direction
if there exists $\ASet\in \TypeSet_\exists$ such that $\ASet$ is a singleton consisting of an element in  $\SttSet'$, and for each
$\BSet\in (\TypeSet_\exists\setminus \{\ASet\})\cup \TypeSet_\forall$, $\BSet$ does not contain elements in $\SttSet'$. The previous notion
is extended to symmetric basic formulas in the obvious way.

Let $\SSet$ and $\SSet'$ be two non-empty sets and $f: \SSet' \rightarrow \SSet$ be a surjective map from $\SSet'$ to $\SSet$.
For each
 one-step interpretation $(\SSet,\ISet)$ over $\SttSet$ with domain $\SSet$, let $\ISet_f$ be the mapping from $\SSet'$ to $2^{\SttSet}$
defined as follows: $\ISet_f(s')=\ISet(f(s'))$.
Moreover, for a one-step interpretation $(\SSet',\ISet')$ over $\SttSet$ with domain $\SSet'$, let $(\ISet')_{f^{-1}}$ be the mapping from $\SSet$ to $2^{\SttSet}$
defined as follows: $(\ISet')_{f^{-1}}(s)=\bigcup_{s'\in f^{-1}(s)}\ISet'(s')$.

We now define a constructive mapping associating to each first-order $\SttSet$-constraint $\theta$, a symmetric first-order
$\SttSet$-constraint $\BI(\theta)$ defined as follows. If $\theta$ corresponds to a constraint $\theta^{=}(\TypeSet_\exists,\TypeSet_\forall)$ in basic form, then $\BI(\theta)$ is the corresponding symmetric constraint in basic form
 $\theta(\TypeSet_\exists,\TypeSet_\forall)$. If instead $\theta$ is a
disjunction $\bigvee_{j\in \JSet}\theta_j$  of basic formulas $\theta_j$, then
$\BI(\theta)$ is defined as $\bigvee_{j\in \JSet}\BI(\theta_j)$.   Otherwise,
by~\cite{CFVZ20}, $\theta$ is effectively equivalent to a disjunction
$\bigvee_{j\in \JSet}\theta_j$  of basic formulas $\theta_j$: thus, in this
case, $\BI(\theta)$ is defined as $\bigvee_{j\in \JSet}\BI(\theta_j)$. Note
that $\BI(\theta)$ is a disjunction of symmetric basic formulas and
by~\cite{CFVZ20}, it exploits all and only the predicates in $\SttSet$
which occur in $\theta$.

We first show the following result.

\begin{lemma}\label{lemma:SimpleFormFO} Let  $\SSet$ and $\SSet'$ be two non-empty sets and $f: \SSet' \rightarrow \SSet$ be a surjective map from $\SSet'$ to $\SSet$ such that
for each $s\in\SSet$, $f^{-1}(s)$ is infinite. Then, for each first-order $\SttSet$-constraint $\theta$, the following holds:
\begin{enumerate}
  \item for each one-step interpretation $(\SSet,\ISet)$ over $\SttSet$, $(\SSet,\ISet)\models \BI(\theta)$ iff
   $(\SSet',\ISet_f)\models  \theta$;
  \item for each one-step interpretation $(\SSet,\ISet)$ over $\SttSet$, $(\SSet,\ISet)\models \widetilde{\BI(\theta)}$ iff
   $(\SSet',\ISet_f)\models \widetilde{\theta}$;
  \item for each one-step interpretation $(\SSet',\ISet')$ over $\SttSet$, if $(\SSet',\ISet')\models \theta$ then
   $(\SSet,(\ISet')_{f^{-1}})\models \BI(\theta)$;
  \item    for each one-step interpretation $(\SSet',\ISet')$ over $\SttSet$, if $(\SSet',\ISet')\models \widetilde{\theta}$ then
   $(\SSet,(\ISet')_{f^{-1}})\models \widetilde{\BI(\theta)}$.
\end{enumerate}
\end{lemma}
\begin{proof}
\noindent \emph{Proof of Property~1}. By definition of $\BI(\theta)$, without loss of generality,  we can assume that $\theta$
is a constraint $\theta^{=}(\TypeSet_\exists,\TypeSet_\forall)$ in basic form. Hence, $\BI(\theta)$
is $\theta(\TypeSet_\exists,\TypeSet_\forall)$.   Let $\TypeSet_\exists =\{\ASet_1,\ldots,\ASet_n\}$ for some $n\geq 0$.
We need to show that $(\SSet,\ISet)\models \theta(\TypeSet_\exists,\TypeSet_\forall)$ iff $(\SSet',\ISet_f)\models \theta^=(\TypeSet_\exists,\TypeSet_\forall)$.

 First, assume that $(\SSet,\ISet)\models \theta(\TypeSet_\exists,\TypeSet_\forall)$. Hence, (i) for each $i\in [1,n]$, there
exists $s_i\in \SSet$ such that $\ASet_i\subseteq \ISet(s_i)$ and (ii) for each $s\in \SSet$, there exists $\BSet\in \TypeSet_\forall$ such that $\BSet\subseteq \ISet(s)$.
Since for each $s\in\SSet$, $f^{-1}(s)$ is infinite, there exist $n$ distinct elements  $s'_1,\ldots, s'_n$ of $\SSet'$ such that $f(s'_i)=s_i$. Thus, since
$\ISet_f(s')=\ISet(f(s))$, it follows that $(\SSet',\ISet_f)\models \theta^=(\TypeSet_\exists,\TypeSet_\forall)$.

For the converse implication, assume that $(\SSet',\ISet_f)\models \theta^=(\TypeSet_\exists,\TypeSet_\forall)$. Hence, (i) there exist $n$ distinct elements
$s'_1,\ldots,s'_n$ of $\SSet'$ such that $\ASet_i\subseteq \ISet(f(s'_i))$ for each $i\in [1,n]$, and (ii) for each $s'\in \SSet'\setminus \{s'_1,\ldots,s'_n\}$,
there exists $\BSet\in \TypeSet_\forall$ such that $\BSet\subseteq \ISet(f(s'))$. Hence, $(\SSet,\ISet)$ satisfies the existential part of $\theta(\TypeSet_\exists,\TypeSet_\forall)$. For the universal part of $\theta(\TypeSet_\exists,\TypeSet_\forall)$, let $s\in \SSet$. Since $f^{-1}(s)$ is infinite, there exists
$s'\in \SSet'\setminus \{s'_1,\ldots,s'_n\}$ such that $f(s')=s$. Hence, $\BSet\subseteq \ISet(s)$ for some $\BSet\in \TypeSet_\forall$, and the result follows.\vspace{0.2cm}

\noindent \emph{Proof of Property~2}. Recall that the complement of $(\SSet,\ISet)$ is the one-step interpretation $(\SSet,\ISet^{c})$, where
$\ISet^{c}(s)=\SttSet\setminus \ISet(s)$ for each $s\in \SSet$. Since $\ISet_f(s')=\ISet(f(s'))$ for each $s'\in \SSet'$, it holds that
$(\ISet_f)^{c}= (\ISet^{c})_f$. Hence, by Property~1,  $(\SSet,\ISet^{c})\not\models \BI(\theta)$ iff
   $(\SSet',(\ISet^{c})_f)\not\models \theta$ iff $(\SSet',(\ISet_f)^{c})\not\models \theta$.
   By Proposition~\ref{prop:CompOneStep}, it holds that (i) $(\SSet,\ISet)\models \widetilde{\BI(\theta)}$ iff
   $(\SSet,\ISet^{c})\not\models \BI(\theta)$, and (ii) $(\SSet',\ISet_f)\models \widetilde{\theta}$
   iff $(\SSet',(\ISet_f)^{c})\not\models \theta$. Hence, we obtain that
   $(\SSet,\ISet)\models \widetilde{\BI(\theta)}$ iff
   $(\SSet',\ISet_f)\models \widetilde{\theta}$ and the result follows.
\vspace{0.2cm}

\noindent \emph{Proof of Property~3}. Let $(\SSet',\ISet')\models \theta$ and $\ISet=(\ISet')_{f^{-1}}$. We need to show that
  $(\SSet,\ISet)\models \BI(\theta)$. By construction, $\ISet(s)=\bigcup_{s'\in f^{-1}(s)}\ISet'(s')$ for each $s\in \SSet$ and
  $\ISet_f(s')=\ISet(f(s'))$ for each $s'\in\SSet'$. Hence, $\ISet'(s')\subseteq \ISet_f(s')$ for each $s'\in\SSet'$. By monotonicity, we obtain that
    $(\SSet',\ISet_f)\models \theta$. Thus, by Property~1, the result follows. \vspace{0.2cm}

\noindent \emph{Proof of Property~4}. The proof of Property~4 is similar to the proof of Property~3, but we apply Property~2 instead of Property~1.
\end{proof}

\begin{definition}\label{def:InvariantBisumulationHFATAMapping}
Let $\AutName = \tuple {\Sigma} {\SttSet} {\trnFun} {\isttElm} {\Family} {\Family_\exists} {\Omega}$ be an \HFTA in normal form.
We denote by  $\BI(\AutName)$ the symmetric \HFTA given by $\BI(\AutName) = \tuple {\Sigma} {\SttSet} {\BI(\trnFun)} {\isttElm} {\Family} {\Family_\exists} {\Omega}$,
where the transition function $\BI(\trnFun)$ is defined as follows for all $(\sttElm,a)\in \SttSet\times \Sigma$:
\begin{itemize}
  \item $\sttElm$ is transient: $\BI(\trnFun)(\sttElm,a)= \BI(\trnFun(\sttElm,a))$. Note that $\BI(\trnFun(\sttElm,a))$ only contain states of order lower than $\sttElm$.
  \item $\sttElm\in \SttSet_i$,  where $\SttSet_i$ is an existential component: since $\AutName$ is in normal form, $\trnFun(\sttElm,a)$ is a non-empty disjunction of
  basic formulas which are $\SttSet_i$-functional in one direction. We set $\BI(\trnFun)(\sttElm,a)= \BI(\trnFun(\sttElm,a))$. Note that $\BI(\trnFun)(\sttElm,a)$ is a disjunction of symmetric basic formulas which are $\SttSet_i$-functional in one direction.
  \item $\sttElm\in \SttSet_i$,  where $\SttSet_i$ is a universal component: since $\AutName$ is in normal form, $\widetilde{\trnFun(\sttElm,a)}$ is a
   disjunction of  basic formulas which are $\SttSet_i$-functional in one direction. We set $\BI(\trnFun)(\sttElm,a)= \widetilde{\theta}$ where $\theta= \BI(\widetilde{\trnFun(\sttElm,a)})$.
\end{itemize}
Note that since $\BI(\theta)$ uses only and only the states in $\SttSet$ occurring in $\theta$, $\BI(\AutName)$ satisfies the first-order existential and first-order
universal requirements of \HFTA.
\end{definition}

We now introduce the notion of $\omega$-expansion of a Kripke
tree.
Let $\DSet$ be a non-empty set of directions. We denote by $f_{\DSet}$ the mapping $f_{\DSet}: \DSet\times \SetN \rightarrow \DSet$
defined as follows: $f_{\DSet}(d,n)=d$ for all $(d,n)\in \DSet\times \SetN$. Note that  $f_{\DSet}$ is surjective.
We extend $f_{\DSet}$ to finite  words $w$ over $\DSet\times \SetN$: $f_{\DSet}(w)=f_{\DSet}(w(0))\ldots f_{\DSet}(w(n-1))$ where $n=|w|$.
For a word $v$ over $\DSet$, $f_{\DSet}^{-1}(v)$ denotes the set of words $w$ over $\DSet\times \SetN$ such that
$f_{\DSet}(w)=v$. We extend the notation $f_{\DSet}^{-1}(v)$ to non-empty sets $\TSet\subseteq \DSet^{*}$ in the obvious way. Note that
if $\TSet$ is a tree, then $f_{\DSet}^{-1}(\TSet)$ is a tree over the set of directions $\DSet\times \SetN$. Moreover, for each path $\pi$ of
$f_{\DSet}^{-1}(\TSet)$, the sequence $f_{\DSet}(\pi(0)) f_{\DSet}(\pi(1))\ldots$ is a path of $\TSet$: with a little abuse of notation,
such a sequence is denoted by $f_{\DSet}(\pi)$.
For a Kripke tree $\LT=(\TSet, \Lab)$ where $\TSet\subseteq \DSet^{*}$, the \emph{omega-expansion} of
$\LT$ is the Kripke tree $\LT_\omega = (f_{\DSet}^{-1}(\TSet),\Lab_{\omega})$ where for each node $w\in f_{\DSet}^{-1}(\TSet)$,
$\Lab_{\omega}(w)=\Lab(f_{\DSet}(w))$.  By construction, the following evidently holds.

\begin{remark}\label{remark:OmegaExpansion} Let $\LT=(\TSet, \Lab)$ be a Kripke tree where $\TSet\subseteq \DSet^{*}$ and
$\LT_\omega = (\TSet_{\omega},\Lab_{\omega})$ be the omega-expansion of $\LT$. Then, $\LT$ and $\LT_{\omega}$ are bisimilar.
Moreover, the binary relation $\RSet\subseteq \TSet_{\omega}\times \TSet$ consisting of the pairs $(w,f_{\DSet}(w))$ such that
$w\in\TSet_{\omega}$ is a bisimulation between $\LT_{\omega}$ and $\LT$.
\end{remark}

 Given a  \HFTA $\AutName$ in normal form, the following lemma establishes the relation between the languages accepted by $\AutName$ and its symmetric
counterpart $\BI(\AutName)$.

\begin{lemma}\label{lemma:ResultOnSymmetricTransform} Let $\AutName$ be an \HFTA in normal form over $2^{\Prop}$,
 $\LT=(\TSet, \Lab)$ be a Kripke tree,  $\LT_\omega = (\TSet_{\omega},\Lab_{\omega})$ be the omega-expansion of $\LT$.  Then,
 \begin{equation}\label{eq:ResultOnSymmetricTransform}
   \LT_{\omega}\in \LangFun(\AutName) \text{ iff } \LT\in\LangFun(\BI(\AutName))
 \end{equation}
 \end{lemma}
 \begin{proof}
Let $\AutName = \tuple {2^{\Prop}} {\SttSet} {\trnFun} {\isttElm} {\Family} {\Family_\exists} {\Omega}$
  and  $\TSet\subseteq \DSet^{*}$ for a given non-empty set $\DSet$ of directions.
  In the proof, we exploit the mapping $f_{\DSet}$. Recall that
  $\TSet_{\omega}= f_{\DSet}^{-1}(\TSet)$ and $\Lab_{\omega}(w)= \Lab(f_{\DSet}(w))$ for each node $w\in \TSet_{\omega}$. \vspace{0.1cm}

 We first prove the left-right direction of (\ref{eq:ResultOnSymmetricTransform}).  Assume that $\LT_{\omega}\in \LangFun(\AutName)$. Hence, there exists an accepting run $r_{\omega}=(\TSet_{r_{\omega}},\Lab_{r_{\omega}})$ of $\AutName$ over
  $\LT_{\omega}$. Let $r=(\TSet_{r_{\omega}},\Lab_r)$ where $\Lab_r$ is the $\SttSet\times \TSet$-labelling defined as follows for each $\tau\in \TSet_{r_{\omega}}$:  if $\Lab_{r_{\omega}}(\tau)= (\sttElm,w)$ (note that $w\in \TSet_{\omega}$), then $\Lab_r(\tau)=(\sttElm,f_{\DSet}(w))$. We show that $r$ is an accepting run of $\BI(\AutName)$ over $\LT$. Hence, the result follows. By construction, it suffices to show that  $r$ is a run of  $\BI(\AutName)$ over $\LT$. Evidently,
  $\Lab_r(\varepsilon)=(\isttElm,\varepsilon)$. Now, let $\tau\in \TSet_{r_{\omega}}$. Then, $\Lab_{r_{\omega}}(\tau)= (\sttElm,w)$ and $\Lab_r(\tau)= (\sttElm,f_{\DSet}(w))$ for some
  $w\in\TSet_{\omega}$ and $\sttElm\in\SttSet$. Let $\SSet'$  be the set of children of $w$ in $\TSet_{\omega}$,   $\SSet$ be the set of children of
  $f_{\DSet}(w)$ in $\TSet$, and $f$ be the mapping assigning to each node $w'\in\SSet'$ the node $f_{\DSet}(w')$ of $\TSet$. By construction, $f$ is a surjective mapping from $\SSet'$ to $\SSet$ such that for each $v\in\SSet$, $f^{-1}(v)$ is infinite.
  Hence, by construction, for each child $\tau'$ of $\tau$ in
  $\TSet_{r_{\omega}}$, $\Lab_r(\tau')$ is of the form $(\sttElm',v)$ for some $v\in \SSet$.
  Let $\ISet'$  be the mapping associating to each node $w'\in \SSet'$, the set of states $\sttElm'\in \SttSet$ such that for some child $\tau'$ of $\tau$ in
  $\TSet_{r_{\omega}}$, $\Lab_{r_{\omega}}=(\sttElm',w')$. Since $r_{\omega}$ is a run of $\AutName$ over $\LT_{\omega}$, the one-step interpretation
  $(\SSet',\ISet')$ is a model of $\trnFun(\sttElm,\Lab_{\omega}(w))$. Let
  $\ISet$  be the mapping associating to each node $v\in \SSet$, the set of states $\sttElm'\in \SttSet$ such that for some child $\tau'$ of $\tau$ in
  $\TSet_{r_{\omega}}$, $\Lab_r=(\sttElm',v)$. We show that the one-step interpretation $(\SSet,\ISet)$ is a model of $\BI(\trnFun)(\sttElm,\Lab_{\omega}(w))$. Hence, being
  $\Lab_{\omega}(w)=\Lab(f_{\DSet}(w))$, the result follows. By construction, $\ISet=(\ISet')_{f^{-1}}$. We distinguish two cases:
  \begin{itemize}
    \item $\sttElm$ is not universal: in this case, $\BI(\trnFun)(\sttElm,\Lab_{\omega}(w))=\BI(\theta)$ where $\theta = \trnFun(\sttElm,\Lab_{\omega}(w))$.
    Thus, since $(\SSet',\ISet')\models \theta$ and $\ISet=(\ISet')_{f^{-1}}$, by Lemma~\ref{lemma:SimpleFormFO}(3), it follows that
    $(\SSet,\ISet)\models \BI(\theta)$, and the result follows.
    \item $\sttElm$ is universal:  in this case, $\BI(\trnFun)(\sttElm,\Lab_{\omega}(w))=\widetilde{\BI(\widetilde{\theta})}$ where $\theta = \trnFun(\sttElm,\Lab_{\omega}(w))$.
    Thus, since $(\SSet',\ISet')\models \theta$ and $\ISet=(\ISet')_{f^{-1}}$, by applying Lemma~\ref{lemma:SimpleFormFO}(4) and being $\widetilde{\widetilde{\theta}}=\theta$,   the result holds in this case too.
  \end{itemize} \vspace{0.2cm}

 For the right-left direction of (\ref{eq:ResultOnSymmetricTransform}), assume that $\LT\in\LangFun(\BI(\AutName))$. Hence, there exists an accepting run $r=(\TSet_r,\Lab_r)$ of  $\BI(\AutName)$ over
  $\LT$. Let $\HSet$ be a non-empty set of directions such that $\TSet_r\subseteq \HSet^{*}$ and
   $\TSet_{r_\omega} = f_{\HSet}^{-1}(\TSet_r)$ (the omega-expansion of $\TSet_r$).
    Moreover, let $\Lab_{r_\omega}$ be any $\SttSet\times \TSet_{\omega}$-labelling of $\TSet_{r_\omega}$ satisfying the following property for each node $\tau\in \TSet_{r_\omega}$:
   \begin{itemize}
    \item if $\Lab_r(f_{\HSet}(\tau))=(\sttElm',w)$ (note that $w\in\TSet$), then $\Lab_{r_\omega}(\tau)=(\sttElm',w')$ for some $w'\in \TSet_{\omega}$
    such that $f_{\DSet}(w')=w$.
    \item if $\Lab_{r_\omega}(\tau)=(\sttElm',w')$, then for each child $\tau'$ of $\tau$ in $\TSet_{r_\omega}$,
     $\Lab_{r_\omega}(\tau')$ is of the form $(\sttElm'',w'')$ for some child $w''$ of $w'$ in $\TSet_{\omega}$.
  \end{itemize}
    By construction, such a labelling $\Lab_{r_\omega}$ exists. Moreover, by reasoning as for left-right direction of (\ref{eq:ResultOnSymmetricTransform}) and by applying
    Properties~1 and~2 of Lemma~\ref{lemma:SimpleFormFO}, we deduce that $r_{\omega}=(\TSet_{r_\omega},\Lab_{r_\omega})$ is an accepting run of $\AutName$ over $\LT_{\omega}$, and the result follows.
  \end{proof}

For a  symmetric basic formula
$\theta(\TypeSet_\exists,\TypeSet_\forall)$ over $\SttSet$, let $\theta_{\GSym}(\TypeSet_\exists,\TypeSet_\forall)$ be the symmetric graded $\SttSet$-constraint
defined as follows, where
\[
\bigwedge_{\ASet\in \TypeSet_\exists}(\DMod \bigwedge_{\sttElm\in\ASet}\sttElm) \wedge \BMod(\bigvee_{\BSet\in  \TypeSet_\forall}\bigwedge_{\sttElm\in\BSet}\sttElm)
\]
note that if $\ASet$ is empty, then the expression $\DMod \bigwedge_{\sttElm\in\ASet}\sttElm$ is for $\true$ (as usual the empty conjunction is $\true$ and
the empty disjunction is $\false$). Moreover, if the Boolean formula $\bigvee_{\BSet\in\TypeSet_\forall}\bigwedge_{\sttElm\in\BSet}\sttElm$ is equivalent to $\true$ (resp., $\false$),
then the expression $\BMod(\bigvee_{\BSet\in \TypeSet_\forall}\bigwedge_{\sttElm\in\BSet}\sttElm)$ is for $\true$ (resp., $\false$). 

\begin{remark}\label{remark:SymmetricBasicFormulas} $\theta(\TypeSet_\exists,\TypeSet_\forall)$ and $\theta_{\GSym}(\TypeSet_\exists,\TypeSet_\forall)$ are equivalent.
Moreover, if  $\theta(\TypeSet_\exists,\TypeSet_\forall)$ is $\SttSet'$-functional in one direction, then $\theta_{\GSym}(\TypeSet_\exists,\TypeSet_\forall)$
is of the form $\DMod \sttElm$ or   $(\DMod \sttElm) \wedge \xi$ such that $\xi$ is a conjunction of atoms which refer only to elements in $\SttSet\setminus \SttSet'$.
\end{remark}

We now establish the following crucial result.

 \begin{proposition}\label{prop:BisimulationInvariantHFATA}  Given an \HFTA $\AutName$ in normal form,  one can construct a symmetric \HGTA $\AutName_S$ such that
 whenever $\LangFun(\AutName)$ is bisimulation-closed, then $\LangFun(\AutName)= \LangFun(\AutName_S)$.
\end{proposition}
\begin{proof}
For a disjunction $\bigvee_{j\in\JSet}\theta_j$ of symmetric basic formulas $\theta_j$,
 $ (\bigvee_{j\in\JSet}\theta^j)_{\GSym}$ denotes the symmetric graded constraint $ \bigvee_{j\in\JSet}\theta^j _{\GSym}$, where
 $\theta^j _{\GSym}$ is the symmetric graded constraint associated with $\theta^j$.
Let $\AutName = \tuple {2^{\Prop}} {\SttSet} {\trnFun} {\isttElm} {\Family} {\Family_\exists} {\Omega}$ (without loss of generality, we assume that the
input alphabet is of the form $2^{\Prop}$). Define
 $\AutName_S = \tuple {2^{\Prop}} {\SttSet} {\trnFun_S} {\isttElm} {\Family} {\Family_\exists} {\Omega}$ where for each $(\sttElm,a)\in\SttSet\times 2^{\Prop}$,
 $\trnFun_S(\sttElm,a)$ is defined as follows:
 \begin{itemize}
  \item $\sttElm$ is transient or existential: by Definition~\ref{def:InvariantBisumulationHFATAMapping}, $\BI(\trnFun)(\sttElm,a)$ is a disjunction of symmetric basic formulas which are $\SttSet_i$-separated in one direction if the component $\SttSet_i$ of $\sttElm$ is existential. In this case, we set
        $\trnFun_S(\sttElm,a)= (\BI(\trnFun)(\sttElm,a))_{\GSym}$.
  \item the component $\SttSet_i$ of $\sttElm$ is universal: by Definition~\ref{def:InvariantBisumulationHFATAMapping},   $\BI(\trnFun)(\sttElm,a)$ is of the form $\widetilde{\theta}$, where
  $\theta$ is   a  disjunction of symmetric basic formulas which are $\SttSet_i$-functional in one direction. In this case,
  we set $\trnFun_S(\sttElm,a)= \widetilde{(\theta)_{\GSym}}$.
\end{itemize}
By construction and Remark~\ref{remark:SymmetricBasicFormulas}, $\AutName_S$ is a symmetric \HGTA such that $\LangFun(\AutName_S)= \LangFun(\BI(\AutName))$.
Let $\LangFun(\AutName)$ be bisimulation-closed. We show that $\LangFun(\AutName)= \LangFun(\BI(\AutName))$. Hence, the result follows.
 Let $\LT=(\TSet, \Lab)$ be a Kripke tree where $\TSet\subseteq \DSet^{*}$ and
$\LT_\omega = (\TSet_{\omega},\Lab_{\omega})$ be the omega-expansion of $\LT$. By Lemma~\ref{lemma:ResultOnSymmetricTransform},
it holds that  $\LT_{\omega}\in \LangFun(\AutName)$ iff $ \LT \in \LangFun(\BI(\AutName))$. Moreover,
since $\LT$ and $\LT_{\omega}$ are bisimilar (Remark~\ref{remark:OmegaExpansion}) and
$\LangFun(\AutName)$ is bisimulation-invariant, it holds that $\LT_{\omega}\in \LangFun(\AutName)$ iff $ \LT \in \LangFun(\AutName)$.
Hence, we obtain that $\LT\in \LangFun(\AutName)$ iff $ \LT \in  \LangFun(\BI(\AutName))$, and by the arbitrariness of
$\LT$, the result follows.
\end{proof}

We can now prove Theorem~\ref{theo:BisimulationInvariantMCL}.

\theoBisimulationInvariantMCL*
\begin{proof}
Let $\varphi$ be an  \MCL sentence which is bisimulation-invariant. By Theorem~\ref{theo:FromMCLtoHFATA} and  Proposition~\ref{prop:BisimulationInvariantHFATA},
one can construct a symmetric \HGTA $\AutName$ such that $\LangFun(\AutName)=\LangFun(\varphi)$. Moreover, being symmetric \HGTA a subclass of \HFTA, by
Theorem~\ref{theo:FromHFATAtoMCL}, a symmetric \HGTA can be effectively converted into an equivalent \MCL sentence. Thus, since \CDL and symmetric \HFTA are bisimulation-invariant and expressively equivalent in a constructive way (Theorem~\ref{theo:FromHGATAtoCDL} and Theorem~\ref{theo:FromCDLtoHGATA}), the result follows.
\end{proof}


\end{document}